\documentclass[a4paper,11pt]{article}
\pdfoutput=1 % if your are submitting a pdflatex (i.e. if you have
\usepackage{jheppub} % for details on the use of the package, please
\usepackage[T1]{fontenc} % if needed

\usepackage{epsfig,amsmath} 
\usepackage{hepnames,hepunits}
\usepackage{bm}

\def\lsim{\mathrel{\rlap{\lower4pt\hbox{\hskip1pt$\sim$}}
    \raise1pt\hbox{$<$}}}                % less than or approx. symbol
\def\gsim{\mathrel{\rlap{\lower4pt\hbox{\hskip1pt$\sim$}}
    \raise1pt\hbox{$>$}}}     \bf          % greater than or approx. symbol

\def\ltap{\raisebox{-.6ex}{\rlap{$\,\sim\,$}} \raisebox{.4ex}{$\,<\,$}} 
\def\gtap{\raisebox{-.6ex}{\rlap{$\,\sim\,$}} \raisebox{.4ex}{$\,>\,$}} 

\def\tmd{TMD merging}

\title{Multi-jet Merging with TMD Parton Branching}

\author[a]{A.~Bermudez~Martinez,}
\author[b,c,d]{F.~Hautmann,}
\author[b]{M.L.~Mangano}

\affiliation[a]{Deutsches Elektronen-Synchrotron DESY, Notkestr. 85, 22607 Hamburg, Germany}
\affiliation[b]{CERN, Theoretical Physics Department, CH 1211 Geneva 23}
\affiliation[c]{Elementaire Deeltjes Fysica, Universiteit Antwerpen, B 2020 Antwerpen}
\affiliation[d]{RAL, Chilton OX11 0QX and University of Oxford, Oxford   OX1 3PU}

\emailAdd{armando.bermudez.martinez@desy.de}
\emailAdd{hautmann@thphys.ox.ac.uk}
\emailAdd{michelangelo.mangano@cern.ch}

\preprint{CERN-TH-2022-131  \;\;\;  DESY-22-133}
\abstract{One of the main theoretical systematic uncertainties in studies of final states with large jet multiplicities at high-energy hadron colliders  
is associated with the merging of QCD parton showers and hard-scattering matrix elements. 
We present a method to incorporate the physics of transverse momentum recoils due to initial-state shower evolution into multi-jet merging algorithms 
by using the concept of transverse momentum dependent (TMD) distributions and the associated parton branching.  
We   investigate the dependence on the merging scale  and    
illustrate the impact of the new method at the level of both exclusive and inclusive final-state observables by studying differential 
jet rates, transverse momentum spectra and multiplicity distributions, using vector boson + jets events at the LHC as a case study.
}

\keywords{Jets, QCD, Electroweak Interaction}

\begin{document} 
\maketitle
\flushbottom
\unboldmath

\section{Introduction} 
\label{Intro}

Events with  large multiplicities of hadronic jets in high-energy $pp$ collisions are among the most spectacular final states  observed  at the Large Hadron Collider (LHC), and will continue to be explored in future collider experiments~\cite{Azzi:2019yne,Mangano:2016jyj}, at ever increasing jet multiplicity and with increasingly high statistics. The accurate theoretical description of multi-jet  events has great practical importance, as multi-jets are used to precisely determine the properties of particles decaying to them (e.g. gauge or Higgs bosons and the top quark), to perform precision tests of strong-interaction dynamics in the Quantum Chromodynamics (QCD) sector of the Standard Model (SM), and to search for signatures of new physics beyond the Standard Model (BSM). 

Theoretical predictions for multi-jet observables have relied for the last twenty years on 
``merging''  
techniques~\cite{Catani:2001cc,Lonnblad:2001iq,Mangano:2002,Mrenna:2003if,Mangano:2006rw,Alwall:2007fs,Frederix:2012ps,Hoeche:2012yf,Lonnblad:2012ix,Bellm:2017ktr}, to 
combine matrix-element and parton-shower event generators. The former 
describe the underlying hard process with bare partons providing the primary sources for widely separated jets,  the latter 
describe the evolution of partons by radiative processes predominantly at small angles, and the two are 
sewn together,  so as to avoid either double counting or missing events, via a ``merging scheme''  and 
merging scale. 

The primary goal of the merging scheme is to ensure that the best possible approximation (fixed-order exact matrix elements or the parton shower evolution) is used at any given point of phase-space. The scheme and its parameters select the approximation to be used, and correct the event weight accordingly. Since neither of the descriptions (matrix element or parton shower) is exact, the choice of merging parameters introduces an important theoretical systematic uncertainty, discussed at 
leading order (LO) in~\cite{Mrenna:2003if,Alwall:2007fs,Mangano:2006rw,Lavesson:2008ah,Hoeche:2009rj,Hamilton:2009ne,Lonnblad:2011xx} 
and next-to-leading order (NLO) in~\cite{Frederix:2012ps,Hoeche:2012yf,Lonnblad:2012ix,Bellm:2017ktr}.  
This systematic uncertainty reflects the mismatch between the matrix-element and the parton-shower weights assigned to a given final state. The larger the mismatch, the larger the uncertainty. The phase-space regions that are mostly affected are those describing final states for which the jet multiplicity can vary under minor changes of the merging parameters. Typically, this happens if a jet is soft or close to another hard jet. A better modeling of the emission probability for such jets by the parton-shower evolution  would reduce the difference with the weight assigned to these events by the matrix-element description, reducing the mismatch and the relative systematic uncertainty.
 
Motivated by this, we explored in an earlier proposal~\cite{Martinez:2021chk} the role of  parton-shower algorithms that go beyond the approximation of small-angle, collinear emissions,  
by taking into account  transverse-momentum recoils in the initial-state  shower  through the use of  
transverse momentum dependent (TMD)~\cite{Angeles-Martinez:2015sea} 
parton distributions. 
The reason for this is twofold. On one hand, multi-jets probe phase-space regions characterized 
by 
multiple momentum scales,  in which 
TMDs control  perturbative QCD resummations~\cite{Luisoni:2015xha} to all orders in the strong  coupling (e.g., in the Sudakov region~\cite{Collins:1984kg} and high-energy 
region~\cite{Catani:1990eg}).  
On the other hand,  transverse momentum recoils 
can influence  the theoretical uncertainty associated with combining matrix-element and parton-shower contributions~\cite{Dooling:2012uw,Hautmann:2013fla},   
 and thus affect  the dependence of multi-jet cross sections on the merging scale. In this paper we further elaborate on the results of Ref.~\cite{Martinez:2021chk}, providing a more detailed account of the algorithm and presenting additional studies to demonstrate its performance.

While advanced formalisms have been developed 
to describe TMD effects in inclusive observables  
(see e.g. the recent 
 next-to-next-to-next-to-leading-order (N$^3$LO) 
calculations~\cite{Ebert:2020yqt,Luo:2019szz}  
of TMD coefficient functions, and 
third-order studies~\cite{Chen:2022cgv,Neumann:2022lft}  
of transverse momentum Drell-Yan (DY) spectrum at the LHC), 
 TMD implications on the exclusive structure of final states with high jet multiplicity 
 have only  just begun to be explored in Ref.~\cite{Martinez:2021chk}, where we introduced a 
systematic merging method to  analyze such effects in multi-jets.  We aim at obtaining results for both inclusive and exclusive multi-jet observables, accounting simultaneously for the contribution of different jet multiplicities in the final state and for the contribution of initial-state TMD evolution. 
To this end, we  employ the parton branching (PB) formulation of TMD evolution set 
out in~\cite{Hautmann:2017fcj}. Any of the existing merging algorithms can then be used, in principle, to combine samples of different parton multiplicity showered using the TMD parton branching. In this work, we focus however on the so-called MLM matching 
prescription~\cite{Mangano:2002,Mrenna:2003if,Alwall:2007fs,Mangano:2006rw}.
First results from this investigation have been presented 
in~\cite{Martinez:2021dwx}.  

The paper is organized as follows. We start in Sec.~\ref{pb-formulation} by briefly recalling 
the main features of the PB formulation of TMD evolution. In Sec.~\ref{sec:pbmlm} we describe the TMD multi-jet merging method and 
illustrate its features by presenting the results of applying the method to differential jet rates. In Sec.~\ref{sec:Zjets} we 
perform a detailed analysis of $Z$-boson production associated with multi-jets at the LHC, computing theoretical predictions for $Z$+jets observables with the 
TMD jet merging method and comparing them with LHC experimental data. In 
Sec.~\ref{sec:syst} we examine the theoretical systematic uncertainty 
 associated with the merging parameters and the dependence of theoretical predictions on the merging scale.  In 
 Sec.~\ref{sec:comp-pythia}  we study the sensitivity of our results to  final-state parton showers, and present  a comparison with {\sc Pythia} Monte Carlo results based on collinear multi-jet merging. 
 In Sec.~\ref{sec:pdfsens} we study the 
 sensitivity of our results to nonperturbative  effects in TMD and collinear parton distribution 
 functions (PDFs). 
 Sec.~\ref{sec:comm-x} includes  remarks on possible off-shell TMD effects in the evaluation of partonic matrix elements. 
 We give our conclusions in Sec.~\ref{sec:concl}.

\section{Parton branching formulation of TMD evolution} 
\label{pb-formulation}
 
 In this section we summarize the basic elements of the PB formulation of 
 TMD evolution.  
 First we recall  the evolution equations; next we discuss 
 the physical picture of  transverse momentum broadening 
 from TMD evolution, and its implications for the 
 production of multiple jets. 
 
\subsection{Evolution equation}
\label{subsec:eveqs}

 The PB method~\cite{Hautmann:2017fcj} uses the unitarity picture of parton 
evolution~\cite{Webber:1986mc}, commonly used in showering algorithms~\cite{Bengtsson:1987kr,Marchesini:1987cf},  
for both collinear and TMD parton distributions. Soft gluon emission and transverse momentum recoils are treated 
by introducing the soft-gluon resolution scale $z_M$~\cite{Hautmann:2017xtx} to separate 
resolvable and non-resolvable branchings. Sudakov form factors $\Delta_j$  are used to 
 express the probability for non-resolvable branching in a given evolution interval from 
one scale $\mu_0$ to a higher scale $\mu$, 
\begin{equation}
\label{eq1}
\Delta_j(\mu^2, \mu_0^2)  = \exp\left[-\sum_\ell \int_{\mu_0^2}^{\mu^2}\frac{\textrm{d}\mu^{\prime 2}}{\mu^{\prime 2}} \int_0^{z_M}\textrm{d}z \  
\ z \ P_{\ell j }^{(R)}\left(z,\alpha_s\right) \right]\;,   
\end{equation}
where $j$, $\ell$ are flavor indices, $z$ is the longitudinal momentum fraction, $\alpha_s$ is the strong coupling, and 
$P_{k j }^{(R)} $ are resolvable splitting functions, computable as power series expansions in $\alpha_s$. 

In this 
approach the TMD evolution equations can be written as 
\begin{eqnarray}
\label{eq2}
{A}_j( x, {\bm k}^2, \mu^2) &=& \Delta_j(\mu^2, \mu_0^{2}){A}_j( x, {\bm k}^2, \mu_0^2)
  \\ \nonumber
 &+& 
 \sum_\ell\int \frac{\textrm{d}^2{\boldsymbol \mu}^{\prime}}{\pi {\mu}^{\prime 2}}\Theta(\mu^{2}-\mu^{\prime 2})\Theta(\mu^{\prime 2}-\mu_0^{ 2})
  \\ \nonumber
 &\times& 
  \int_x^{1}{ {\textrm{d}z} \over z} \ \Theta(z_M - z )
  {  { \Delta_j(\mu^2, \mu_0^2  )  } \over 
  { \Delta_j(\mu^{\prime 2}, \mu_0^2 ) } } \ 
  \\ \nonumber
 &\times& 
  P_{j \ell }^{(R)}(z,\alpha_s((1-z)^2\mu^{\prime 2})){A}_\ell( x/z,  | {\bm k} + (1-z){\boldsymbol \mu}^\prime |^2, \mu^{\prime 2}) 
  \nonumber
  \; , 
\end{eqnarray} 
where ${A}_j( x, {\bm k}^2, \mu^2)$ is the 
TMD distribution of flavor $j$ carrying the longitudinal momentum  fraction $x$ of the hadron's momentum and  transverse momentum ${\bm k}$
at the evolution scale $\mu$,  $\mu_0$ is the initial evolution scale, and  $ \mu^\prime = \sqrt{ {\boldsymbol \mu}^{\prime 2}}$ is 
the momentum scale at which the branching occurs. 
% The functions $a(z)$ and $b(z)$  in Eq.~(\ref{eq2})  specify the  ordering variable  
%and choice of scale in the strong 
%coupling.  
The branching 
probabilities are evaluated using the form of the strong coupling 
according to angular ordering, with the scale in $\alpha_s$ given 
by the emitted transverse momentum at the branching, $q_T=(1-z) \mu$.
It was observed in~\cite{Hautmann:2017fcj,Hautmann:2017xtx} that 
%taking the angular-ordered~\cite{Webber:1986mc,Marchesini:1987cf,Catani:1990rr} form for $a(z)$ and $b(z)$ 
% gives rise 
angular 
ordering~\cite{Webber:1986mc,Marchesini:1987cf,Catani:1990rr,Hautmann:2019biw}  leads 
to well-prescribed TMDs, i.e., stable  with respect to variations of the soft-gluon resolution scale $z_M$. 
In this work we will follow this observation and will consider 
angular-ordered evolution. 

 Collinear limits  may be obtained by integrating Eq.~(\ref{eq2})  over all transverse momenta. 
%For  soft-gluon resolution  
%$z_M \to 1$,   integrating  Eq.~(\ref{eq2}) over all transverse momenta 
%yields    collinear PDFs satisfying  
%DGLAP evolution equations~\cite{Gribov:1972ri,Altarelli:1977zs,Dokshitzer:1977sg}. 
%The convergence to DGLAP 
For $z_M \to 1$ and $\alpha_s \to \alpha_s(\mu^{\prime 2})$,  
the convergence to 
collinear PDFs satisfying  DGLAP evolution equations~\cite{Gribov:1972ri,Altarelli:1977zs,Dokshitzer:1977sg}
 has  been 
verified numerically~\cite{Hautmann:2017xtx} 
at LO and NLO against the evolution program~\cite{Botje:2010ay} 
  at the    level of better   than 1\% over a range of five orders of magnitude both in $x$ and in $\mu$.
%On the other hand,  by using angular ordering with $ \mu^\prime$-dependent soft-gluon 
%resolution $z_M (\mu^\prime) = 1 - q_0 / \mu^\prime$~\cite{Hautmann:2019biw}, with $q_0$ being the minimum transverse momentum with which a parton can be resolved, 
% Eq.~(\ref{eq2})  returns,  once it 
%is integrated over all transverse momenta,    the  coherent-branching equation of~\cite{Catani:1990rr}. 

Besides the collinear limits, Eq.~(\ref{eq2}) can be used at unintegrated level for Monte Carlo event simulation 
of TMD effects~\cite{Baranov:2021uol}.   
 The PB method enables the 
explicit calculation of the kinematics at every branching 
vertex  
once the evolution scale is specified in terms of kinematic variables. 
If the TMD distribution $ {A}_j( x, {\bm k}^2, \mu^2)$ evaluated at the scale 
$\mu^{2}$ is known, the corresponding TMD parton shower can be 
generated by backward evolution, 
starting from the hard scattering 
process. 

The basic elements of the initial state shower in the backward evolution scheme are described in Ref.~\cite{Baranov:2021uol}. 
By a method analogous to that used in the case of collinear 
showers~\cite{Bengtsson:1986gz,Gieseke:2003rz,Platzer:2011dq},  
the Sudakov form factor for backward evolution is obtained 
from the solution of Eq.~(\ref{eq2}). This Sudakov factor  
depends on the ratio of two TMD distributions. 
The kinematical mapping between the branching variables 
and the physical variables is dictated by the angular ordering, 
$q_T = \mu (1 - z) $, where $\mu$ and $z$ are the branching 
scale and longitudinal momentum transfer, and $q_T$ is the 
final-state transverse momentum~\cite{Hautmann:2017fcj}. The 
evolution of the 
transverse momentum in the backward shower follows that  
 of the TMD distribution. One of the features characterizing the 
 TMD shower compared to collinear shower algorithms, in 
 particular,  concerns  the treatment of the 
soft-gluon resolution scale $z_M$. The resolution scale 
in the shower matches the resolution scale  in the TMD 
evolution equation (\ref{eq2}). This can be contrasted with 
the collinear shower case, in which  parton distributions 
evolve via DGLAP equations, while  the resolution scale affects 
the shower: see discussions in Refs.~\cite{Nagy:2020gjv,Nagy:2022bph,Gellersen:2020}. 

A TMD final-state shower is not available yet. Current 
applications of the PB method 
use standard  {\sc Pythia}~\cite{Sjostrand:2014zea,Sjostrand:2006za} or {\sc Herwig}~\cite{Bellm:2015jjp,Corcella:2002jc} final-state showering algorithms. 
These are applied both to the final-state partons 
and to the timelike splittings of partons radiated from the 
initial state. A study of final-state showering  in 
PB calculations is carried out in Ref.~\cite{Yang:2022qgk}. 

The TMD forward evolution (\ref{eq2}) with 
backward shower for the initial state and the  
final-state shower 
will be used 
for multi-jet merging in this work, as will be described in the next section. 

Numerical solutions to Eq.~(\ref{eq2}) 
have been used in~\cite{Martinez:2018jxt} 
to obtain  TMD densities at NLO from fits to precision deep inelastic scattering (DIS) HERA data~\cite{Abramowicz:2015mha}, 
performed 
using the   fitting platform   \verb+xFitter+~\cite{xFitter:2022zjb,Alekhin:2014irh} and the numerical techniques  
developed in~\cite{Hautmann:2014uua} to treat the transverse momentum dependence in the fitting procedure. 
In~\cite{Martinez:2018jxt} two sets of PB TMD distributions are described: 
Set~1, which uses the evolution scale as argument  in the running coupling $\alpha_s$, similar to what 
is used in HERAPDF 2.0 NLO~\cite{Abramowicz:2015mha}, and Set~2, which uses the transverse momentum in the evolution of $\alpha_s$.
These PB TMDs have been combined 
 with  NLO calculations of Drell-Yan (DY) production in the 
  {\sc MadGraph5\_aMC@NLO}~\cite{Alwall:2014hca} framework    
to compute  vector-boson transverse momentum spectra at LHC energies~\cite{Martinez:2019mwt} 
and fixed-target energies~\cite{Martinez:2020fzs}.  The theoretical predictions thus obtained are shown to provide a 
good description of DY transverse momentum measurements over a wide range in energies and masses, from the 
LHC~\cite{Aad:2015auj,Sirunyan:2019bzr} to 
PHENIX~\cite{Aidala:2018ajl} to R209~\cite{Antreasyan:1981eg} to NuSea~\cite{Webb:2003ps,Webb:2003bj}.  

By the same NLO + PB TMD method, angular correlations have been examined in 
di-jet~\cite{Abdulhamid:2021xtt} and DY + jet~\cite{Yang:2022qgk} final states  
at large transverse momenta. A good description of the LHC angular 
measurements~\cite{CMS:2017cfb,CMS:2019joc} is found.  

The PB approach has recently 
been used~\cite{Martinez:2022gsz} to 
make a determination   of the nonperturbative 
rapidity evolution kernel (see e.g.~\cite{Hautmann:2021ovt})  
in the 
factorization formalism~\cite{Collins:1981va,Collins:1984kg} for DY at 
low transverse momenta. The calculation is based on the event 
generator described in Ref.~\cite{Baranov:2021uol}. The results may be directly compared with 
extractions of the rapidity kernel 
from DY experimental data, e.g.~\cite{Hautmann:2020cyp},  
and with determinations from lattice calculations, e.g.~\cite{Ebert:2019tvc}. 

An extension of the PB evolution equations has been proposed 
in~\cite{Hautmann:2022xuc} to take into account the 
transverse momentum dependence  of splitting functions at each branching, 
as defined from high-energy factorization~\cite{Catani:1994sq}. 
This extension is relevant for phenomenology at the highest energy 
frontier, where the small-$x$ (Regge) phase space~\cite{Catani:1990eg} 
opens up and high-energy resummations are called for. 
See Refs.~\cite{Jung:2010si,Jung:2000hk,Andersen:2011zd,Hoeche:2007hlb,Golec-Biernat:2007tjf,Chachamis:2015zzp,Andersson:1995jt,Orr:1997im,Schmidt:1996fg} for shower Monte Carlo algorithms including the 
small-$x$ region. 
In the present 
work, we will not consider this generalized evolution, and we will 
limit ourselves to using PB evolution in  the 
form discussed in Refs.~\cite{Hautmann:2017fcj,Hautmann:2017xtx}.

\subsection{Transverse momentum broadening}
\label{subsec:broadening}

\begin{figure}[hbtp]
  \begin{center}
	\includegraphics[width=.49\textwidth]{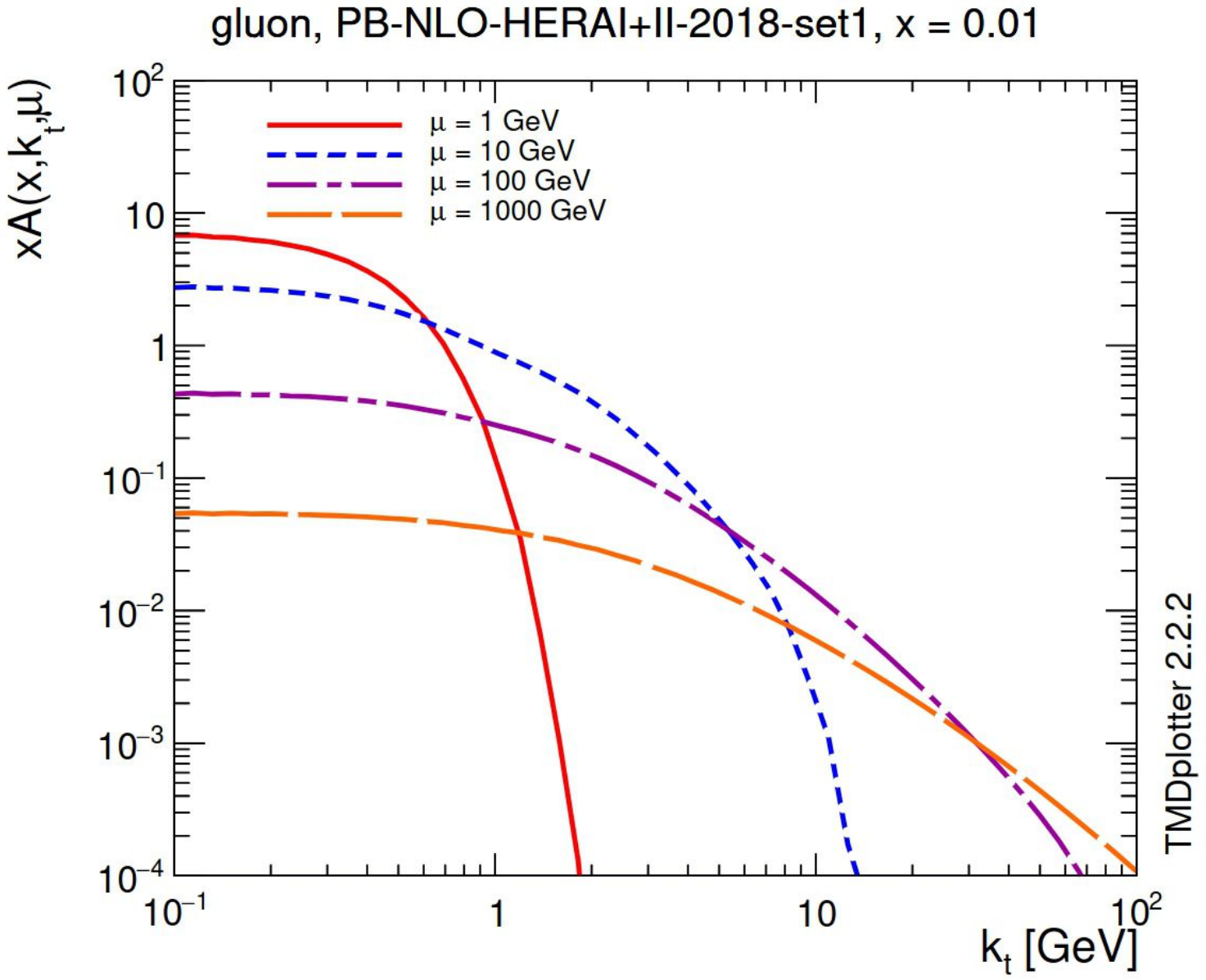}
  \caption{TMD  gluon distributions versus  $k_T$ 
  for momentum fraction $x= 10^{-2}$ and different values of the evolution scale $\mu$. 
  The PB TMD Set~1 of Ref.~\cite{Martinez:2018jxt} is used.}
\label{fig-gTMD}
\end{center}
\end{figure}

The evolution equations  (\ref{eq2})  imply  that a 
TMD distribution characterized by a given width in transverse momentum 
($k_T = | {\bm k} |$) at the scale $\mu_0$ will be subject to a broadening 
in $ k_T $ as the evolution scale $\mu$ increases, as a result of 
multi-gluon emissions. An example, taken from the TMD parton distribution  
library~\cite{Abdulov:2021ivr,Hautmann:2014kza}, is shown 
in Fig.~\ref{fig-gTMD}.

Our  primary interest in this work is to analyze the  consequences of the 
large transverse momentum tails, induced by TMD evolution,  on 
the structure of  final states with multiple jets in high-energy collisions. 

Consider a final state characterized by hard scale $\mu$, e.g., the 
transverse momentum of the hardest jet in  the event. 
To assess the contribution to the production of an extra jet with 
transverse momentum 
$p_T <  \mu$ from  the high-$k_T$ tail of the TMD distribution 
evolved to scale $\mu$, it is useful to introduce 
 integral TMD distributions $a_j$, obtained from 
${A}_j$   in  Eq.~(\ref{eq2})  by $k_T$-integration as follows 
 \begin{equation} 
\label{eq2integraltmd}
a_j ( x, {\bm k}^2, \mu^2) = 
\int    { {d^2 {\bm k }^\prime} \over \pi} \  {\cal A}_j ( x , {\bm k }^{\prime 2} , \mu^2) \  \Theta ( {\bm k }^{\prime 2}-{\bm k }^{ 2} ) 
\; . 
\end{equation} 
The distribution $a_j$ evaluated at $k_T = 0$ gives the fully integrated initial-state distribution. The fractional contribution to 
$a_j$  from the tail  above transverse momentum $k_T$, with $k_T$ of the order of 
the jet $p_T$,  is given  by the ratio 
\begin{equation}
\label{normalized-itmd} 
R_j ( x, {\bm k}^2, \mu^2) = a_j ( x, {\bm k}^2, \mu^2) / a_j ( x, { 0}, \mu^2)  \; .  
\end{equation} 

\begin{figure}[hbtp]
  \begin{center}
	\includegraphics[width=.49\textwidth]{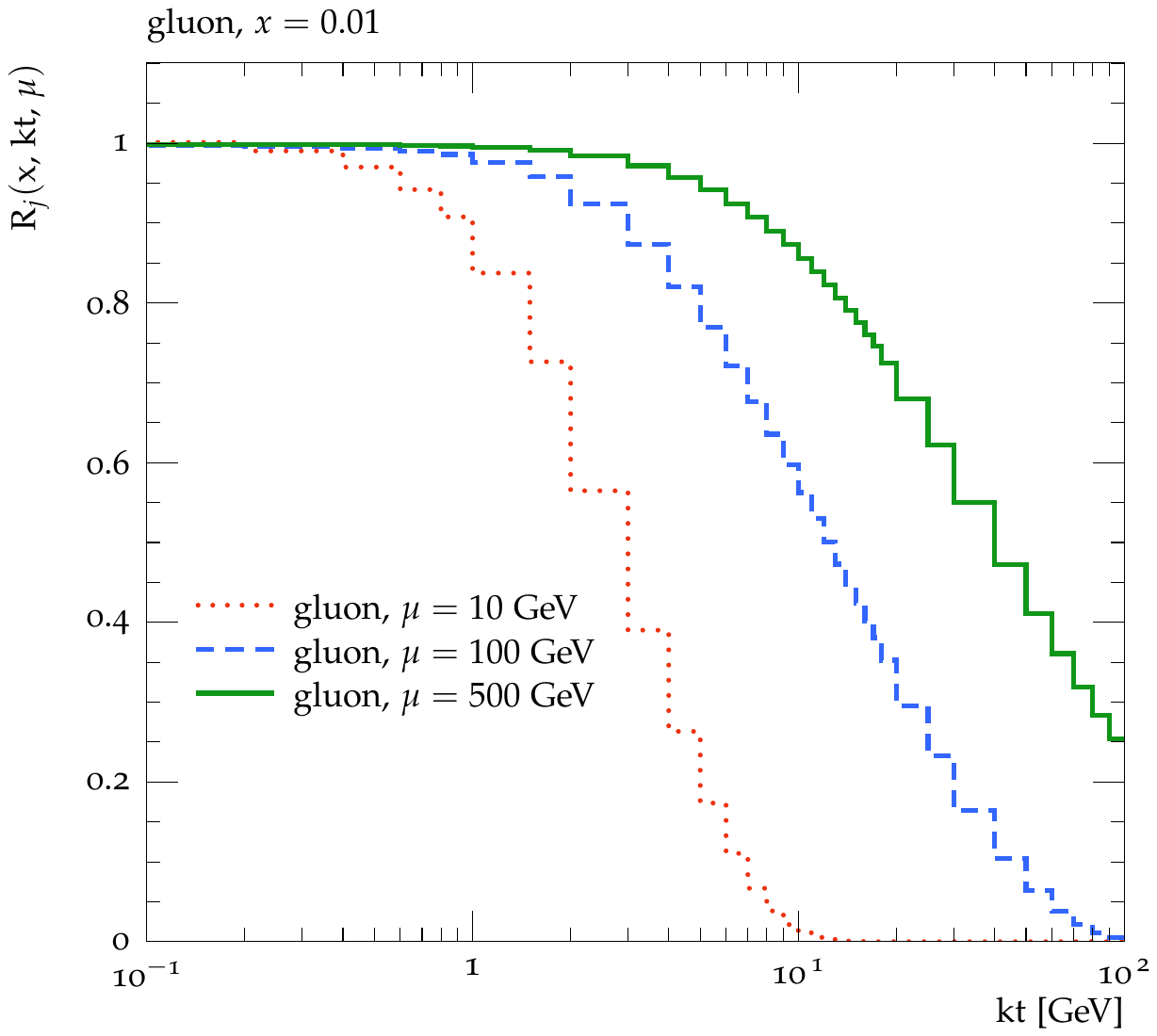}
  \caption{The $| {\bm k} |$ spectrum of the integral  
  TMD  gluon distribution in Eq.~(\ref{eq2integraltmd}), normalized to ${\bm k}=0$,   
  for momentum fraction $x= 10^{-2}$ and different values of the evolution scale $\mu$. 
  The PB TMD Set~2~\cite{Martinez:2018jxt} is used.}
\label{fig-iTMD-vs-kTmin}
\end{center}
\end{figure}

In Fig.~\ref{fig-iTMD-vs-kTmin} we illustrate the $k_T$ dependence of Eq.~(\ref{eq2integraltmd}) 
by  showing the  integral TMD gluon        
 distribution $a_g ( x, {\bm k}^2, \mu^2) $, normalized to    ${\bm k}=0$,  
 obtained from the PB TMD Set~2~\cite{Martinez:2018jxt} for $x=10^{-2}$ and various values of $\mu$. We note, for example, that for $\mu=100$ (500) GeV, there is a 30\% probability that the gluon has developed a transverse momentum larger than 20 (80) GeV.

The results in  Fig.~\ref{fig-iTMD-vs-kTmin}  illustrate that,  
while the distribution 
is falling off at large $k_T$,  for the jet transverse scales 
observed in  the LHC  kinematics  the contribution from the 
region $p_T \ltap k_T \ltap \mu$ is non-negligible  compared  
to the contribution of an extra parton perturbatively emitted 
via  hard-scattering matrix elements.  
Both contributions need to be taken into account. 
To avoid the double counting between 
the extra jet emission induced 
by the TMD initial-state evolution and 
that arising from the inclusion of a higher-order matrix element, 
a merging methodology  is needed. 
 This is the subject of the next section.

\section{TMD multi-jet merging}
\label{sec:pbmlm}

In this section we describe the basic elements of the 
TMD multi-jet merging method.  
We illustrate its main features  and apply it to the computation of 
 differential jet rates (DJRs). 

\subsection{Basic elements}
\label{subsec:elements}

Parton showers are able to simulate multi-jet topologies. If an $n$-jet configuration is available at the matrix element level,  the corresponding $(n+1)$-jet configuration can be generated by a parton shower emission. However,  the corresponding accuracy  is limited to  emissions   in the soft and collinear phase space regions. A hard, wide-angle emission from an $n$-jet configuration will be better described by an $(n+1)$-jet matrix element calculation. The naive sum of the $n$- and $(n+1)$-jet calculations would not be correct due to regions of the $(n+1)$-jet phase space which  would be doubly populated, both by the $(n+1)$-jet matrix elements and by the parton shower emissions off the $n$-jet configuration. Furthermore the $(n+1)$-jet partial cross section would be unstable, with its value  strongly depending on the phase space cut which  prevents the extra parton from approaching the divergent soft or collinear regions. 

The main purpose of a jet merging algorithm is to enforce exclusivity of the $(n+m)$-jet matrix element calculations above a merging scale $\mu_m$, except for the highest available multiplicity which will remain inclusive. For instance, if the $n$-jet matrix element configuration is made exclusive, double counting with the $(n+1)$-jet matrix element configuration will be avoided. This can be achieved by means of Sudakov form factors which will suppress emissions from the $n$-jet configuration. When properly applied to the $(n+1)$-jet matrix element configuration,  the Sudakov factors will also suppress the divergent phase space region, making the partial $(n+1)$-jet cross section stable.

Using the Sudakov factor in Sec.~\ref{pb-formulation}, 
the $(n+1)$-jet cross section $d\sigma_{n+1}^{PS}$ given by the parton shower approximation from an $n$-jet configuration can be expressed as 
\begin{equation}
\label{eq3}
d\sigma_{n+1}^{PS} = \sigma_n d\phi_{rad} \mathcal{R}^{PS}(\mu^2)\Delta(\mu_{max}^{2},\mu^{2}),
\end{equation} 
where $\mu^2$ refers to the scale of the emission, $d\phi_{rad}=d\mu^2/\mu^2 dz /z$ is the radiation phase space, $\mu^2_{max}$ is the maximum scale in the process (set at the value of the renormalization scale selected for a given process and final state, a parameter labeled as SCALUP in the conventional Les Houches file format~\cite{Alwall:2006yp}), 
$\mathcal{R}^{PS}$ is the parton shower emission probability. For simplicity the flavor indices and the $z_M$ parameter are not shown in Eq.~(\ref{eq3}). Similarly one 
can construct the $(n+1)$-jet cross section $d\sigma_{n+1}$ using the $(n+1)$-jet matrix element, 
\begin{equation}
\label{eq4}
d\sigma_{n+1} = \sigma_n d\phi_{rad} \mathcal{R}(\mu^2)\Delta(\mu_{max}^{2},\mu^{2}),
\end{equation}    
where $\mathcal{R}=\sigma_{n+1}(\mu^2)/\sigma_n$ is the real emission probability. A merging algorithm would consist in using Eq.~(\ref{eq3}) for emissions below an arbitrary merging scale $\mu_m$ and Eq.~(\ref{eq4}) for emissions above $\mu_m$. The residual discontinuity at the merging scale can be further reduced by reweighting the strong coupling in 
Eq.~(\ref{eq4}) to the corresponding shower history.

The inclusion of TMD evolution effects in a merging algorithm can be performed 
according to the procedure that we describe next. This  extends 
the standard 
MLM matching~\cite{Mangano:2002,Mrenna:2003if,Mangano:2006rw,Alwall:2007fs} 
(which we recall here for completeness).   The merging procedure consists of 
the following items. 
\begin{enumerate}
	\item Evaluate the jet cross sections $\sigma_{n}$ with $n=0,1,..,N$, and generation cut $\mu_c<\mu_m$. The generation cut $\mu_c$ gives the 
lower threshold for the outgoing partons' transverse momentum.   In the calculations that follow we  take the generation cut to be $\mu_c=15$ GeV.  
	Parton samples are generated in any given kinematic configuration (e.g. within a certain pseudorapidity range) according to the matrix elements, with a probability proportional to the respective cross section.
	\item Reweight the strong coupling in the matrix elements according to the values from the corresponding shower history, as prescribed in Ref.~\cite{Catani:2001cc} and implemented in the MLM algorithm~\cite{Alwall:2007fs}. 
	\item For each of the two initial state partons of a given event, extract values of ${\bm k_i}$ ($i=1,2$) distributed according to the evolution Eq.~(\ref{eq2}), setting $\mu^2=\mu^2_{max}$. If $ {\bm k_i}^2 \geq \mu^2_{min}$ for any $i=1,2$, the event is rejected, and its contribution to the sample cross section is subtracted. $\mu^2_{min}$ corresponds to the minimum energy scale for the event, defined by $\mu^2_{min}={\mbox{min}}\{ p^2_{ti}, p^2_{tij} \}$ where $i,j=1,..,n$,  $p_{ti}$ is the transverse momentum of  parton $i$, and $p^2_{tij}$ ($i\neq j$) measures the relative transverse momentum between partons $i$, $j$.  In the case of $Z$+0 jets, we selected $\mu_{min}=m_Z$.  The overall kinematics of the final state is then reconstructed by including the transverse boost induced by the transverse momentum ${\bm k}={\bm k_1}+{\bm k_2}$.
	\item Initial-state partons of the generated events are showered using the backward shower evolution that 
	corresponds to Eq.~(\ref{eq2}). Final-state partons are showered using 
	the standard  {\sc Pythia}~\cite{Sjostrand:2014zea,Sjostrand:2006za} or {\sc Herwig}~\cite{Bellm:2015jjp,Corcella:2002jc} showers. 
	\item The MLM prescription~\cite{Mangano:2006rw,Alwall:2007fs}   is applied by matching the kinematics of the showered event and that of the parton-level event after the ${\bm k}$ boost. This differs from the standard MLM procedure, where the final state of the parton-level event has no overall transverse momentum. We recall here the steps of the MLM matching:  
	     \begin{itemize}
	          \item Partons produced in the showered events are clustered with a jet algorithm (e.g. the anti-$k_t$ one~\cite{Cacciari:2008gp}) defined by a ``cone size'' and a transverse energy. The resulting jets are characterized by the merging parameters that should be kept fixed for all samples: the cone size $R_\text{clus}$, minimum transverse energy $E_{clus}$ (merging scale) and  maximum pseudo-rapidity within which jets are recostructed, $\eta_{clusmax}$. The systematic uncertainty of the method can be estimated by varying these parameters.
	          \item For a given multiplicity $n$, start from the hardest parton and select its closest jet in $R$. The parton and the jet are matched if $R< c R_\text{clus}$. We take $c = 1.5$.  After removing the matched jet from the list the matching procedure proceeds for the remaining jets and partons. The event is matched when each parton is matched to a corresponding jet. Events are rejected if not matched.
	          \item Exclusivity is enforced for the samples with $n < N$ by requiring the corresponding matched events not to have extra jets in order to be accepted. When $n=N$, additional jets are kept, provided their transverse momentum is smaller than that of all matched jets. 
	     \end{itemize}
\end{enumerate}
We will refer to the LO method described above as the  TMD merging method.   
We have described the method using the MLM merging prescription.   
Analogous LO versions of the method can in principle be derived using other merging 
prescriptions, such as CKKW-L, and adding the corresponding steps outlined at the points 3 and 4 above. 

\subsection{From MLM to TMD merging}   
\label{subsec:djr}
To start the assessment of the impact of TMD corrections to multi-jet merging, we consider in this section the  parton-level differential jet rates (DJRs) in $Z$+jets final states. The DJRs represent the distributions of the variable $d_{n,n+1}$, the square of the energy scale at which an $n$-jet event is resolved as an $(n+1)$-jet event, with parton-level jets reconstructed by the $k_\text{t}$ jet-clustering~\cite{Catani:1993hr,Ellis:1993tq}. Since the DJRs provide the splitting scales in the jet-clustering algorithm, they follow closely the measure used in the definition of the merging scale. Therefore, they have always been considered a powerful means to test the consistency and systematic uncertainties of multi-jet merging 
algorithms~\cite{Mrenna:2003if}. 

For this study, we use {\sc MadGraph5\_aMC@NLO}~\cite{Alwall:2014hca} to generate 
$Z   + 0,1,2,3$ jet samples at LO with generation cuts for the partons given by: $p_{T}>\mu_{c}=15$ GeV, $\vert\eta\vert<2.5$, $\Delta R>0.4$. We consider $pp$ collisions at a center-of-mass energy $\sqrt{s} =8$ and 13~TeV, for later comparisons against published LHC experimental results~\cite{Aad:2015auj,Aaboud:2017hbk,Khachatryan:2016crw,Sirunyan:2018cpw}.  
The PB TMD evolution is implemented 
in the event generator  {\sc Cascade}~\cite{Baranov:2021uol}. We use this to generate the TMD backward shower. We use 
{\sc Pythia}6.4~\cite{Sjostrand:2006za} 
 for the final-state shower. 
 We  apply the parton distributions obtained from DIS fits in~\cite{Martinez:2018jxt} with $\alpha_s(M_Z) = 0.118$. The nominal value for the merging scale is chosen to be $\mu_{m}= 23$ GeV.

\begin{figure}[hbtp]
  \begin{center}
        \includegraphics[width=.49\textwidth]{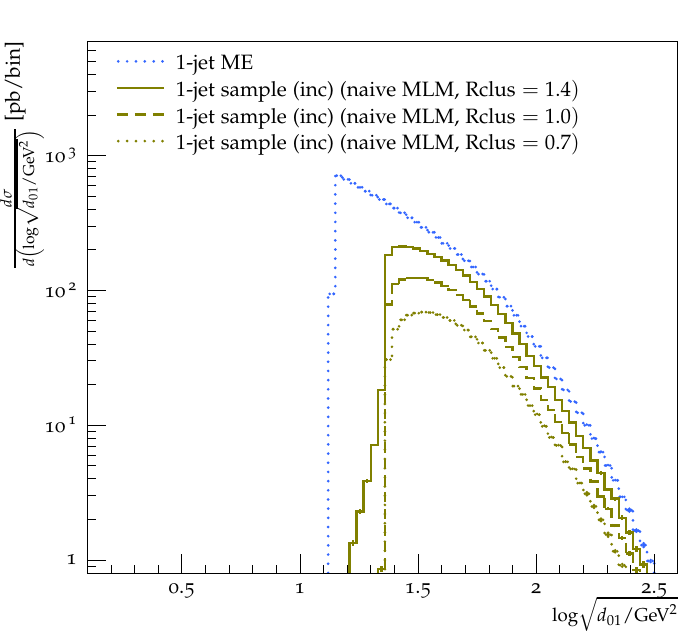}
	       \includegraphics[width=.49\textwidth]{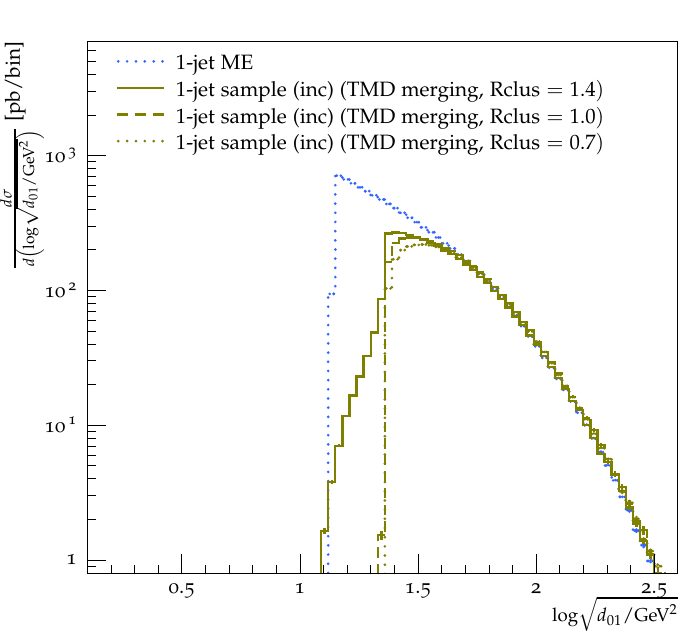}
  \caption{The $d_{01}$ spectrum at parton level, where $d_{01}$ represents the energy-square scale at which a $0$-jet event is resolved as a $1$-jet event in the k$_\perp$ jet-clustering algorithm. The dotted blue curve  represents the ME level of the $Z+$1 jet calculation, while 
  the green curves represent the $Z+$1 jet result using the ``naive MLM'' (left) and 
  TMD merging (right) methods for different values of the $R_\text{clus}$ parameter. The predictions are computed in inclusive (inc) mode.}
  \label{fig1}
  \end{center}
\end{figure}

We begin by computing the lowest-order DJR   $d_{0,1}$, giving 
the energy-square scale at which a $0$-jet event is resolved as a $1$-jet event.  
  In Fig.~\ref{fig1} (right) we present  
 the result for  $d_{0,1}$   from the  TMD merging method  introduced in 
 Subsec.~\ref{subsec:elements}    for different values of the $R_\text{clus}$ parameter. In Fig.~\ref{fig1} (left)
 we compare this result   with the result one would obtain by changing item 5 in  the 
method of Subsec.~\ref{subsec:elements} to 
a naive implementation of the MLM merging approach, 
in which the jets that are reconstructed after the parton shower has taken place are matched to the original ME partons (produced in item 1 of the 
method in Subsec.~\ref{subsec:elements}) instead of the partons resulting from the 
PB-TMD  evolution 
(as in item 3 of Subsec.~\ref{subsec:elements}). We stress that this is not the canonical MLM implementation, in which no ${\bm k}$ evolution and boost is applied. We introduce this ``naive'' version of item 5 in order to highlight the relevance of properly adapting the parton-level event to the new kinematics induced by the TMD evolution.
 
The blue curves in Fig.~\ref{fig1} represent the ME level of the prediction.  The green curves   
 in Fig.~\ref{fig1}   
 represent the result of calculations using the ``naive MLM'' (left) and TMD merging  (right) methods for different values of  $R_\text{clus}$. 
We observe in the results of Fig.~\ref{fig1} (left), based on 
``naive MLM'',   a strong dependence of the inclusive $Z+$1 jet contribution to the $d_{01}$ distribution on the merging procedure, specifically 
the $R_\text{clus}$ parameter. Moreover the dependence is also present at high values of $d_{01}$,  where instead the ME accuracy should be preserved. The reason for this loss of rate relative to the input parton-level distribution is the angular mismatch between the initial parton, prior to the ${\bm k}$ boost, and the jet reconstructed after radiation, whose direction is modified by the TMD boost. 
On the other hand, in the results of Fig.~\ref{fig1} (right) we observe that the TMD merging method of Subsec.~\ref{subsec:elements} 
 gives a contribution  to the $d_{01}$ distribution   
 from the $Z+$1 jet calculation 
 that preserves 
 the ME accuracy at high scales.  Furthermore,   the effect from varying $R_\text{clus}$ is localized around the merging scale, as with the standard matching algorithms.

\begin{figure}[hbtp]
  \begin{center}
	  \includegraphics[width=.49\textwidth]{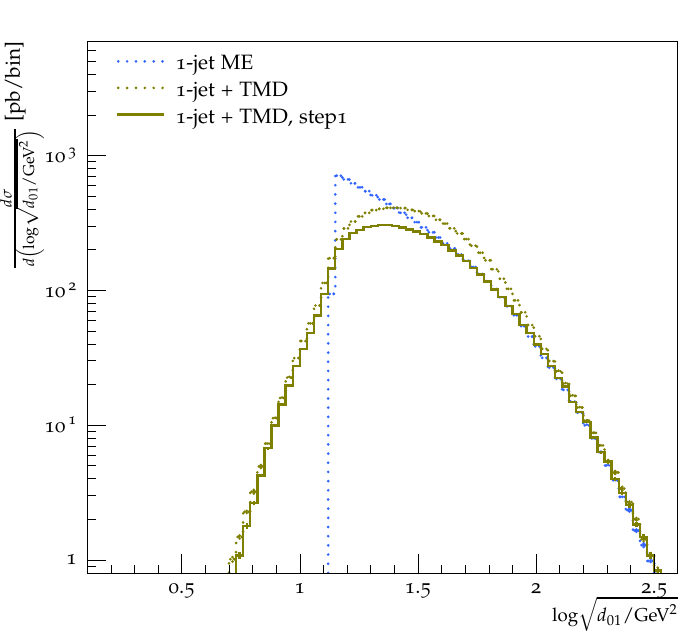}
      \includegraphics[width=.49\textwidth]{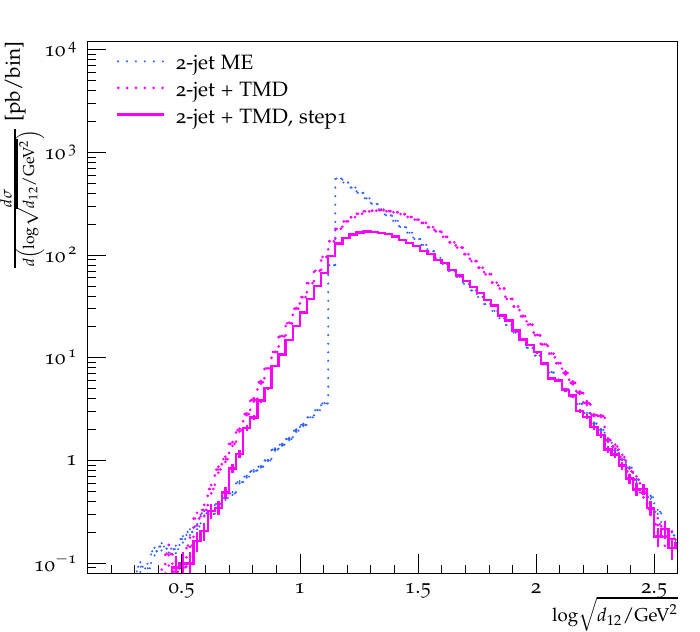}
  \caption{The $d_{n,n+1}$ spectra for $n=0$ (left) and $n=1$ (right) at parton level, where $d_{n,n+1}$ represents the energy-square scale at which an $(n+1)$-jet event is resolved as an $n$-jet event in the k$_\perp$ jet-clustering algorithm. The dotted blue curve represents the ME level of the $Z+$1 (left) and $Z+$2 (right) jet calculations. The dotted green and magenta curves represent the $Z+$1 and $Z+$2 jet ME computations respectively, after the PB-TMD evolution is applied. The solid green and magenta curves represent the $Z+$1 and $Z+$2 jet computations respectively, after step1 of the  
  TMD merging  method is applied, where step1 corresponds to the items 1, 2, 3 and 4 of the procedure in Subsec.~\ref{subsec:elements}.}
  \label{fig2}
  \end{center}
\end{figure}

We  next  focus on   features     encoded  in  items 1, 2, 3 and  4  
 of the TMD merging   method introduced in  Subsec.~\ref{subsec:elements}.  
We refer to the part of the algorithm consisting of items 1, 2, 3 and 4  as the 
first step (step1) of the    TMD merging  method. As  mentioned earlier 
in Sec.~\ref{pb-formulation}, both the TMD (forward) evolution and backward shower 
are used for the multi-jet merging method. They are represented, respectively, by item 3 and 4 
of the TMD merging  algorithm in  Subsec.~\ref{subsec:elements}.  Item 3, in particular, implies 
the condition $ {\bm k_i}^2 \leq \mu^2_{min}$ applied on the forward evolution of each initial-state parton.   

\begin{figure}[hbtp]
  \begin{center}
	  \includegraphics[width=.70\textwidth]{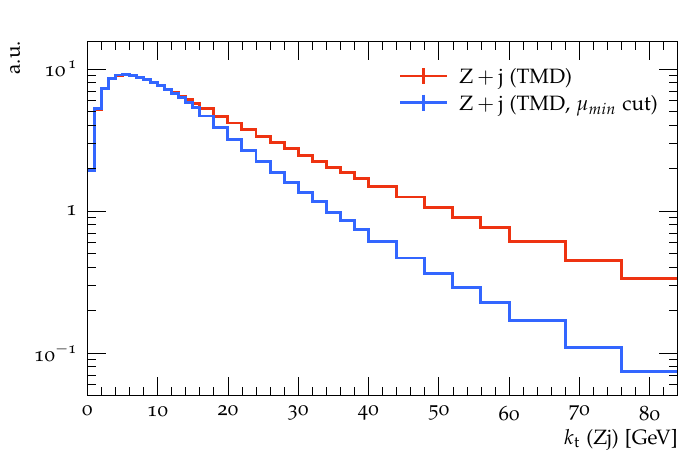}
  \caption{Transverse momentum distribution of a $Z$+jet partonic system induced by the TMD evolution, with (blue) and without (red) the ${\bm k}^2_i<\mu^2_{min}$ cut described in item 3 of the matching algorithm.}
  \label{ktzj}
  \end{center}
\end{figure}

Fig.~\ref{fig2} presents results for the 
$d_{01}$ and $d_{12}$  DJRs corresponding to step1, illustrating  explicitly  the role of the $  \mu^2_{min}$ condition, which was introduced in point 3. of Section~\ref{subsec:elements}.  
  The dotted blue curves show the ME level (1,2-jet ME) of the predictions. 
  The dotted green and magenta curves show the result at the level in which the ME are evolved forward using Eq.~(\ref{eq2}) (1,2-jet $+$ TMD) but without applying 
  $ {\bm k_i}^2 \leq \mu^2_{min}$. 
 The solid green (left) and magenta (right) 
 curves in  Fig.~\ref{fig2} show the contribution  to the $d_{01}$ (left) and $d_{12}$ (right) distributions,  
 from the $Z+$1 and $Z+$2  jet calculations     respectively,  resulting from step1 (1,2-jet $+$ TMD, step1).  In each case, the difference between the dotted and solid curves 
 shows explicitly the $  \mu^2_{min}$  effect. A more explicit representation of the impact of the TMD evolution and of the $  \mu^2_{min}$ cut is given in Fig.~\ref{ktzj}. The red line shows the transverse momentum $k_t(Zj)$ of a parton-level Z+jet sample, following the TMD evolution that gives, by construction, $k_t(Zj) = \vert {\bm k}\vert $. The  $\mu_{min}$ cut, defined in item 3 of the merging algorithm, suppresses events where the transverse momentum built up by the evolution of either of the two initial-state partons exceeds the jet $p_T$, leading to the blue curve. 

\begin{figure}[hbtp]
  \begin{center}
	 \includegraphics[width=.49\textwidth]{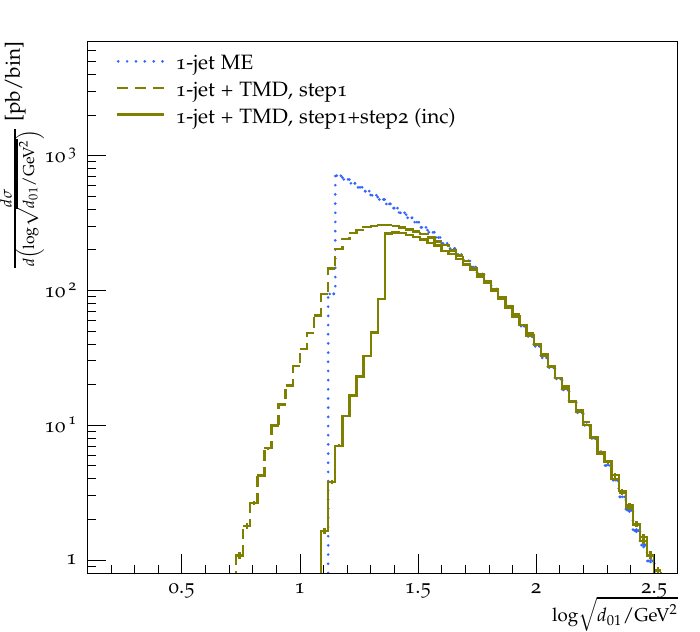}
     \includegraphics[width=.49\textwidth]{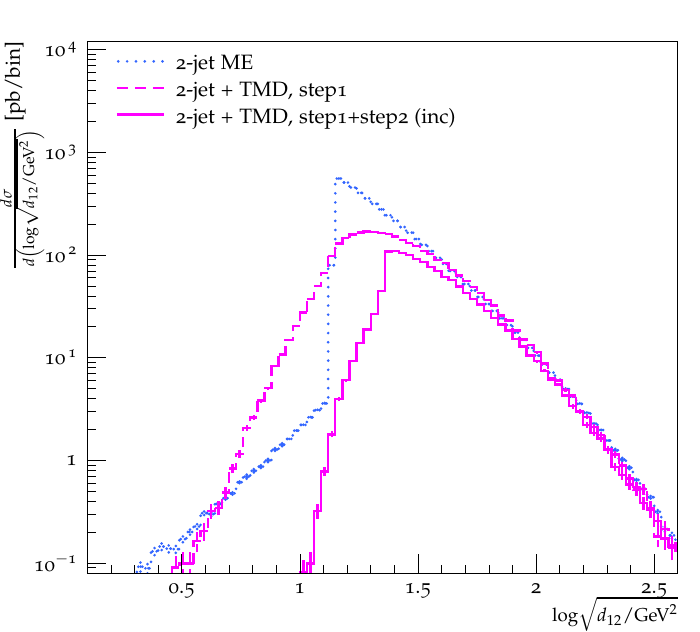}
  \caption{The $d_{n,n+1}$ spectra for $n=0$ (left) and $n=1$ (right) at parton level, where $d_{n,n+1}$ represents the energy-square scale at which an $(n+1)$-jet event is resolved as an $n$-jet event in the k$_\perp$ jet-clustering algorithm. The dotted blue curves represents the ME level of the $Z+$1 (left) and $Z+$2 (right) jet calculations. The dashed green and magenta curves represent the $Z+$1 and $Z+$2 jet computations respectively after step1 of the TMD merging method is applied where step1 corresponds to the items 1, 2, 3 and 4 of the procedure. The solid green and magenta curves represent the $Z+$1 and $Z+$2 jet computations respectively when the full TMD merging method is applied (step1 + step2, where step1 corresponds to the items 1, 2, 3 and 4 of the procedure while step2 corresponds to item 5). The predictions represented by the solid lines are calculated in inclusive (inc) mode.}
  \label{fig3}
  \end{center}
\end{figure}

We now move on to study the effect of the second step (step2) of the 
TMD merging algorithm, consisting of item 5 in 
 Subsec.~\ref{subsec:elements}. In Fig.~\ref{fig3} we show the contribution from the $Z+$1 (left) and $Z+$2 (right) jet calculations resulting from step1 (1,2-jet $+$ TMD, step1) and step2 (1,2-jet $+$ TMD, step1+step2), to the $d_{01}$ and $d_{12}$ distributions respectively. In addition we show the computations at the ME level (1,2-jet ME). The dotted blue curves represent the ME level of the prediction ($\sigma_n$), the dashed  curves represent the calculation resulting from step1, and the solid  curves are the result after step2 is applied. 

In Fig.~\ref{fig3} we observe that step2 introduces the merging scale cut at 23 GeV. Furthermore, when implemented in inclusive mode, step2 applies a small correction around the merging scale to the $Z+$1 and $Z+$2 jets calculations resulting from step1. The resulting correction ($\sim 25 \%$) is larger for the $Z+$2 jets computation than for the $Z+$1 case ($\sim 5 \%$) because the suppression includes not only the regions in phase space for which the partons are soft, but also the region for which the two partons are collinear, which is not present in the $Z+$1 calculation. 

Figures~\ref{fig2} and~\ref{fig3} show that at all steps of the   TMD merging algorithm the predictions for the largest multiplicities agree with the ME level, which is in accordance with the expectation that fixed-order contributions dominate higher scales, while TMD effects become important at low and intermediate scales. 

\begin{figure}[hbtp]
  \begin{center}
         \includegraphics[width=.49\textwidth]{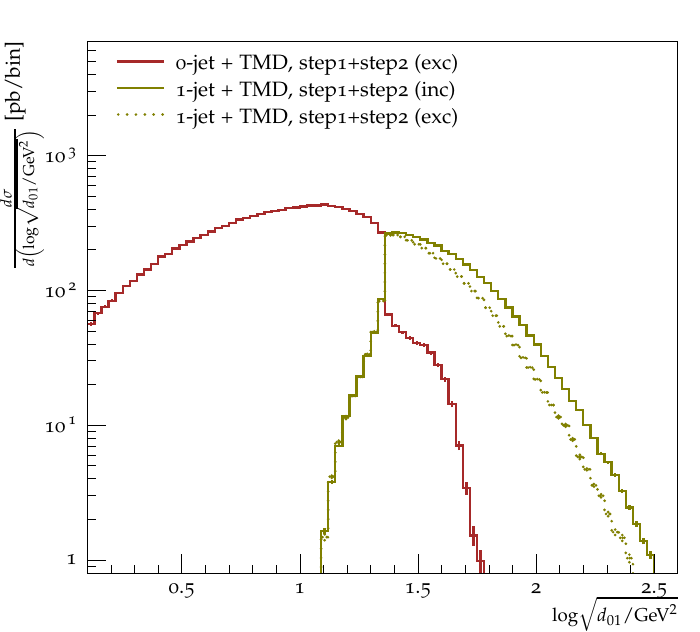}
         \includegraphics[width=.49\textwidth]{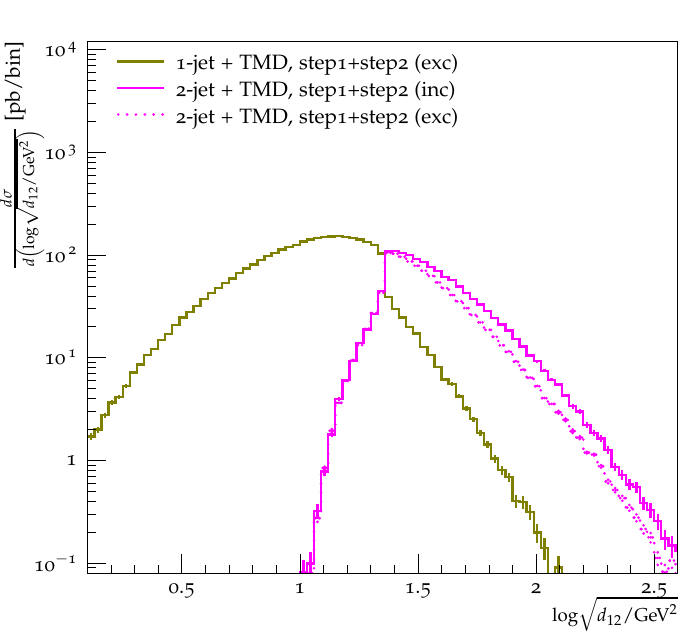}
  \caption{The $d_{n,n+1}$ spectra for $n=0$ (left) and $n=1$ (right) at parton level, where $d_{n,n+1}$ represents the energy-square scale at which an $(n+1)$-jet event is resolved as an $n$-jet event in the k$_\perp$ jet-clustering algorithm. The predictions represent the computations of the $Z+$0,1 (left) and  $Z+$1,2 (right) jet multiplicities when the full 
  TMD merging  method is applied (step1 + step2), where step1 corresponds to the items 1, 2, 3 and 4 of the procedure while step2 corresponds to item 5. The $Z+$1 contribution to $d_{01}$ (left) as well as the $Z+$2 contribution to $d_{12}$ (right) are calculated in both exclusive (exc) and inclusive (inc) modes. }
  \label{fig4}
  \end{center}
\end{figure}

In the following we study the effect of applying step2 in exclusive mode compared to inclusive mode. In Fig.~\ref{fig4} we show the contribution from the $Z+$0,1 (left) and $Z+$1,2 (right) jet calculations resulting from step2, to the $d_{01}$ and $d_{12}$ distributions respectively. The largest multiplicity in each subfigure is calculated both in inclusive and exclusive mode, while the lowest multiplicity in each subfigure is calculated only in exclusive mode. 

\begin{figure}[hbtp]
  \begin{center}
	\includegraphics[width=.49\textwidth]{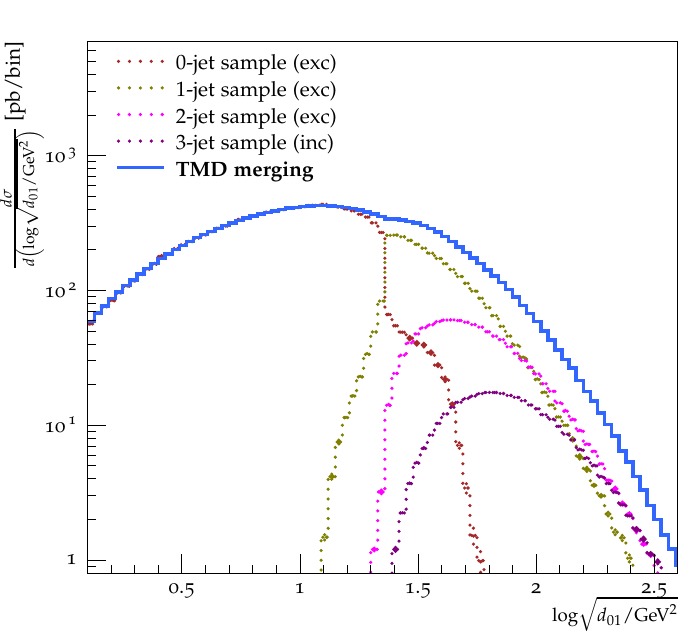}
	\includegraphics[width=.49\textwidth]{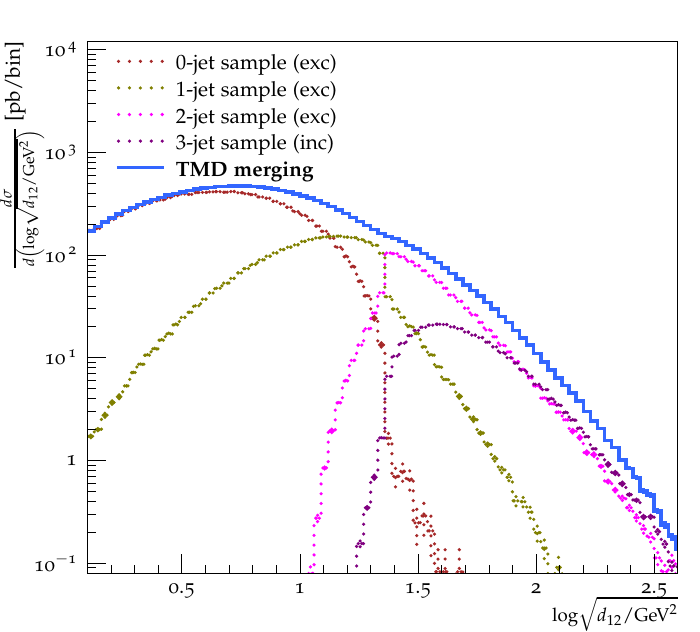}
    \includegraphics[width=.49\textwidth]{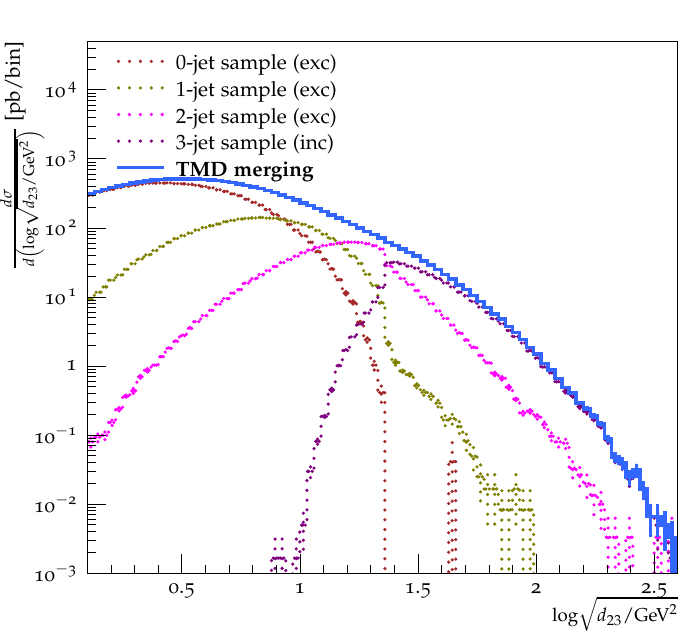} 
  \caption{The $d_{n,n+1}$  spectra for $n=0,1,2$  at parton level, where $d_{n,n+1}$ represents the energy-square 
  scale at which an $(n+1)$-jet event is resolved as an $n$-jet event in the 
  k$_\perp$ jet-clustering algorithm. The dotted curves represent the  contributions of the single-multiplicity samples while the solid curve corresponds to  their sum. For each panel all jet multiplicities are obtained in exclusive (exc) mode except for the highest multiplicity which is calculated in inclusive (inc) mode.}
  \label{fig5}
  \end{center} 
\end{figure}

Figure~\ref{fig4} shows that the suppression from the exclusive mode of the 
TMD merging method can be of the order of 0.5 at large scales while it does not affect the region close to the merging scale compared to the inclusive mode. The difference between the inclusive and exclusive $Z+$1 jet calculations in Fig.~\ref{fig4} (left) and between inclusive and exclusive $Z+$2 jet calculations in Fig.~\ref{fig4} (right) would be covered by the inclusion of higher multiplicities as will be seen in Fig.~\ref{fig5}. 

We have observed in Figs.~\ref{fig2},~\ref{fig3}, and~\ref{fig4} the effect of the different steps of the merging algorithm, with the predictions from the different steps approaching each other and the ME at large $d_{n,n+1}$ values. This is a consequence of the Sudakov-like suppression resulting from the merging algorithm which follows the Sudakov factor as in Eqs.~(\ref{eq3}) and~(\ref{eq4}). The performance of the merging procedure can be better appreciated in Fig.~\ref{fig5},  where a clear separation between the different jet samples is seen at the merging scale value while the resulting overall prediction behaves rather smoothly. A small glitch remains visible at values of the $d_{n,n+1}$ variables around the merging scale. This is a common feature of all matching algorithms (see e.g.~\cite{Mrenna:2003if,Alwall:2007fs}), and reflects the residual systematic error in the matching between shower and matrix elements. In principle the matching parameters, which are varied in order to establish a range for the systematic uncertainty, can be tuned to smooth out further the transition. We return to this point in Section~\ref{sec:syst}, and in particular Fig.~\ref{fig15}, where we study the matching uncertainty for the $d_{n,n+1}$ variables. In Fig.~\ref{fig5} the predictions include $Z+$1,2,3 jets computations, where all the multiplicities are calculated in exclusive mode except for the highest multiplicity which is calculated in inclusive mode. The dotted curves represent the $n$-jet sample contributions, while the solid curve corresponds to  their sum.       
 
%\newpage      

%\newpage

\section{$Z$-boson + jets production}
\label{sec:Zjets}
We next apply the TMD jet-merging method to show predictions for a variety of final-state observables 
in the production of $Z$-bosons in association with multiple jets at the LHC. This study is not meant to be a thorough phenomenological analysis of the existing data~\cite{Aad:2015auj,Aaboud:2017hbk,Khachatryan:2016crw,Sirunyan:2018cpw,AbellanBeteta:2016ugk}, but rather a first proof of the overall quality and robustness of the proposed TMD 
merging  approach. In particular, no effort was made to fine-tune the merging parameters of the MLM merging algorithm: the reference parameters chosen here closely reflect the defaults used in the original 2007 study of the merging algorithm applied to $W$+jets final states~\cite{Alwall:2007fs} ($p_{T,min}=15$~GeV, with $E_{T,clus}=20$~GeV there vs $E_{T,clus}=23$~GeV here). 

%For the calculations, we use {\sc MadGraph5\_aMC@NLO}~\cite{Alwall:2014hca} to generate $Z$ $+0,1,2,3$ jet samples at LO with a generation cut $q_{cut}=15$ GeV in $pp$ collisions at a center-of-mass energy $\sqrt{s} =$  13 TeV. We implement the PB TMD evolution in the generator  {\sc Cascade}~\cite{Jung:2010si}. We use this to generate the PB TMD shower, and for all calculations we  apply the parton distributions obtained from DIS fits in~\cite{Martinez:2018jxt} with $\alpha_s(M_Z) = 0.118$. The nominal value for the merging scale is chosen to be 23 GeV. 

\subsection{$Z$-boson transverse momentum and $\phi^{\ast}$ distributions}   
\label{sec:3-1}

\begin{figure}[hbtp]
  \begin{center}
	\includegraphics[width=.49\textwidth]{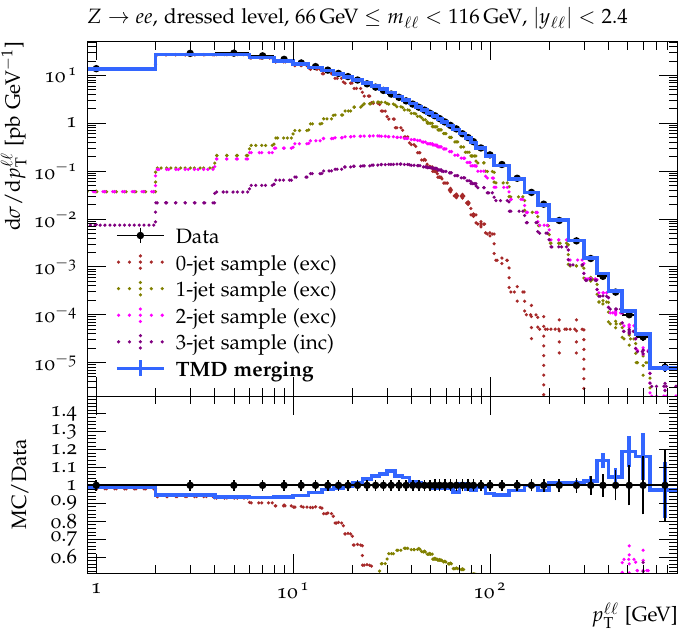}
	\includegraphics[width=.49\textwidth]{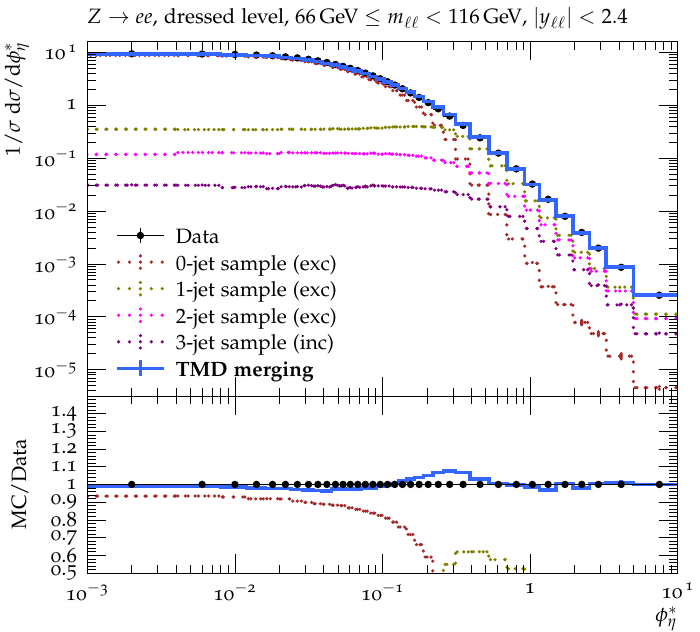}	
  \caption{Transverse momentum $p_T$  spectrum (left) and $\phi^{\ast}$ 
  normalized distribution (right)  of  DY lepton pairs from $Z$-boson decays. Experimental measurements by ATLAS~\protect\cite{Aad:2015auj} at 
  $\sqrt{s} = 8$ TeV are compared to predictions using the \tmd{} calculation. Separate contributions from 
  the different jet samples are shown. All jet multiplicities are obtained in exclusive (exc) mode except for the highest multiplicity which is calculated in 
  inclusive (inc) mode.}
  \label{fig6}
  \end{center}
\end{figure}

In Fig.~\ref{fig6} (left) we show the transverse momentum $p_T$ spectrum of  DY lepton pairs from $Z$-boson decays,     
normalized to the inclusive DY NNLO cross section. In addition, in Fig.~\ref{fig6} (right) we show the $\phi^{\ast}_{\eta}$~\cite{Vesterinen:2008hx,Banfi:2010cf} normalized distribution of  DY lepton pairs from $Z$-boson decays, where $\phi^{\ast}_{\eta}$ is defined as
\begin{equation}
\label{eq5}
\phi^{\ast}_{\eta} = \tan\left( \dfrac{\pi-\Delta\phi}{2}\right) \cdot\sin(\theta^{\ast}_{\eta}).
\end{equation} 
The angle $\Delta\phi$ is the azimuthal separation between the two leptons, and $\theta^{\ast}_{\eta}$ represents the scattering angle of the leptons with respect to the proton beam direction in the rest frame of the dilepton system.
The results of the \tmd{} calculation are compared to the  ATLAS measurements~\protect\cite{Aad:2015auj} at $\sqrt{s} = 8$ TeV. 
The analysis is performed using   {\sc Rivet}~\cite{Buckley:2010ar}.  
Separate contributions from the different jet samples are shown. All jet multiplicities are obtained in exclusive (exc) mode except for the highest multiplicity which is calculated in inclusive (inc) mode.

We observe that the $Z$+0 jet sample constitutes the main contribution at low transverse momentum $p_T$  while the impact of larger jet multiplicities gradually increases with increasing $p_T$. A similar behavior is observed for the $\phi^{\ast}$ distribution. The merged prediction provides a very good description of the whole DY $p_T$ spectrum 
as well as $\phi^{\ast}$ distribution, with deviations of at most 10\% around the points corresponding to the 23~GeV merging scale.  

The results of Fig.~\ref{fig6} may be compared with the results 
obtained in~\cite{Martinez:2019mwt} by applying PB-TMD evolution matched with 
the NLO $Z$-production matrix element, without including higher jet 
multiplicities.  The TMD merging prediction of Fig.~\ref{fig6} 
retains the good description of the low-$p_T$ region already obtained 
in~\cite{Martinez:2019mwt} (see also the 
recent analysis~\cite{CMS:2022ubq}),  and 
improves the behavior~\cite{Martinez:2019mwt}  in 
the high-$p_T$ region by merging TMD showers with higher multiplicities. 

%\newpage

\subsection{Inclusive and exclusive jet multiplicities}   
\label{sec:3-2}

\begin{figure}[hbtp]
  \begin{center}
	\includegraphics[width=.49\textwidth]{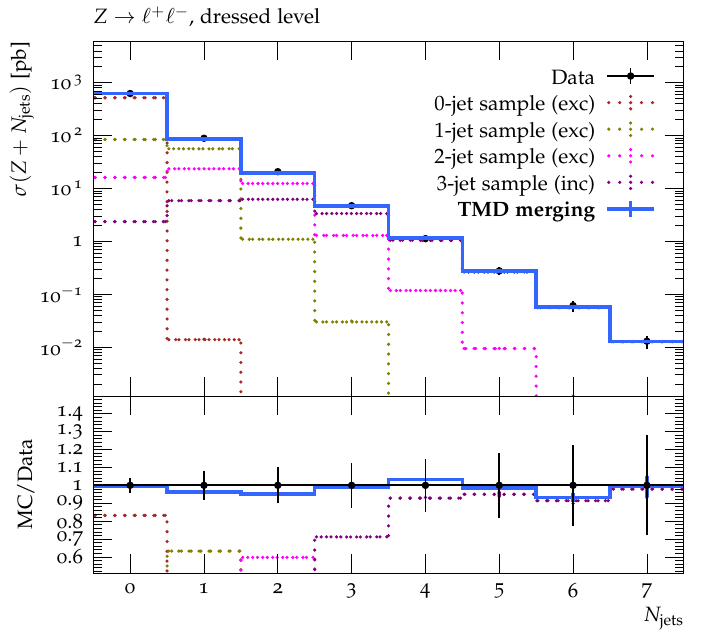}
	\includegraphics[width=.49\textwidth]{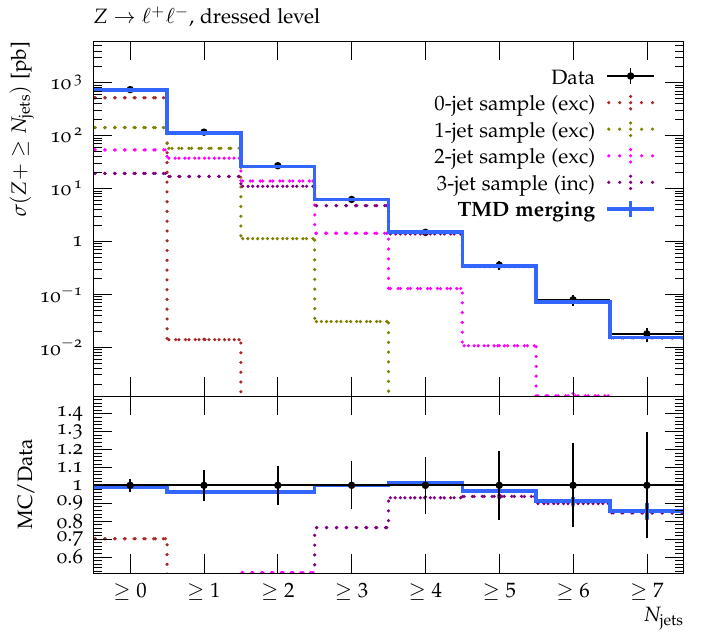}
  \caption{Exclusive (left) and inclusive (right) jet multiplicity distributions in the production of a $Z$-boson in association with jets. Experimental measurements by ATLAS~\protect\cite{Aaboud:2017hbk} at $\sqrt{s} = 13$ TeV are compared to predictions using the \tmd{} calculation. Separate contributions from the different jet samples are shown. All the jet multiplicities are obtained in exclusive (exc) mode except for the highest multiplicity which is calculated in inclusive (inc) mode.}
  \label{fig7}
  \end{center}
\end{figure}

In Fig.~\ref{fig7} we show the results for the exclusive (left) and inclusive (right) jet multiplicities in $Z$+jets events in $pp$ collisions at $\sqrt{s} = 13$ TeV. For illustration, in the rest of this section we compare the predictions with the ATLAS measurements~\cite{Aaboud:2017hbk}. Analogous measurements of $Z$+multijet final states have also been performed by CMS~\cite{Sirunyan:2018cpw}. 

\begin{figure}[hbtp]
  \begin{center}
	\includegraphics[width=.49\textwidth]{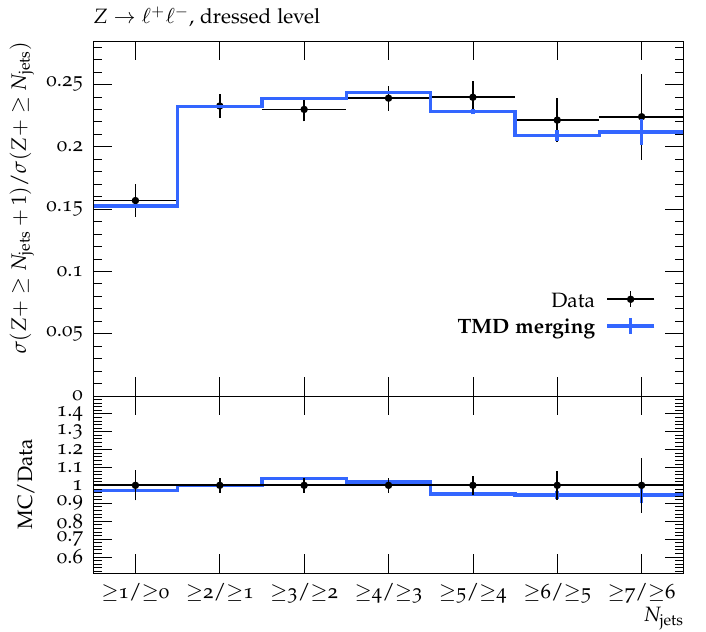}
  \caption{Ratio for successive inclusive jet multiplicities in the production of a $Z$-boson in association with jets. Experimental measurements by ATLAS~\protect\cite{Aaboud:2017hbk} at $\sqrt{s} = 13$ TeV are shown. 
%  We plot the results of the \tmd{} calculation and the contributions from the different jet samples.
  }
  \label{fig7_1}
  \end{center}
\end{figure}

The very good agreement of the prediction with the experimental measurements in Fig.~\ref{fig7} illustrates that not only the whole DY $p_T$ spectrum is described by the \tmd{} calculation, as seen in Fig.~\ref{fig6},  but also the number of jets which result into the lepton pair $p_T$ imbalance. What is particularly remarkable is that the agreement holds up to multiplicities much larger than the maximum number of jets (three) for which the exact LO matrix-element calculation is performed. This underscores the potential benefit of the TMD evolution in better describing hard and non-collinear emissions, compared to the standard collinear evolution.

We next compare in Fig.~\ref{fig7_1} the \tmd{} calculation of the ratio of consecutive inclusive jet multiplicities to the measurement by ATLAS~\protect\cite{Aaboud:2017hbk} at $\sqrt{s} = 13$ TeV. This measurement is sensitive to the strong coupling and is also very well described by the \tmd{} calculation.

\subsection{Jet transverse momentum distributions}  
\label{sec:3-3}

\begin{figure}[hbtp]
  \begin{center}
	\includegraphics[width=.49\textwidth]{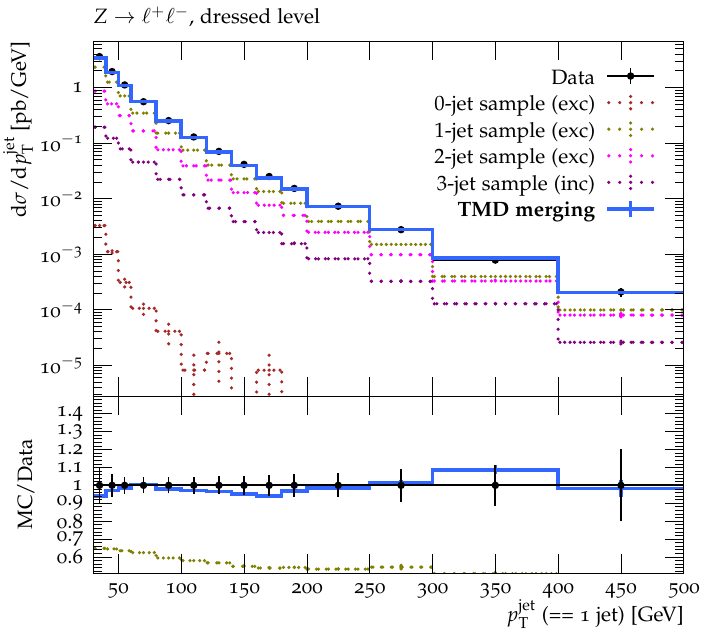}
	\includegraphics[width=.49\textwidth]{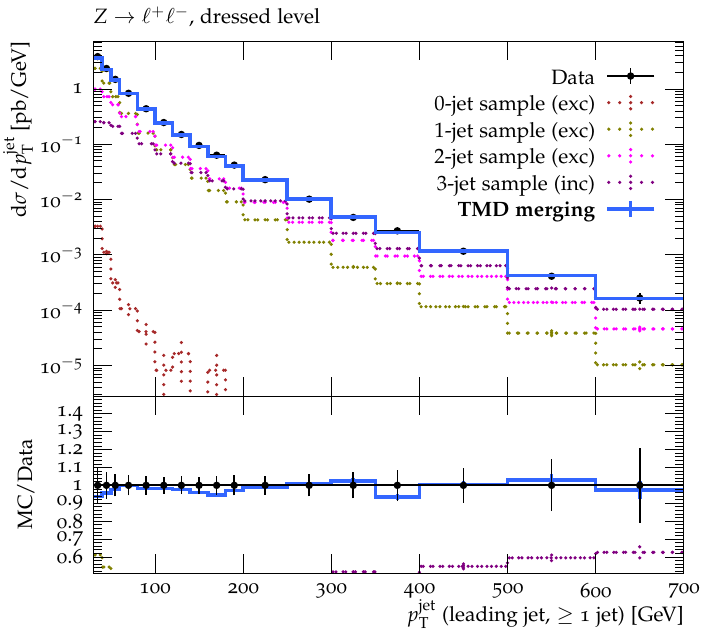}
  \caption{Exclusive (left) and inclusive (right) leading jet $p_T$ spectrum in events with a $Z$-boson in association with jets. Experimental measurements by ATLAS~\protect\cite{Aaboud:2017hbk} at $\sqrt{s} = 13$ TeV are compared to predictions using the \tmd{} calculation. Separate contributions from the different jet samples are shown. All the jet multiplicities are obtained in exclusive (exc) mode except for the highest multiplicity which is calculated in inclusive (inc) mode.}
  \label{fig8}
  \end{center}
\end{figure}

In Fig.~\ref{fig8} we calculate the exclusive (left) and inclusive (right) leading jet $p_T$ spectrum in $Z+$jets events. The calculations are compared with experimental measurements by ATLAS~\protect\cite{Aaboud:2017hbk} at 13 TeV. The contributions from the different jet multiplicities to the final prediction are shown separately.

\begin{figure}[hbtp]
  \begin{center}
	\includegraphics[width=.49\textwidth]{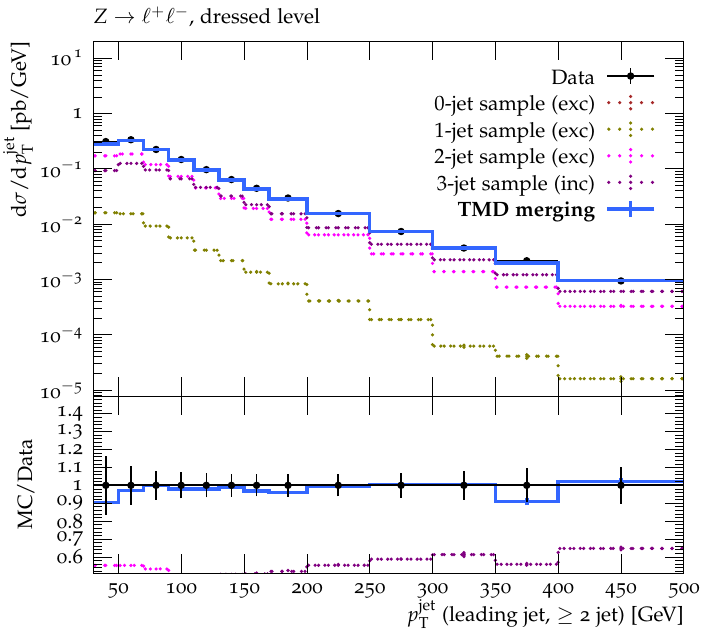}
	\includegraphics[width=.49\textwidth]{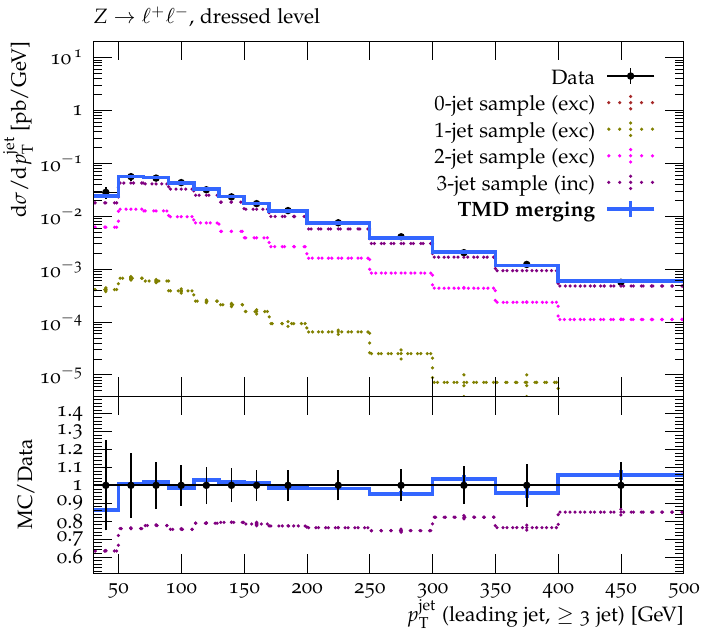}
	\includegraphics[width=.49\textwidth]{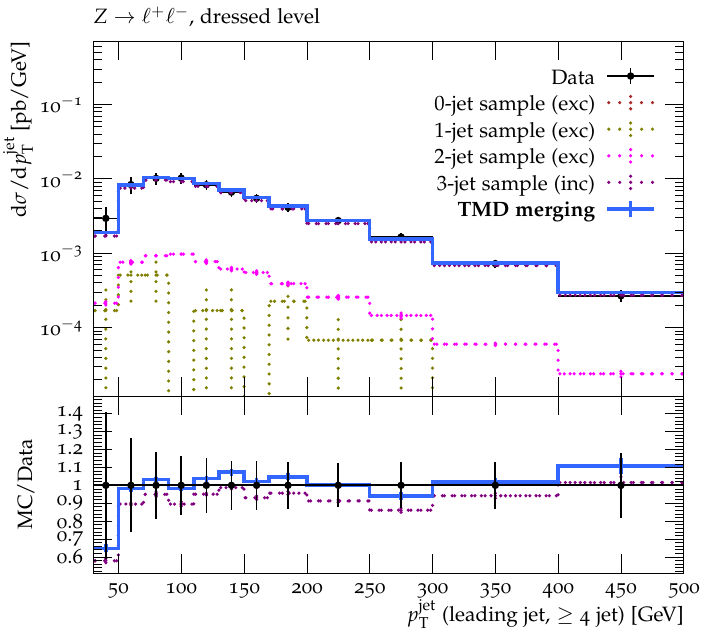}
  \caption{Leading jet $p_T$ spectrum in inclusive $Z+$2 (top left), 3 (top right), and 4 (bottom) jets. 
  Experimental measurements by ATLAS~\protect\cite{Aaboud:2017hbk} at $\sqrt{s} = 13$ TeV are compared to predictions using the \tmd{} calculation. Separate contributions from 
  the different jet samples are shown. All the jet multiplicities are obtained in exclusive (exc) mode except for the 
     highest multiplicity which is calculated in inclusive (inc) mode.}
  \label{fig9}
  \end{center}
\end{figure}

While for the exclusive leading jet $p_T$ spectrum the main contribution comes from the $Z+$1 jet multiplicity, in the inclusive case the $Z+$1 jet multiplicity is only important at low $p_T$, with larger multiplicities giving the main contributions at large $p_T$ of the jet. A similar very good description of the data is observed for both observables, 
notwithstanding the very different multiplicity contributions to the predictions.   

Additionally, in Fig.~\ref{fig9} we calculate the leading jet $p_T$ spectrum in inclusive $Z+$2 (top left), $Z+$3 (top right), and $Z+$4 (bottom) jet events. The calculations are also compared with experimental  measurements by ATLAS~\protect\cite{Aaboud:2017hbk} at 13 TeV.

The very good agreement of the predictions in Figs.~\ref{fig8} and~\ref{fig9} with the experimental measurements  
illustrates that not only the DY $p_T$ spectrum and the jet multiplicity distribution are described by the \tmd{} calculation (Figs.~\ref{fig6} and~\ref{fig7}), but 
also the $p_T$ of the leading jet contribution to $p_T$ imbalance is well described, 
both in the single-jet events and in the multi-jet events. 

%\newpage

\subsection{Di-jet azimuthal separation in $Z$-boson events}  
\label{sec:3-4}

\begin{figure}[hbtp]
  \begin{center}
	\includegraphics[width=.49\textwidth]{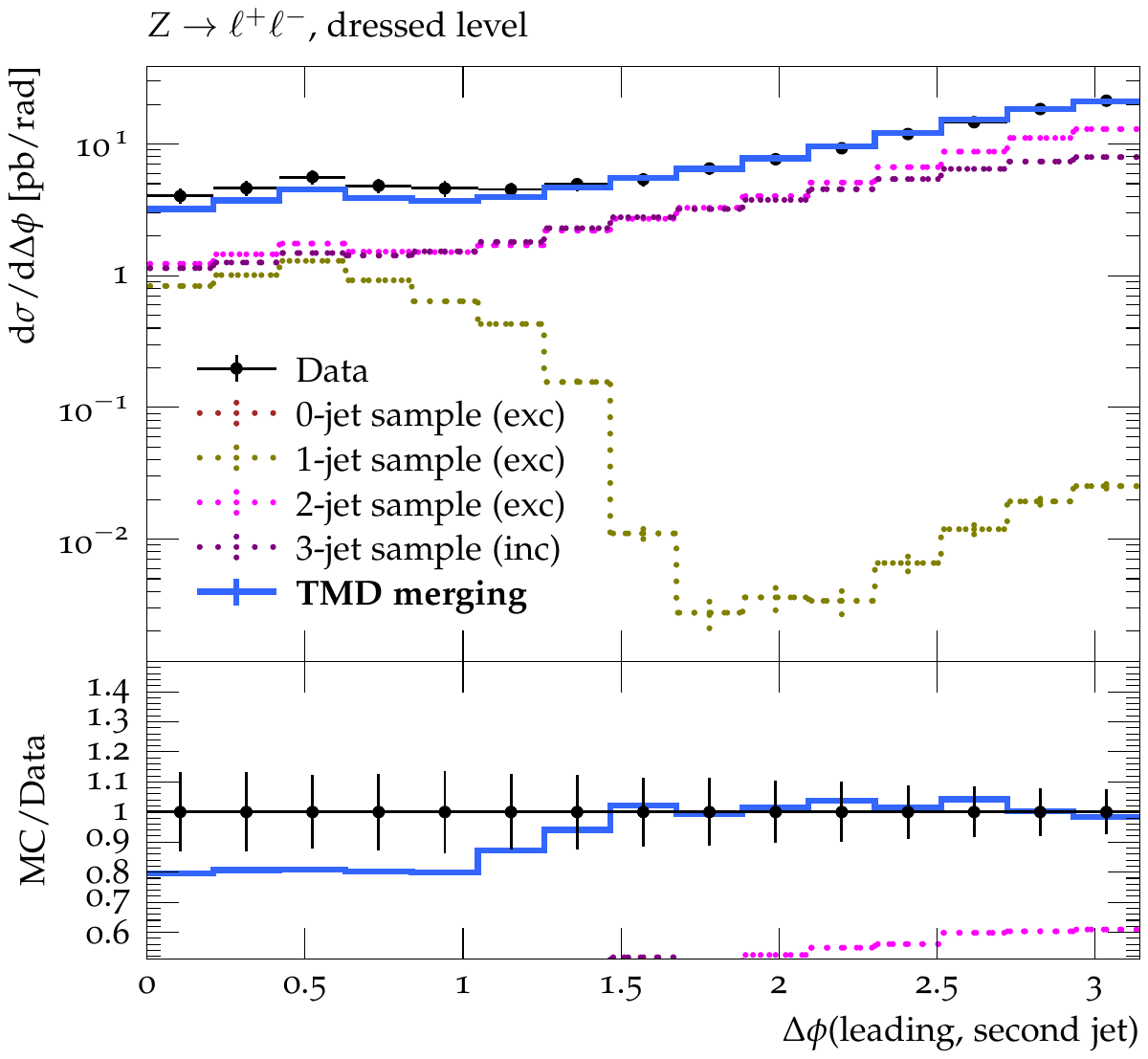}
  \caption{Di-jet azimuthal separation distribution for $Z+ \geq$2 jets events. Experimental measurements by ATLAS~\protect\cite{Aaboud:2017hbk} at $\sqrt{s} = 13$ TeV are compared to predictions using the \tmd{} calculation. Separate contributions from the different jet samples are shown. All the jet multiplicities are obtained in exclusive (exc) mode except for the highest multiplicity which is calculated in inclusive (inc) mode.}
  \label{fig10}
  \end{center}
\end{figure}

In Fig.~\ref{fig10} we show the distribution in 
di-jet azimuthal separation $\Delta\phi$  for $Z+ \geq$2 jets events. The calculation is compared with experimental measurements by ATLAS~\protect\cite{Aaboud:2017hbk} at 13 TeV. The contributions from the different jet multiplicities to the final prediction are shown separately.

Fig.~\ref{fig10} shows that the agreement 
of the TMD merging prediction with the experimental  data is very good in the high $\Delta \phi $ region 
while a 20\% deficit is observed  
in the low $\Delta\phi$ region. We expect the deficit in the low $\Delta\phi$ region 
to be due to missing multi-parton interaction (MPI) 
contributions. Inclusively, $Z+$2 jets processes constitute the leading contribution to the $\Delta\phi$ between the two jets, assuming a single scattering. If a double parton scattering is assumed instead, the leading MPI contribution at low $\Delta\phi$ corresponds to a $Z+$1 jet interaction together with a QCD 2-jet second interaction.
To explore this, we make an estimate 
of the MPI impact on the $\Delta \phi$ distribution 
in Fig.~\ref{fig10}, based on the  
 {\sc Pythia}8~\cite{Sjostrand:2014zea}
MPI parameter tunes CUETP8M1 \cite{CMS:2015wcf} 
and CP5 \cite{CMS:2019csb} obtained by the CMS collaboration. Our estimate is not intended as a systematic study of MPI effects in $Z$ + jets distributions, but simply as a proof of principle that MPI is at the origin of the behavior seen in Fig.~\ref{fig10}.

\begin{figure}[hbtp]
  \begin{center}
	\includegraphics[width=.49\textwidth]{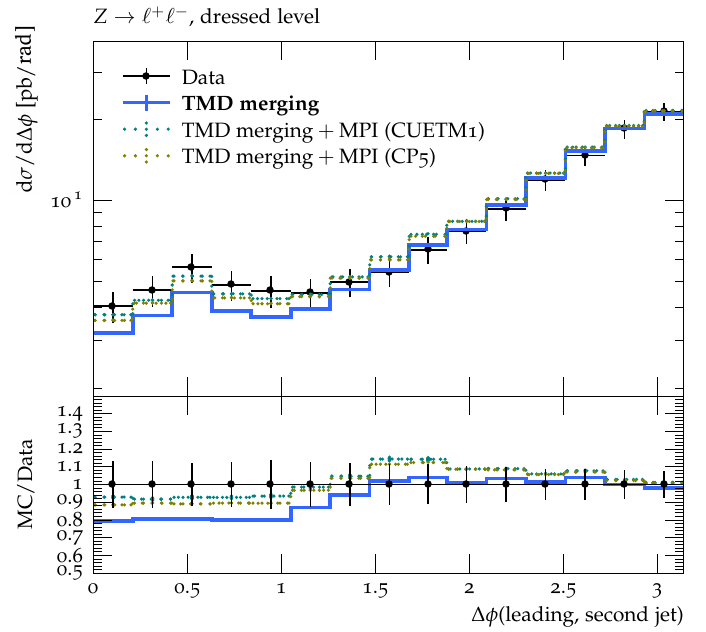}
  \caption{Di-jet azimuthal separation distribution for $Z+ \geq$2 jets events. Experimental measurements by ATLAS~\protect\cite{Aaboud:2017hbk} at $\sqrt{s} = 13$ TeV are compared to predictions using the \tmd{} calculation. The results including the MPI correction, estimated using the {\sc Pythia}8 tunes CUETP8M1 \cite{CMS:2015wcf} and CP5 \cite{CMS:2019csb}, are shown with dashed lines.}
  \label{fig:mpifordeltaphi}
  \end{center}
\end{figure}

For the tunes used to perform the estimate, the effective cross section assuming two independent hard scatterings was determined in \cite{CMS:2015wcf} and \cite{CMS:2019csb} to be, respectively, $\sigma_{\rm{eff}} = $ 27.9 mb (CUETP8M1)  and $\sigma_{\rm{eff}} = $ 25.3 mb (CP5). For each set of 
parameter tunes, we obtain the MPI correction as the difference between the $Z+$1 jet computation of the $\Delta \phi$ distribution including the simulation of parton showers and MPI, and the analogous computation without the simulation of MPI. In 
Fig.~\ref{fig:mpifordeltaphi} we add the resulting correction to the TMD merged 
prediction. The solid blue curve 
in Fig.~\ref{fig:mpifordeltaphi} is the same 
result as in Fig.~\ref{fig10}, while 
the dashed dark-green and light-green 
curves are the result of adding to this the MPI 
contribution obtained with the parameter tunes CUETP8M1 and CP5, respectively. 
We see that, while the region of the 
largest $\Delta \phi$ is not affected significantly  
by MPI, at low $\Delta \phi$ the MPI correction contributes a 10 - 15\% increase to the prediction, which supports our starting hypothesis.

%\newpage

\subsection{Di-jet mass distributions}  
\label{sec:3-5}

\begin{figure}[hbtp]
  \begin{center}
	\includegraphics[width=.49\textwidth]{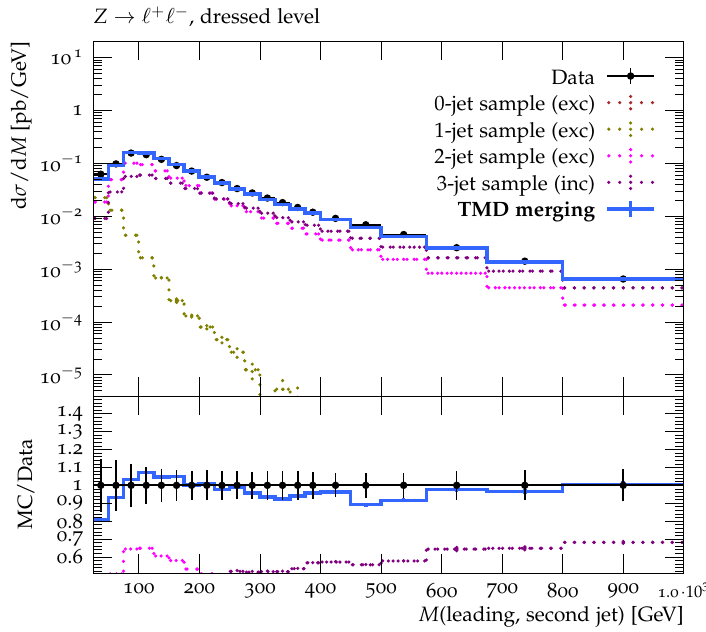}
  \caption{Di-jet mass distribution for $Z+ \geq$2 jet events. Experimental measurements by ATLAS~\protect\cite{Aaboud:2017hbk} at $\sqrt{s} = 13$ TeV are compared to predictions using the \tmd{} calculation. Separate contributions from the different jet samples are shown. All the jet multiplicities are obtained in exclusive (exc) mode except for the highest multiplicity which is calculated in inclusive (inc) mode.}
  \label{fig11}
  \end{center}
\end{figure}

In Fig.~\ref{fig11} we show the di-jet mass distribution for $Z+\geq 2$ jet events. The calculation is compared with experimental measurements by ATLAS~\protect\cite{Aaboud:2017hbk} at 13 TeV. The contributions from the different jet multiplicities to the final prediction are shown separately.

\begin{figure}[hbtp]
  \begin{center}
	\includegraphics[width=.49\textwidth]{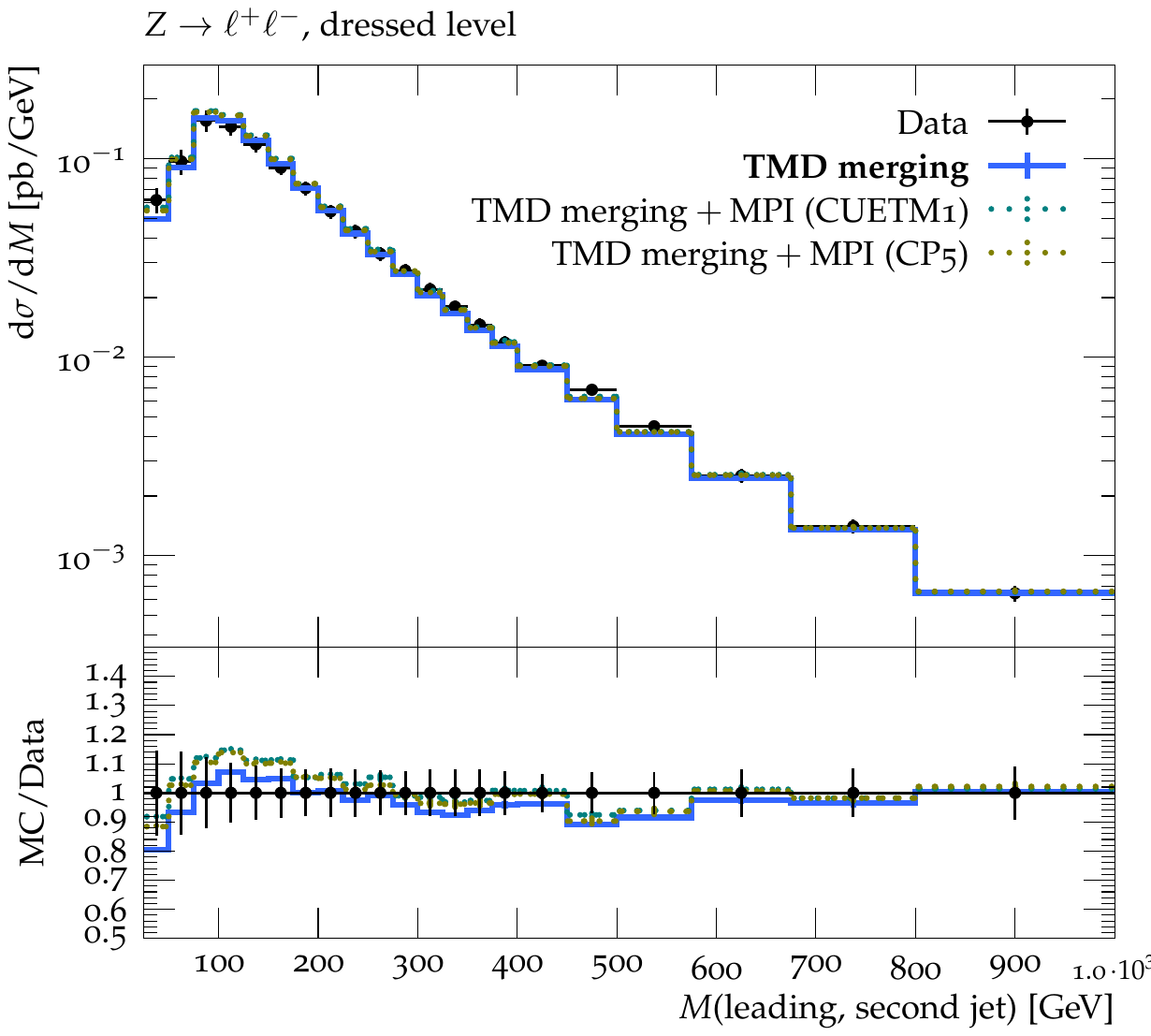}
  \caption{Di-jet mass distribution for $Z+ \geq$2 jet events. Experimental measurements by ATLAS~\protect\cite{Aaboud:2017hbk} at $\sqrt{s} = 13$ TeV are compared to predictions using the \tmd{} calculation. The results including the MPI correction, estimated using the {\sc Pythia}8 tunes CUETP8M1 \cite{CMS:2015wcf} and CP5 \cite{CMS:2019csb}, are shown with dashed lines.}
  \label{fig:mpiformjj}
  \end{center}
\end{figure}

The agreement of the prediction 
with the experimental data is good. 
Also for the di-jet mass we estimate the 
MPI effects, as described in the case of the $\Delta \phi$ 
distribution in the previous subsection. 
The results are shown in 
Fig.~\ref{fig:mpiformjj}.   The MPI contributions do  not affect significantly the region of high di-jet masses, while they 
are non-negligible at low di-jet masses, particularly in the first bin of Fig.~\ref{fig:mpiformjj}.

%We observe  good agreement with the experimental data. We observe that the $Z+2$ partons contribution is  important at low di-jet masses, while at large di-jet masses the $Z+3$ partons contribution dominates.

%\newpage

\subsection{Scalar sum of transverse momenta in $Z$-boson events} 
\label{sec:3-6}

\begin{figure}[hbtp]
  \begin{center}
	\includegraphics[width=.49\textwidth]{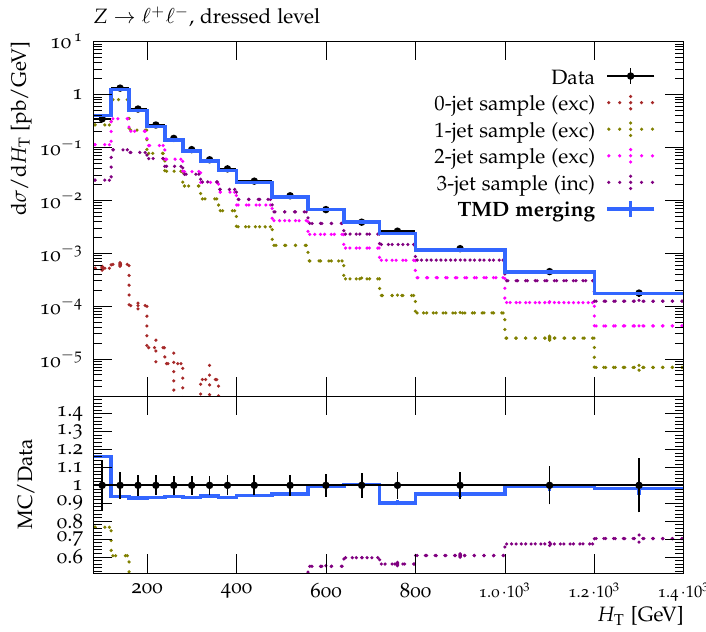}
  \caption{Scalar sum $H_T$ of the transverse momenta of leptons and jets for $Z+$ jets events. 
  Experimental measurements by ATLAS~\protect\cite{Aaboud:2017hbk} at $\sqrt{s} = 13$ TeV are compared to predictions using the \tmd{} calculation. Separate contributions from the different jet samples are shown. All the jet multiplicities are obtained in exclusive (exc) mode except for the highest multiplicity which is calculated in inclusive (inc) mode.}
  \label{fig12}
  \end{center}
\end{figure}

In Fig.~\ref{fig12} we show the 
 scalar sum $H_T$ of the transverse momenta of leptons and jets for $Z+$ jets events. The calculation is compared with experimental measurements by ATLAS~\protect\cite{Aaboud:2017hbk} at 13 TeV. The contributions from the different jet multiplicities to the final prediction are shown separately.

We observe good agreement of the \tmd{} calculation with the experimental data. While at low values of $H_T$ the $Z+$2 parton contribution plays the main role, the higher multiplicity becomes increasingly important as $H_T$ increases.

%\newpage

\subsection{Jet rapidity distribution}  
\label{sec:3-7}

\begin{figure}[hbtp]
  \begin{center}
	\includegraphics[width=.49\textwidth]{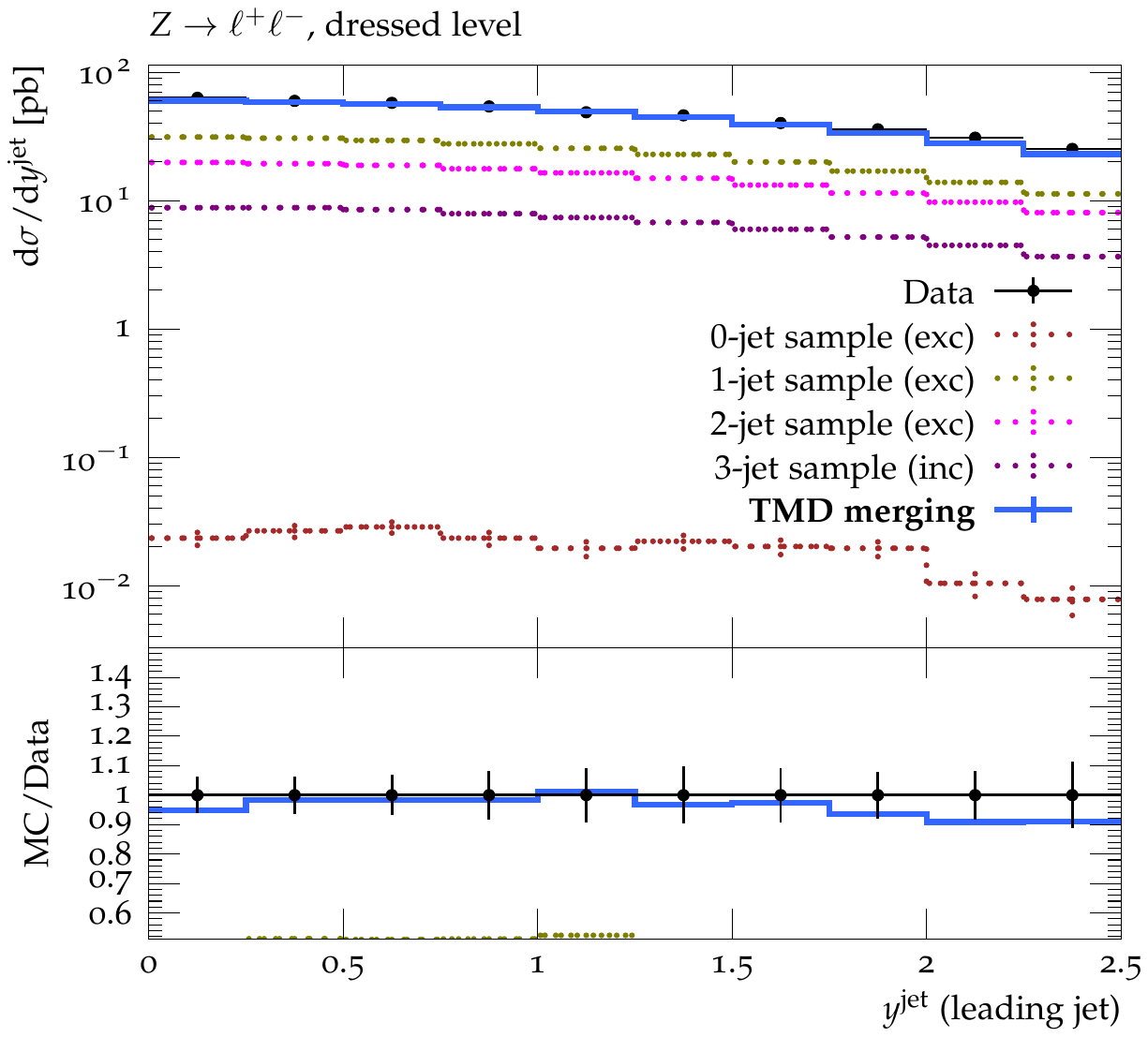}
  \caption{Leading jet rapidity distribution for $Z+$jets events. Experimental measurements by ATLAS~\protect\cite{Aaboud:2017hbk} at $\sqrt{s} = 13$ TeV are compared to predictions using the \tmd{} calculation. Separate contributions from the different jet samples are shown. All the jet multiplicities are obtained in exclusive (exc) mode except for the highest multiplicity which is calculated in inclusive (inc) mode.}
  \label{fig13}
  \end{center}
\end{figure}

In Fig.~\ref{fig13} we show the  leading jet rapidity distribution for $Z+$jets events. The calculation is compared with experimental measurements by ATLAS~\protect\cite{Aaboud:2017hbk} at 13 TeV. The contributions from the different jet multiplicities to the final prediction are shown separately.

We observe good agreement with the experimental data. The $Z+$1 jet multiplicity constitutes the main contribution to the leading jet rapidity distribution, as expected.

%\newpage  

\section{Theoretical systematic uncertainty and merging scale dependence}
\label{sec:syst} 

In this section we 
investigate  
the theoretical uncertainties associated with the TMD 
multi-jet merging algorithm. 
The inclusion of transverse momentum recoils through TMD evolution 
influences the  systematic uncertainty of the merging when 
matrix-element and parton-shower contributions 
are combined. Here we focus in particular on the 
systematic effects occurring through the dependence on the merging scale.

\begin{table}[htb!h]
  \centering
  \begin{tabular}{|c|c|c|c|c|c|}
    \hline\hline
    Merging scale & $\sigma[\text{tot}]$ & $\sigma[\geq 1\text{ jet}]$ & $\sigma[\geq 2\text{ jet}]$ & $\sigma[\geq 3\text{ jet}]$ & $\sigma[\geq 4\text{ jet}]$ \\
    $[\text{GeV}]$ & $[\text{pb}]$ & $[\text{pb}]$ & $[\text{pb}]$ & $[\text{pb}]$ & $[\text{pb}]$ \\
    \hline \hline
    23.0 & 573 & 87.25 & 20.27 & 4.84 & 1.18 \\
    \hline
    33.0 & 563 & 86.15 & 20.48 & 4.86 & 1.19 \\
    \hline\hline
  \end{tabular}
  \caption{Multi-jet rates from the \tmd{} merging algorithm as a function of the merging scale.}
  \label{tab1}
\end{table}

\begin{figure}[hbtp]
  \begin{center}
	\includegraphics[width=.49\textwidth]{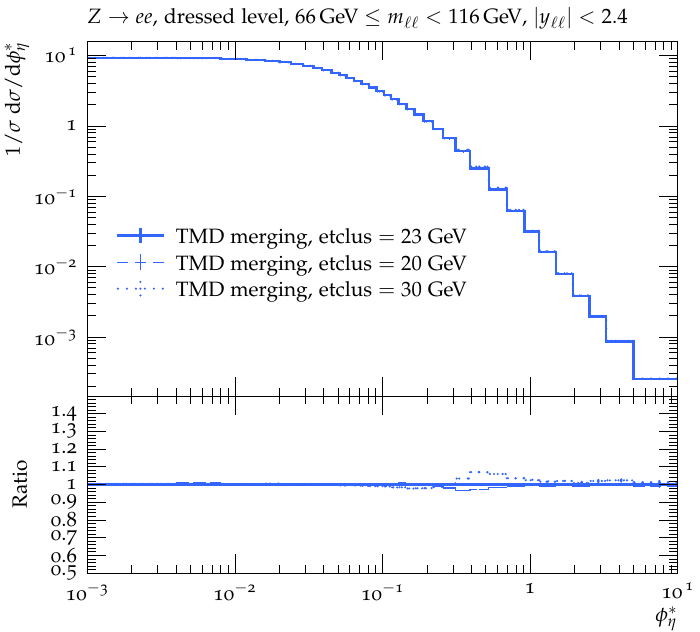} 
	\includegraphics[width=.49\textwidth]{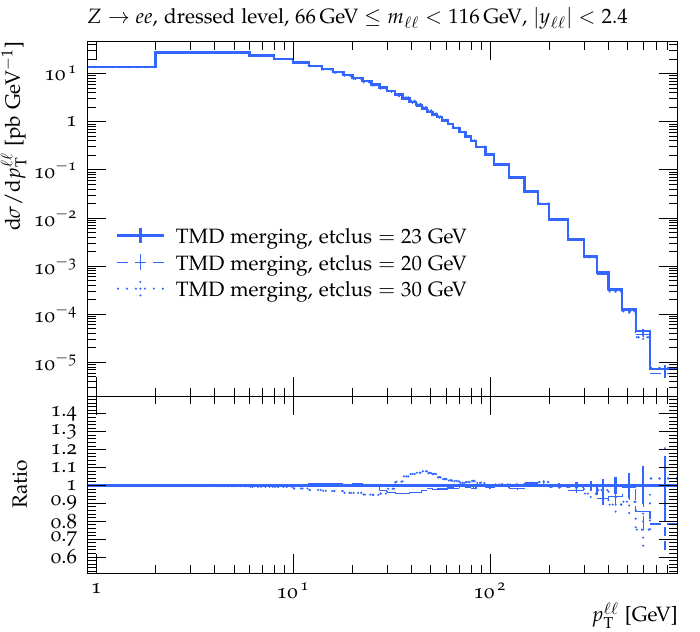}
  \caption{\tmd{} systematic uncertainty at the LHC at $\sqrt{s}=8$ TeV. The transverse momentum (right) and normalized $\phi^{\ast}$ (left) distributions of Drell-Yan lepton pairs are calculated for three different values of the merging scale, where the solid line represents the default setting. The phase space for the calculation follows the one in \cite{Aad:2015auj}.}    	  \label{fig14}
  \end{center}
\end{figure}

In Tab.~\ref{tab1} we  show  the multi-jet rates computed with 
TMD merging for different multiplicities 
for a 10 GeV variation of the merging scale. The results are obtained 
for $ p p $ collisions at 13 TeV, 
with the phase space selection and cuts of~\cite{Aaboud:2017hbk}. The rates shown 
are absolute, i.e., they are not rescaled to the NNLO total cross section. 
We observe that a 10 GeV variation of the merging scale results in less than 2\% variation for all the jet multiplicities considered. This 
systematic uncertainty is significantly smaller than what was found with 
 standard algorithms of collinear merging in 
Ref.~\cite{Alwall:2007fs}, where the variation of the jet multiplicities was found to be 
about 10\% for a 10 GeV change in the merging scale.

%\newpage

\begin{figure}[hbtp]
  \begin{center}
	\includegraphics[width=.49\textwidth]{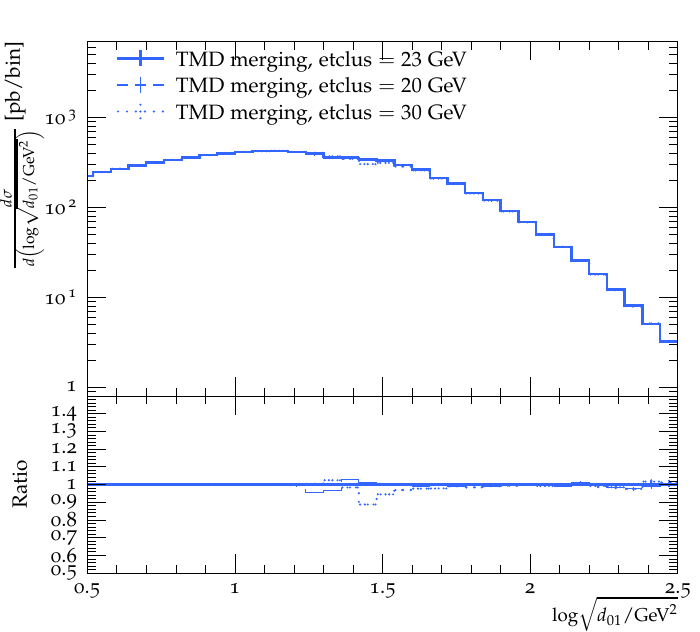} 
	\includegraphics[width=.49\textwidth]{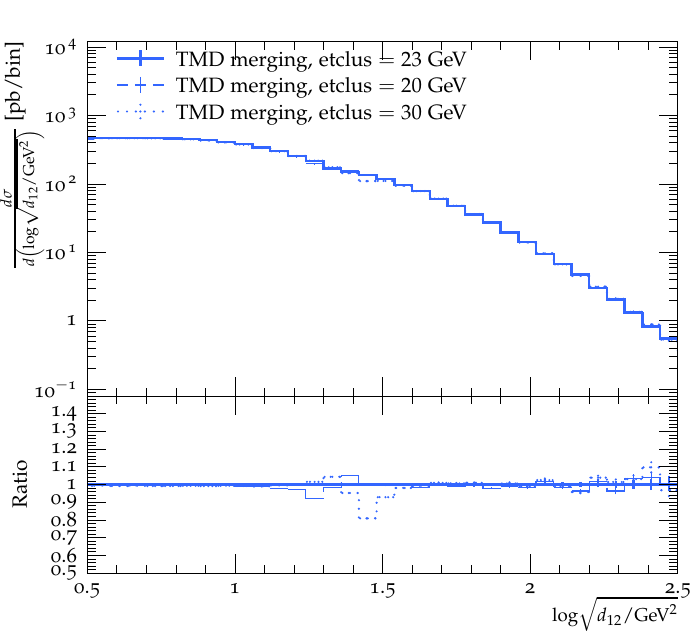}
    \includegraphics[width=.49\textwidth]{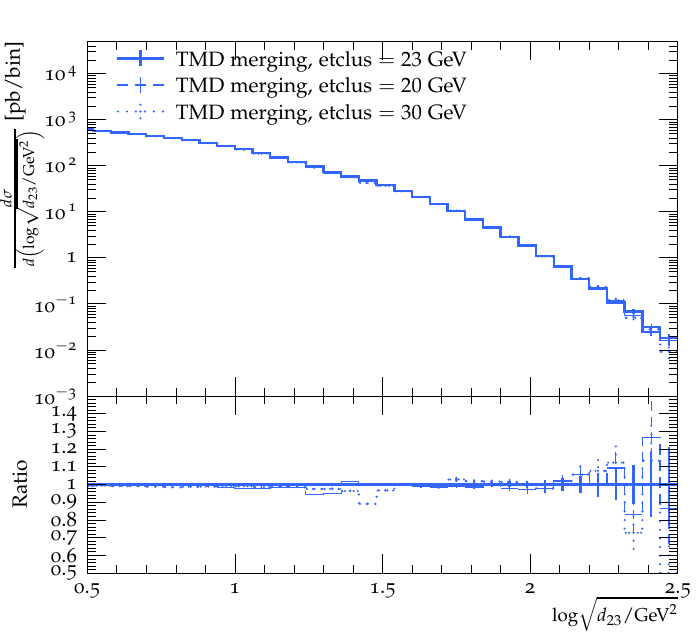} 
  \caption{\tmd{} systematic uncertainty at the LHC at $\sqrt{s}=8$ TeV. The $d_{n,n+1}$ spectra for $n=0,1,2$, where $d_{n,n+1}$ represents the energy-square scale at which an $(n+1)$-jet event is resolved as an $n$-jet event in the k$_\perp$ jet-clustering algorithm, are calculated for three different values of the merging scale. The solid line represents the default setting.}	
  \label{fig15}
  \end{center}
\end{figure}

Besides the effect on the total jet rates, we next examine the merging systematic uncertainty
in the case of  differential distributions. We use the TMD merging algorithm 
with the default value of the merging scale at 23 GeV, as in the previous calculations, 
and consider  variations  of the merging scale around the default value to 
20 GeV and 30 GeV.

%\newpage

\begin{figure}[hbtp]
  \begin{center}
	\includegraphics[width=.49\textwidth]{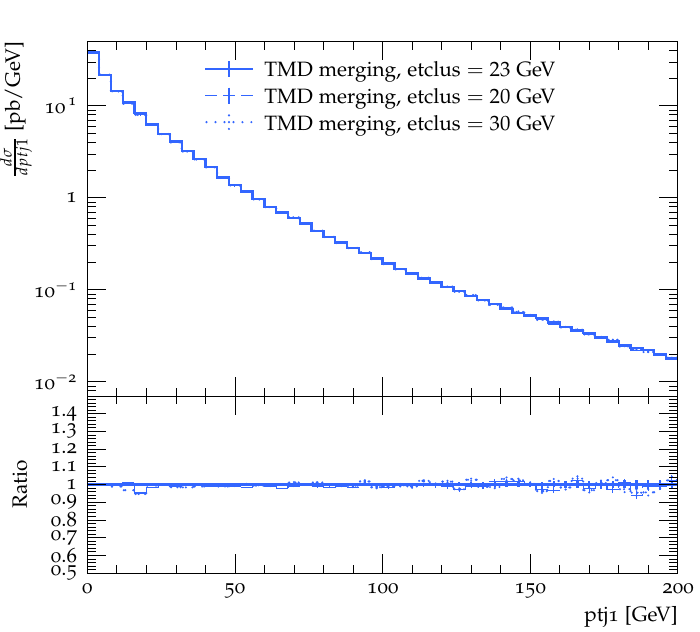} 
	\includegraphics[width=.49\textwidth]{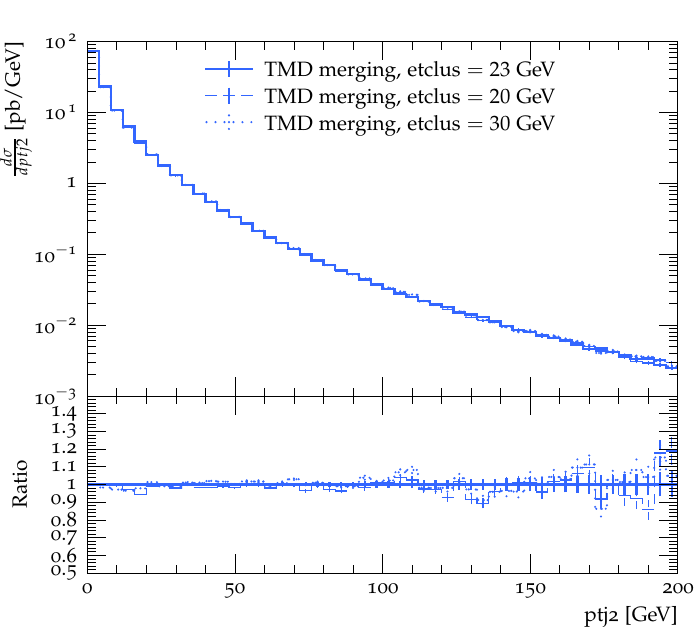}
	\includegraphics[width=.49\textwidth]{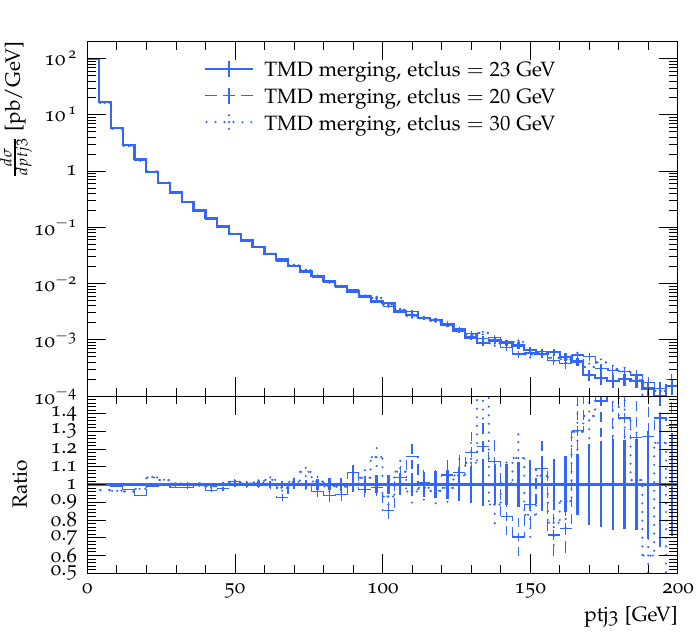}	
	\caption{\tmd{} systematic uncertainty at the LHC at $\sqrt{s}=8$ TeV. The transverse momentum spectra of the leading (top left), second (top right), and third leading (bottom) jets are calculated for three different values of the merging scale, where the solid line represents the default setting.} 	
  \label{fig16}
  \end{center}
\end{figure}

In Fig.~\ref{fig14} we show the transverse momentum (right) and normalized $\phi^{\ast}$ (left) distributions of DY lepton pairs in $pp$ collisions at 8 TeV. We observe that the effect of the merging scale variation is localized around the 
merging scale and it is lower than 10\%.

In Fig.~\ref{fig15} we show the DJRs  for $d_{01}$ (top left), $d_{12}$ (top right), $d_{23}$ (bottom) distributions respectively for a 10 GeV variation of the merging scale. The calculations are performed for DY lepton pair production in $pp$ collisions at 8 TeV. As discussed in previous sections the $d_{n,n+1}$ distribution is very sensitive to the merging scale choice. We observe that  the effect of the merging scale variation is localized around the merging scale, where small kinks in the spectra appear. The effects are however small, lower than 20\%, reducing the systematic uncertainty observed by all merging algorithms discussed in Ref.~\cite{Alwall:2007fs}.
In Fig.~\ref{fig16} we show the transverse momentum spectra of the leading (top left), second (top right), and third  (bottom) jets for DY production in association with jets in $pp$ collisions at 8 TeV. We observe that the effect of the merging scale variation is localized around the merging scale and it is lower than 8\%.

The above results for the total multi-jet rates,  the DJRs and  the transverse 
momentum spectra together build a picture 
 indicating  that the systematic uncertainties from multi-jet merging are reduced when  the transverse momentum recoils in the shower evolution are treated   
    through TMD distributions.

%\newpage

\section{Comparison with collinear multi-jet merging}
\label{sec:comp-pythia} 

In this section we present a comparison of our  results,  based 
on TMD jet merging,  with results from collinear  jet merging.  

To be specific, we compare the TMD merging calculation with a 
calculation in which  the initial-state TMD shower evolution 
is replaced by collinear shower evolution and the TMD merging  is 
replaced by the MLM merging, while the matrix element and 
final-state shower evolution are kept the same in the two calculations. 
Since the final-state parton shower in our \tmd{} calculation 
uses the parton shower routine PYSHOW of {\sc Pythia}6~\cite{Sjostrand:2006za},  we 
compare our \tmd{} results with the results which we obtain by  using 
 {\sc Madgraph}+{\sc Pythia}6  with MLM merging. 

\begin{figure}[hbtp]
  \begin{center}
	\includegraphics[width=.49\textwidth]{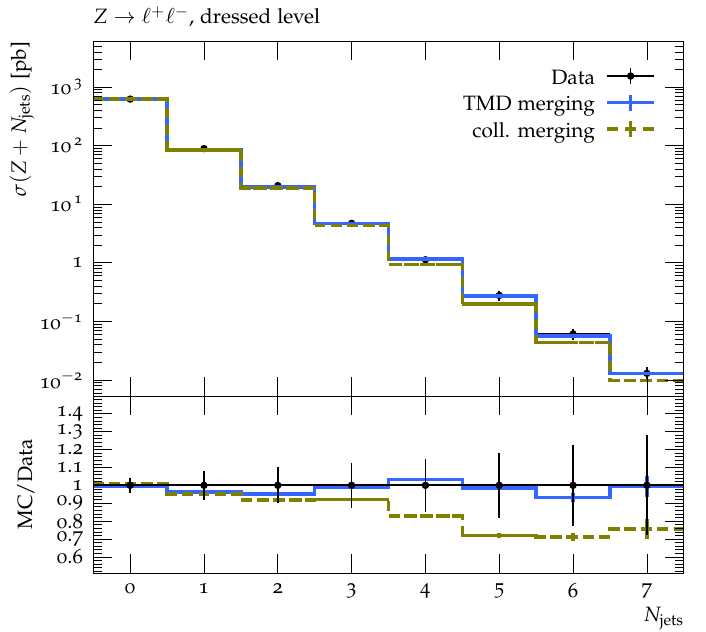}
	\includegraphics[width=.49\textwidth]{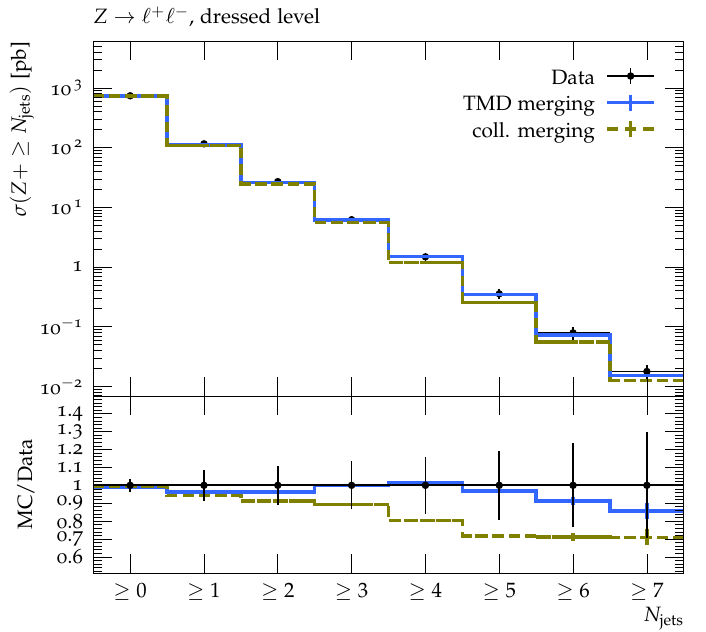}
  \caption{Predictions obtained using  {\sc Madgraph}+{\sc Pythia}6 with MLM merging and the \tmd{} framework are compared for exclusive (left) and inclusive (right) jet multiplicity distributions in the production of a $Z$-boson in association with jets, at $\sqrt{s}=13$~TeV. The phase space for the calculation follows the one in \cite{Aaboud:2017hbk}, whose data are included in these and subsequent plots.}	
  \label{fig32}
  \end{center}	
\end{figure}

\begin{figure}[hbtp]
  \begin{center}
	\includegraphics[width=.49\textwidth]{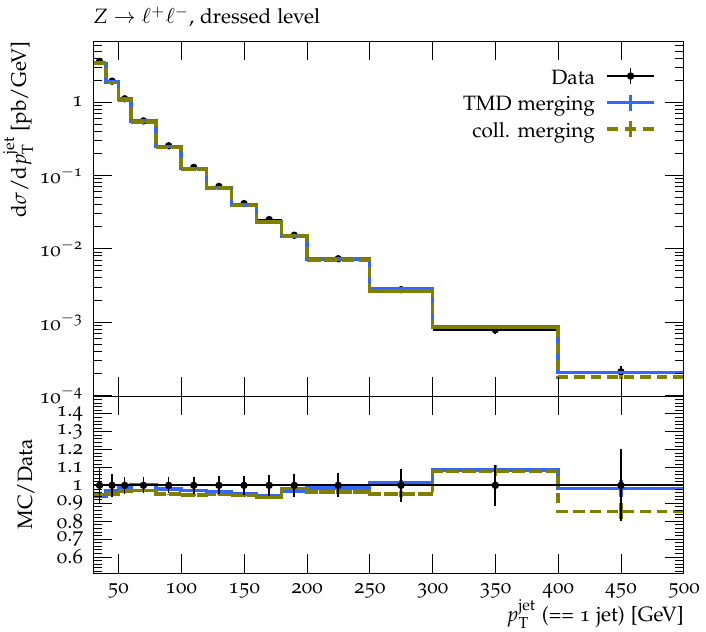}
	\includegraphics[width=.49\textwidth]{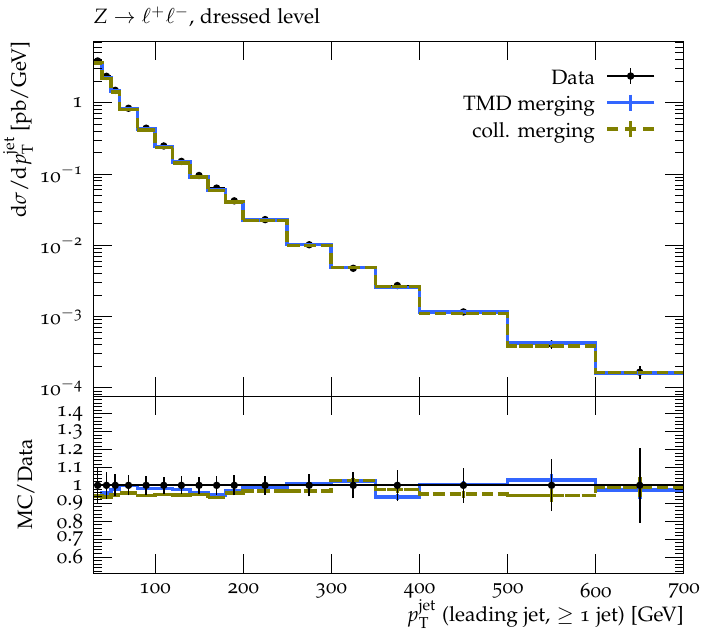}
  \caption{Predictions obtained using  {\sc Madgraph}+{\sc Pythia}6  with MLM merging and the \tmd{} framework are compared for exclusive (left) and inclusive (right) leading jet $p_T$ spectra in the production of a $Z$-boson in association with jets. The phase space for the calculation follows the one in \cite{Aaboud:2017hbk}.}	
  \label{fig33}
  \end{center}	
\end{figure}

\begin{figure}[hbtp]
  \begin{center}
	\includegraphics[width=.49\textwidth]{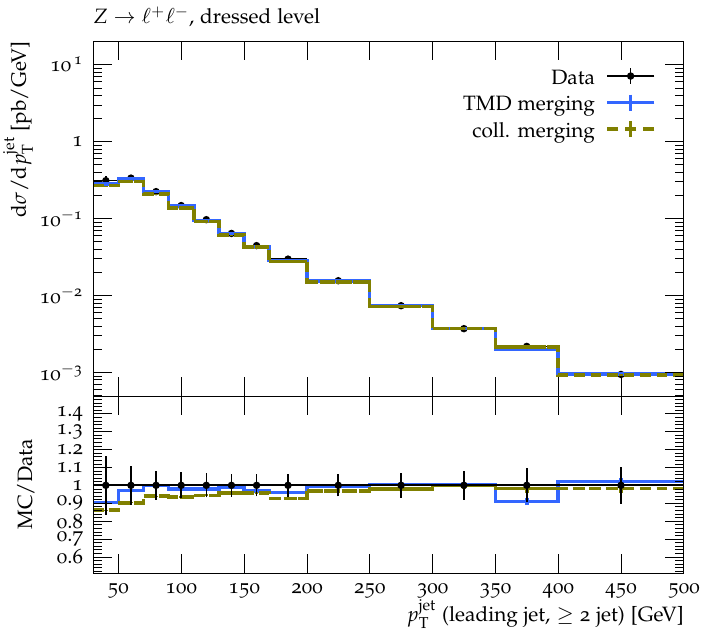}
	\includegraphics[width=.49\textwidth]{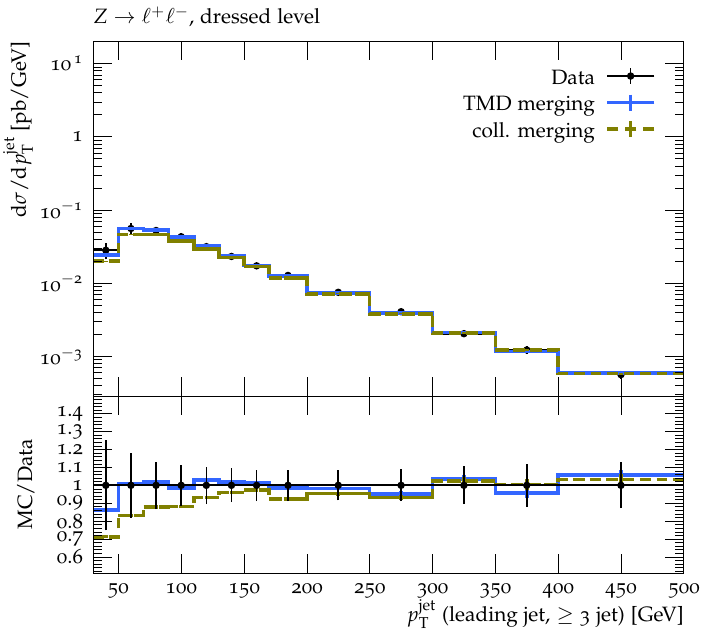}
	\includegraphics[width=.49\textwidth]{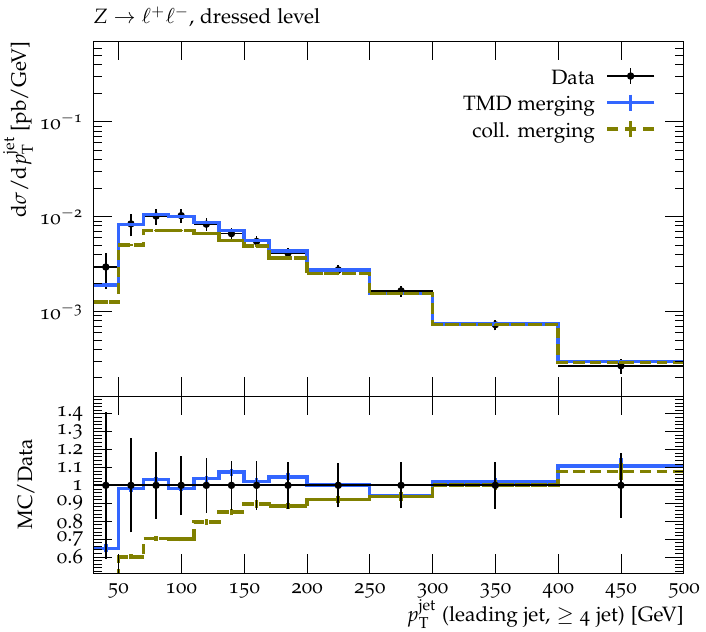}	
  \caption{Predictions obtained using  {\sc Madgraph}+{\sc Pythia}6 with MLM merging, and the \tmd{} framework are compared for the leading jet $p_T$ spectrum in inclusive $Z+$2 (top left), 3 (top right), and 4 (bottom) jets. The phase space for the calculation follows the one in \cite{Aaboud:2017hbk}.}	  
  \label{fig34}
  \end{center}
\end{figure}

\begin{figure}[hbtp]
  \begin{center}
	\includegraphics[width=.49\textwidth]{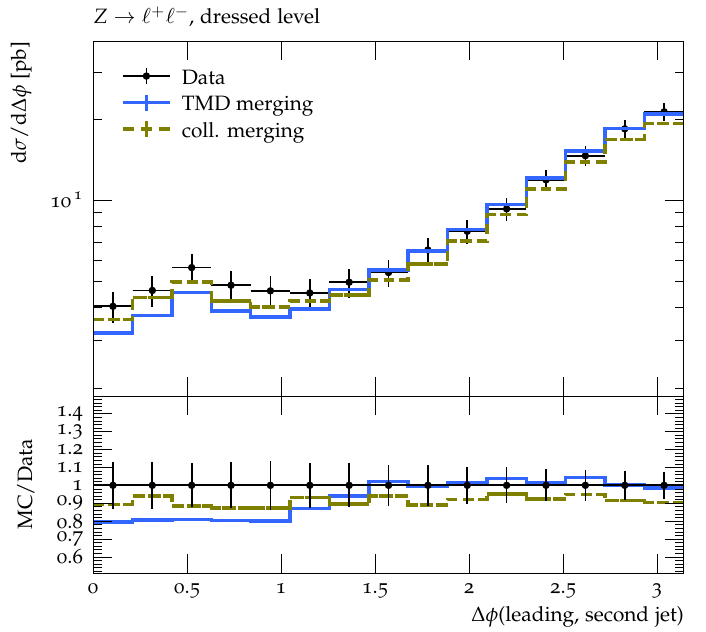}
	\includegraphics[width=.49\textwidth]{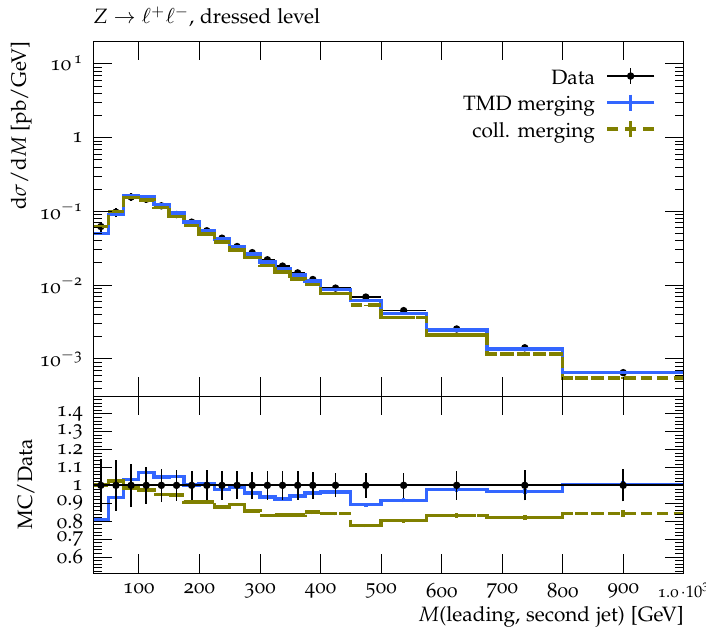}
  \caption{Predictions obtained using  {\sc Madgraph}+{\sc Pythia}6 with MLM merging and the \tmd{} framework are compared for the di-jet azimuthal separation (left) and di-jet mass (right) distributions for $Z+ \geq$2 jets events. The phase space for the calculation follows the one in \cite{Aaboud:2017hbk}.}	  
  \label{fig35}
  \end{center}
\end{figure}

\begin{figure}[hbtp]
  \begin{center}
	\includegraphics[width=.49\textwidth]{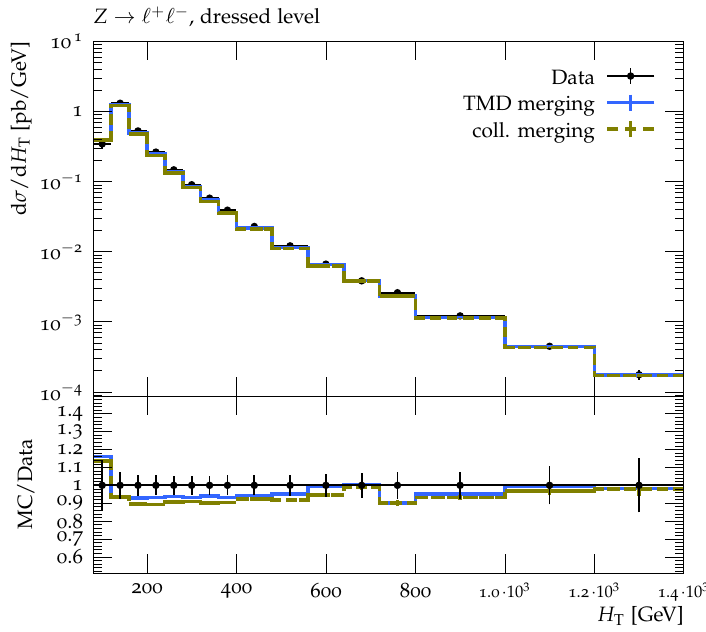}
	\includegraphics[width=.49\textwidth]{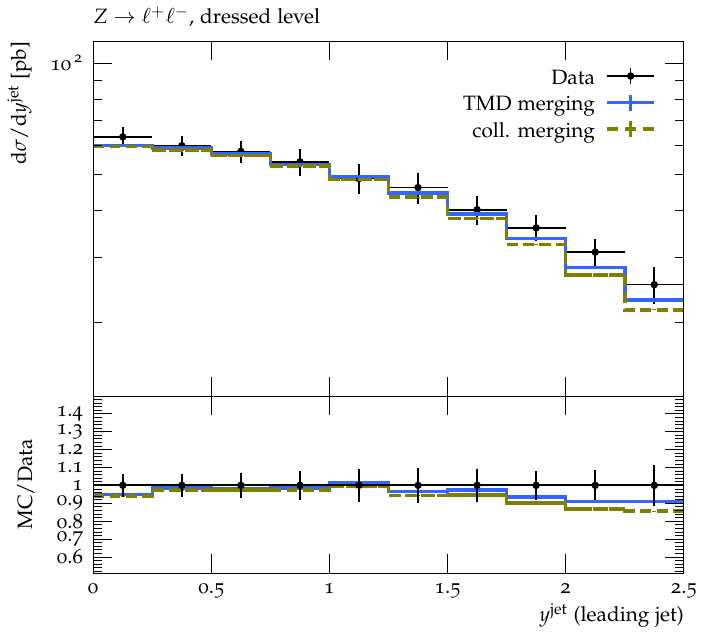}
  \caption{Predictions obtained using  {\sc Madgraph}+{\sc Pythia}6  with MLM merging and the \tmd{} framework are compared for the scalar sum $H_T$ of the transverse momenta of leptons and jets (left), and leading jet rapidity (right) distributions in $Z+$ jets events. The phase space for the calculation follows the one in \cite{Aaboud:2017hbk}.}	  
  \label{fig36}
  \end{center}
\end{figure}

We have first verified that, as expected,  at the matrix element (ME) level 
our results coincide with those of {\sc Madgraph} when using the same generation cuts. 
Next we have verified that,  when considering only the final-state parton shower 
(without including any PB-TMD evolution and initial-state shower),  
our \tmd{} results agree  with the results obtained with  
{\sc Madgraph}+{\sc Pythia}6 computations. 
We have finally 
compared the full \tmd{} and the {\sc Madgraph} + {\sc Pythia}6  calculations. 
Results are reported in Figs.~\ref{fig32}-\ref{fig36}. 


Clear differences emerge in distributions that are most sensitive to higher-order shower emissions, in particular the overall jet multiplicity, shown in Fig.~\ref{fig32}, and the $p_T$ spectrum of the leading jet in final states with at least 4 jets, shown in Fig.~\ref{fig34}. The better agreement of the \tmd{} calculation with data, relative to the canonical MLM-matching procedure implemented in the {\sc Madgraph}+{\sc Pythia}6 
 result, together with the observation made previously that the two approaches are equivalent when limited to the final-state showers only, reinforce the conclusion 
 that the TMD initial-state evolution leads to a better description of 
 higher-order, non-collinear emissions. 

%\newpage

\newpage

\section{Studies of PDF and intrinsic $k_T$ dependence}
\label{sec:pdfsens} 

In the  previous sections  we have studied multi-jet merging methods and we have concentrated in particular on the implications of 
TMD evolution (see Figs.~\ref{fig-gTMD} and \ref{fig-iTMD-vs-kTmin}) on multi-jet observables, examining in detail the example of 
$Z$ + jets production at the LHC.  One may wonder about the role  of the nonperturbative TMD parameters at the  initial (low) scale of the 
TMD evolution in the analysis of the multi-jet observables, and also about the role of PDFs, i.e., of the nonperturbative collinear parameters.  

In traditional applications of TMD physics such as the low-$p_T$ region of DY vector-boson spectra, both 
nonperturbative TMD contributions and PDFs  play an essential and intricate role, as recently discussed e.g. in~\cite{Hautmann:2020cyp,Bury:2022czx}.  
Analogous scenarios are explored 
for processes to be measured by future DIS lepton-hadron experiments in~\cite{Proceedings:2020eah,LHeC:2020van}. 
In this section we see that the case of multi-jet physics is significantly different in this  respect. That is, at the jet scales of  the final states 
investigated in  Secs.~\ref{sec:Zjets}, \ref{sec:syst} and \ref{sec:comp-pythia},   one is  sensitive to the large-$k_T$ tails  induced by TMD evolution  
(illustrated in Fig.~\ref{fig-iTMD-vs-kTmin}), while having very little  sensitivity to   the intrinsic $k_T$  distributions at low evolution scales and to 
differences among  PDF sets. This means that, on one hand,  the large-$k_T$ tails  
due to  multiple QCD radiation embodied by TMD evolution 
 contribute  to the  improved description of high-multiplicity final states, 
 as we have  seen in Secs.~\ref{sec:Zjets} and \ref{sec:comp-pythia}, and to the 
reduction of the merging systematic uncertainty, as we have seen   in Sec.~\ref{sec:syst}; on the other hand, 
these predictions do not  really depend  on   intrinsic $k_T$  distributions or PDFs.

\begin{figure}[hbtp]
  \begin{center}
	\includegraphics[width=.49\textwidth]{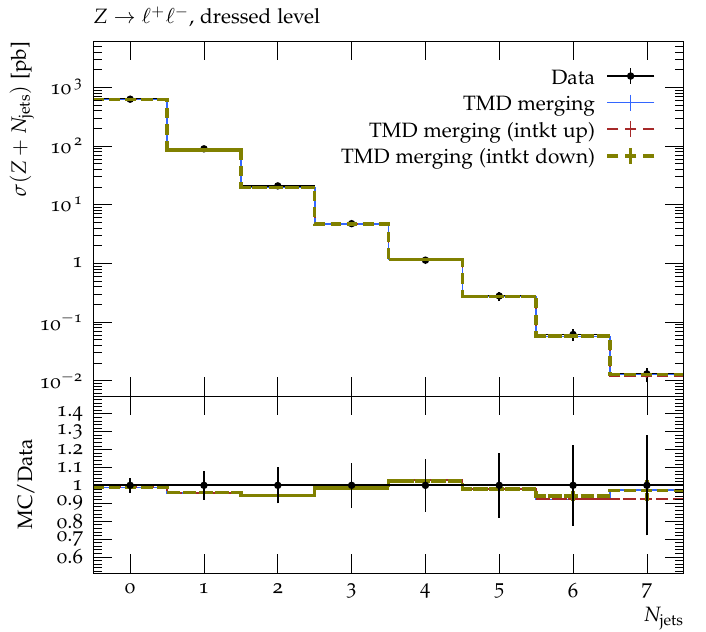}
	\includegraphics[width=.49\textwidth]{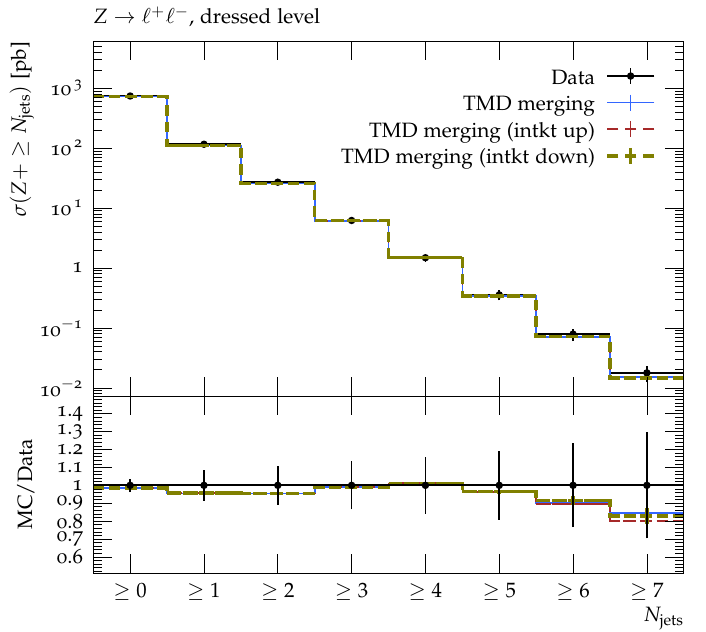}
  \caption{Exclusive (left) and inclusive (right) jet multiplicity distributions in $Z$+jets:  intrinsic-$k_T$ sensitivity.}	
  \label{fig17intkt}
  \end{center}	
\end{figure}	
\begin{figure}[hbtp]
  \begin{center}
	\includegraphics[width=.49\textwidth]{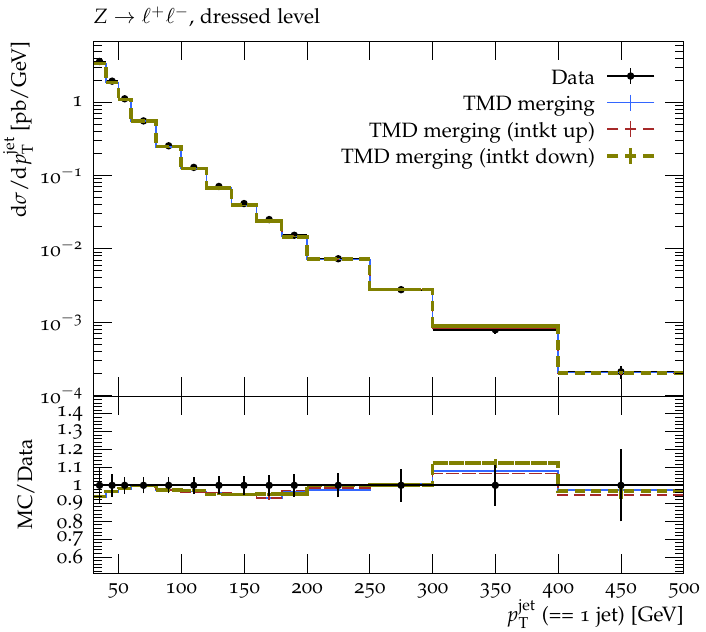}
	\includegraphics[width=.49\textwidth]{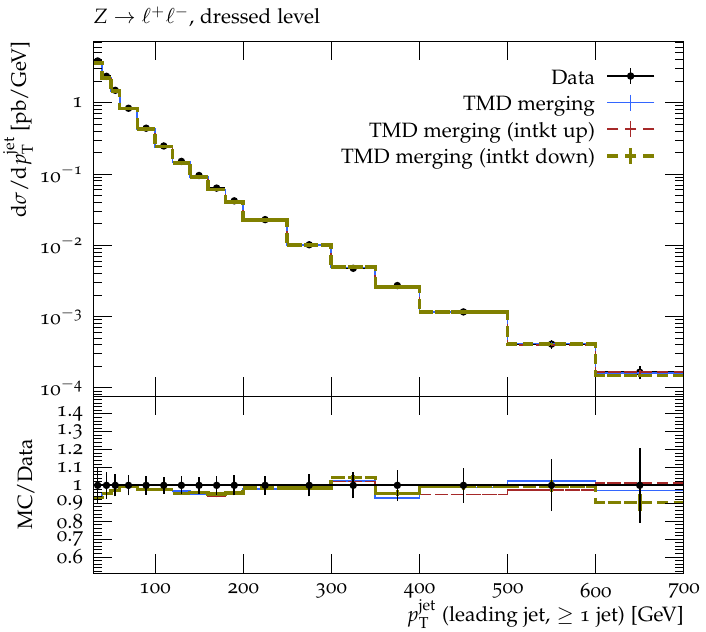}
	\caption{Exclusive (left) and inclusive (right) leading-jet $p_T$ spectra in $Z$+jet events:  intrinsic-$k_T$ sensitivity.}
  \label{fig18intkt}
  \end{center}	
\end{figure}

\begin{figure}[hbtp]
  \begin{center}
	\includegraphics[width=.49\textwidth]{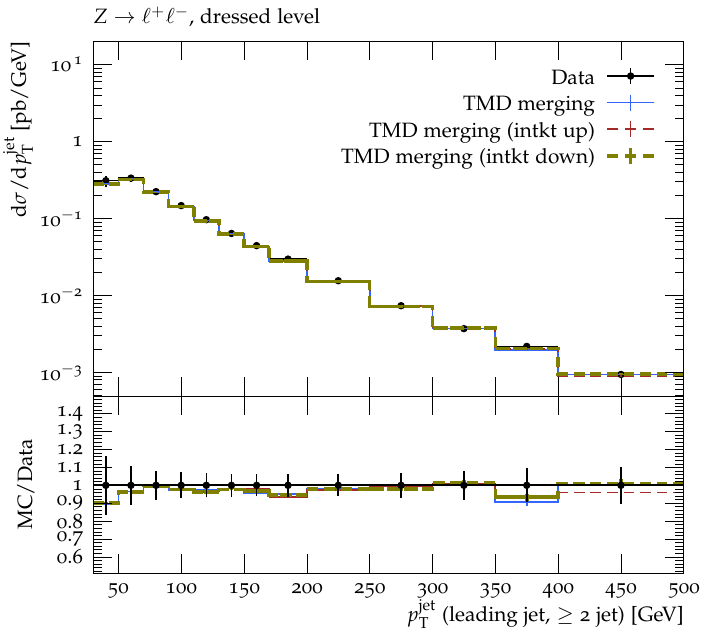}
	\includegraphics[width=.49\textwidth]{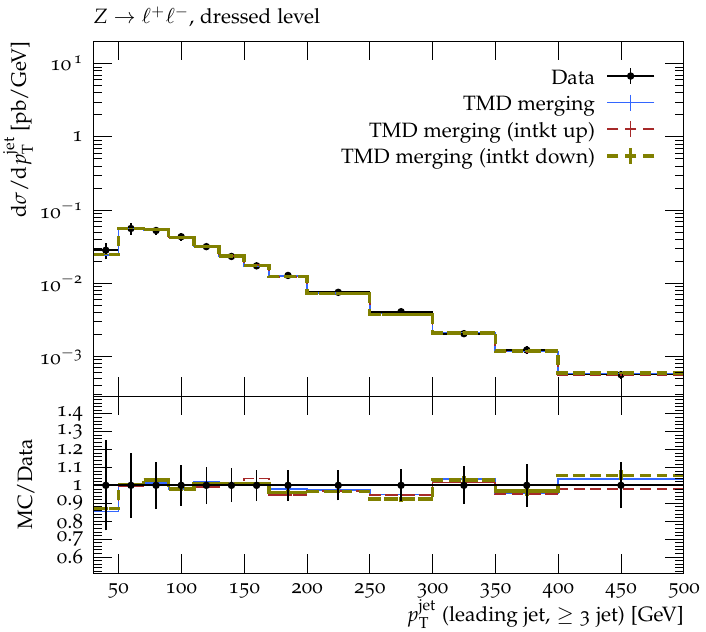}
	\includegraphics[width=.49\textwidth]{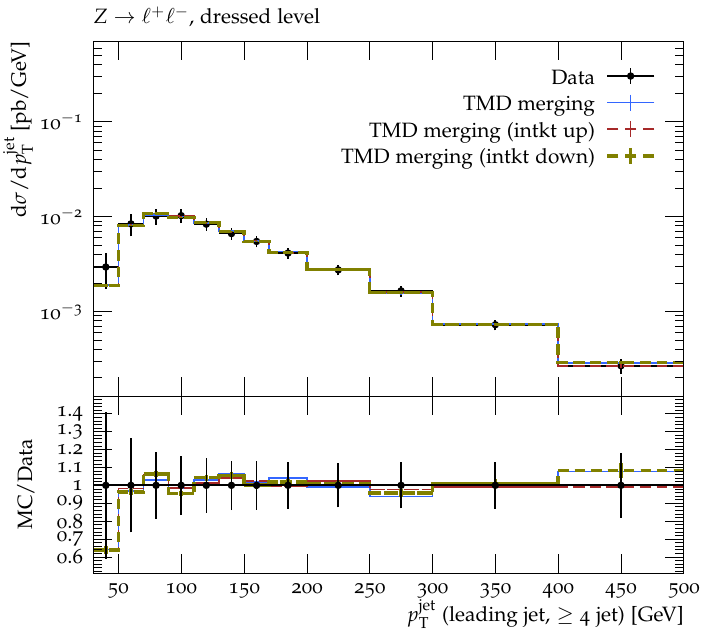}
	\caption{Leading jet $p_T$ spectrum in inclusive $Z+$2 (top left), 3 (top right), and 4 (bottom) jets: intrinsic-$k_T$ sensitivity.}  
  \label{fig19intkt}
  \end{center}
\end{figure}

\begin{figure}[hbtp]
  \begin{center}
	\includegraphics[width=.49\textwidth]{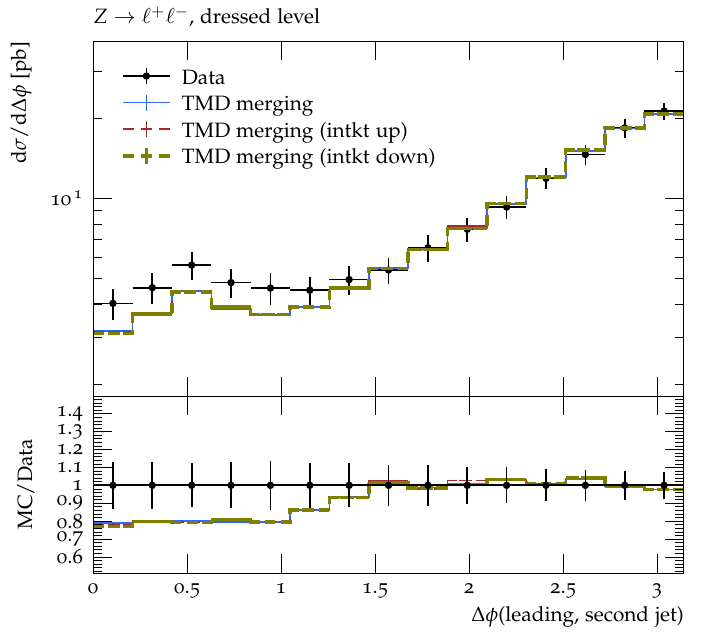}
	\includegraphics[width=.49\textwidth]{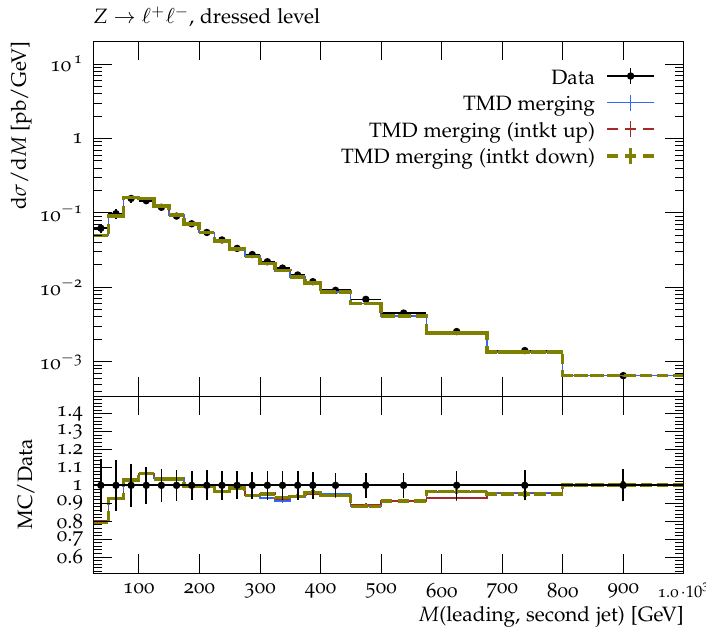} 
	\caption{Di-jet azimuthal separation (left) and di-jet mass (right) distributions for $Z+ \geq$2 jets events: intrinsic-$k_T$ sensitivity.}	 
  \label{fig20intkt}
  \end{center}
\end{figure}

\begin{figure}[hbtp]
  \begin{center}
	\includegraphics[width=.49\textwidth]{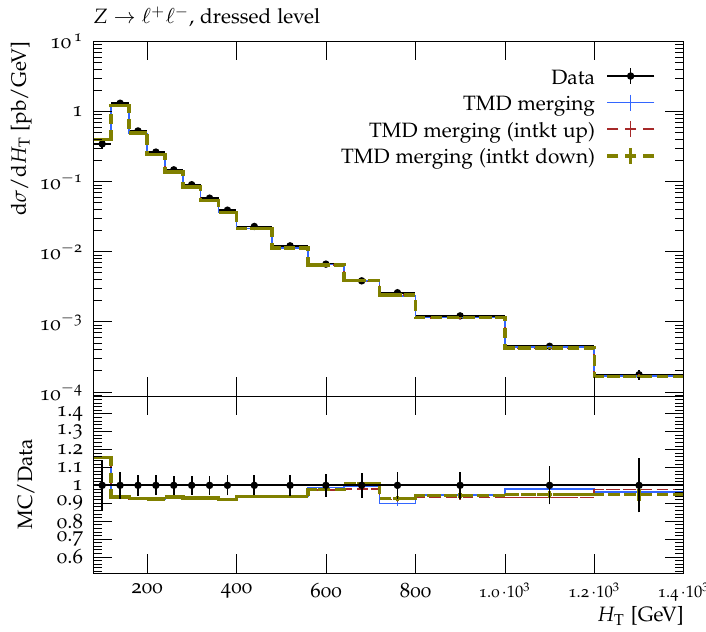}
	\includegraphics[width=.49\textwidth]{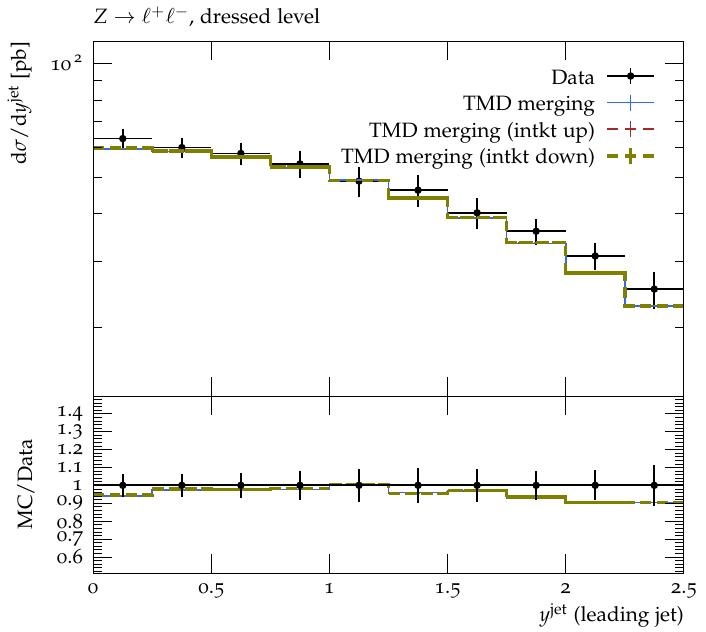} 
	\caption{Scalar sum $H_T$ of the transverse momenta of leptons and jets (left), and leading jet rapidity (right) distributions for $Z+$ jets events: intrinsic-$k_T$  sensitivity.}	 
  \label{fig21intkt}
  \end{center}
\end{figure}

In Figs.~\ref{fig17intkt}-\ref{fig21intkt} we  study the  intrinsic $k_T$ dependence.  
We show results for multi-jet observables obtained in three different 
 intrinsic-$k_T$ scenarios. 
The solid blue curves correspond to the  TMD merging results of 
Sec.~\ref{sec:Zjets} obtained with the TMD parton distributions of PB-TMD Set 2  
extracted in~\cite{Martinez:2018jxt} from fits to DIS data.  This TMD set has 
an intrinsic $k_T$ distribution at the initial evolution scale $\mu_0 = 1.4$ GeV 
described by a gaussian parameterization 
with width $\sigma = 355$ MeV.  
The dashed red and green curves are obtained by varying this 
intrinsic-$k_T$ value by a factor of 2 up and down.  This amounts to 
a very significant distortion  in the $k_T$ distribution at the initial scale.  
Similar variations are studied in the case of DY transverse momentum 
distributions  in Refs.~\cite{Martinez:2019mwt,Martinez:2020fzs}. The  results  in 
Figs.~\ref{fig17intkt}-\ref{fig21intkt} 
show that the predictions for multi-jet observables 
change little under such variations, with the changes being 
 smaller than the experimental uncertainties.

\clearpage

\begin{figure}[hbtp]
  \begin{center}
	\includegraphics[width=.49\textwidth]{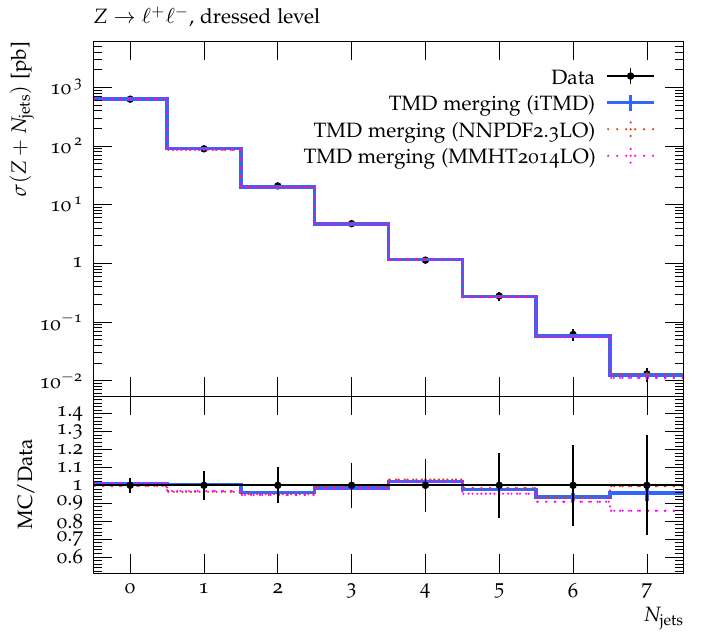}
	\includegraphics[width=.49\textwidth]{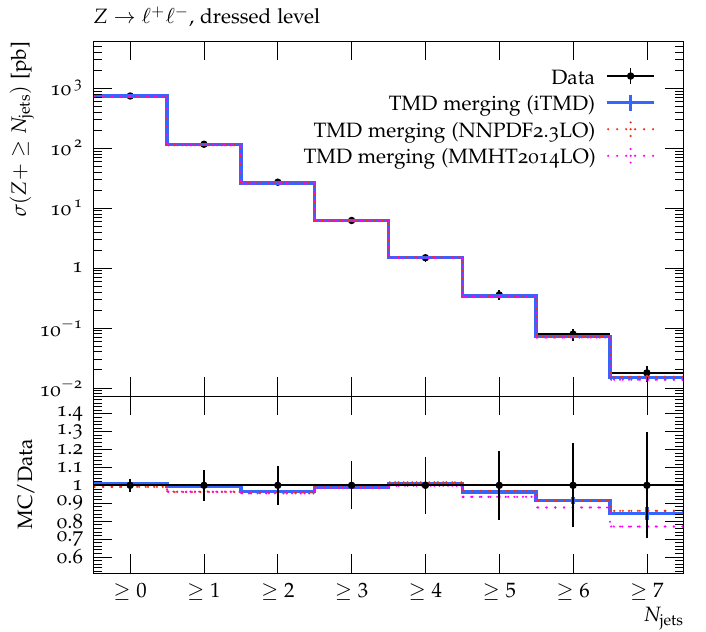}
  \caption{Exclusive (left) and inclusive (right) jet multiplicity distributions in $Z$+jets: PDF sensitivity.}	
  \label{fig17}
  \end{center}	
\end{figure}

\begin{figure}[hbtp]
  \begin{center}
	\includegraphics[width=.49\textwidth]{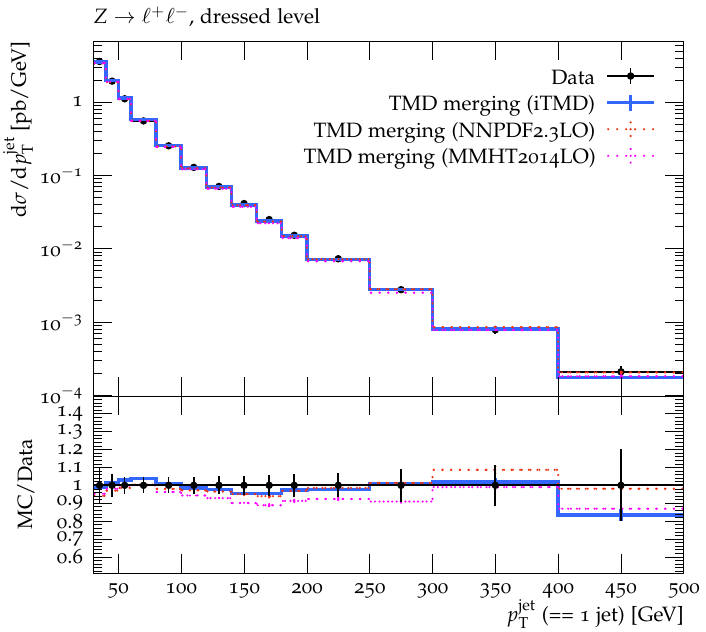}
	\includegraphics[width=.49\textwidth]{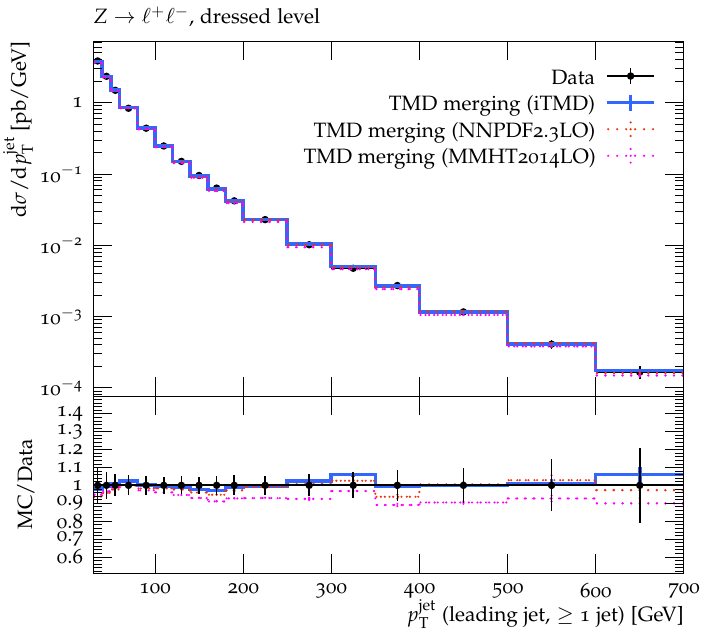}
	\caption{Exclusive (left) and inclusive (right) leading-jet $p_T$ spectra in $Z$+jet events: PDF sensitivity.}
  \label{fig18}
  \end{center}	
\end{figure}

\begin{figure}[hbtp]
  \begin{center}
	\includegraphics[width=.49\textwidth]{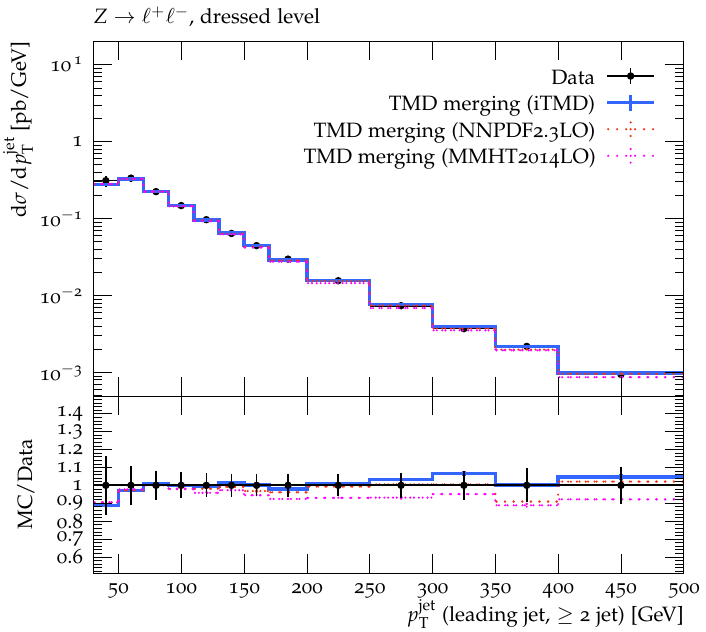}
	\includegraphics[width=.49\textwidth]{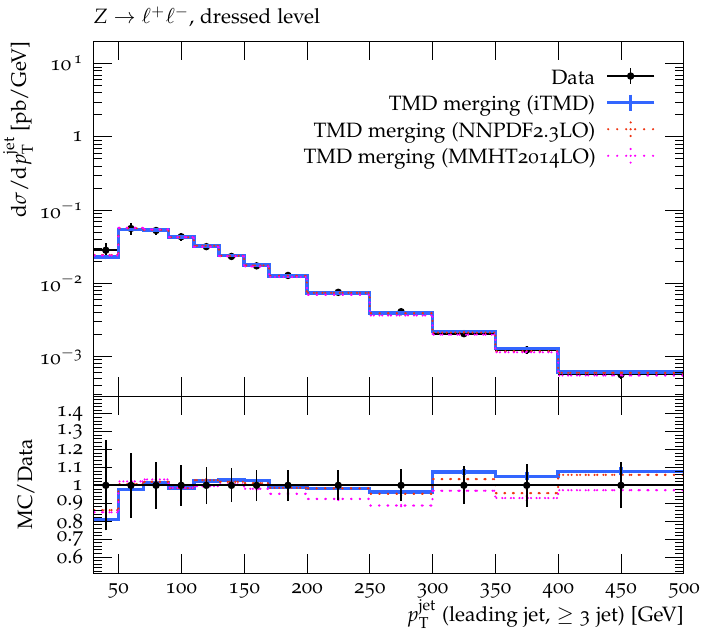}
	\includegraphics[width=.49\textwidth]{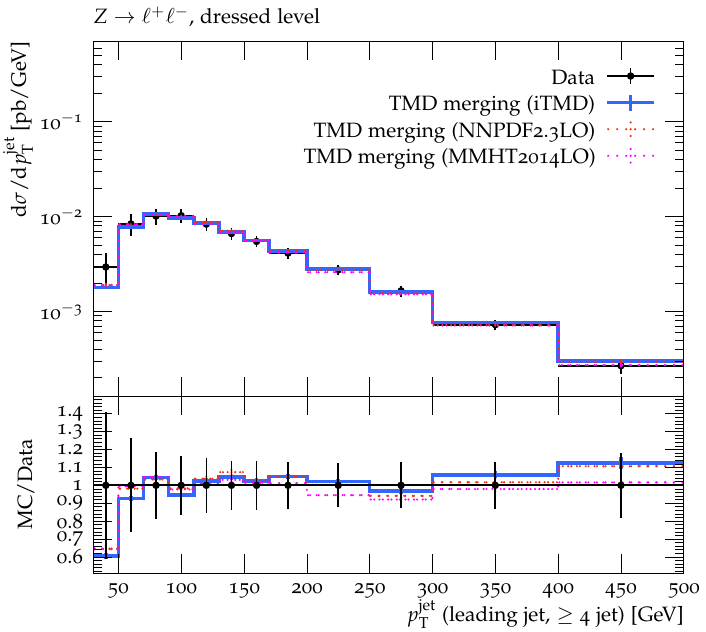}
	\caption{Leading jet $p_T$ spectrum in inclusive $Z+$2 (top left), 3 (top right), and 4 (bottom) jets: PDF sensitivity.}  
  \label{fig19}
  \end{center}
\end{figure}

\begin{figure}[hbtp]
  \begin{center}
	\includegraphics[width=.49\textwidth]{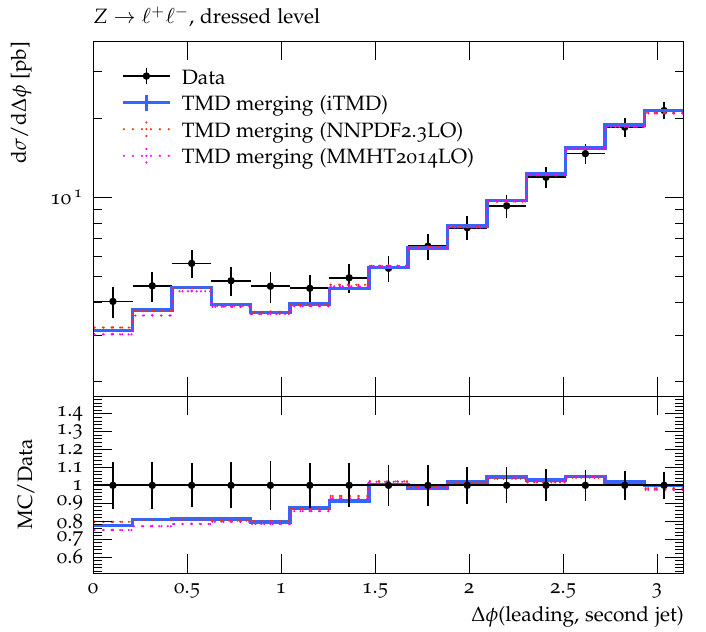}
	\includegraphics[width=.49\textwidth]{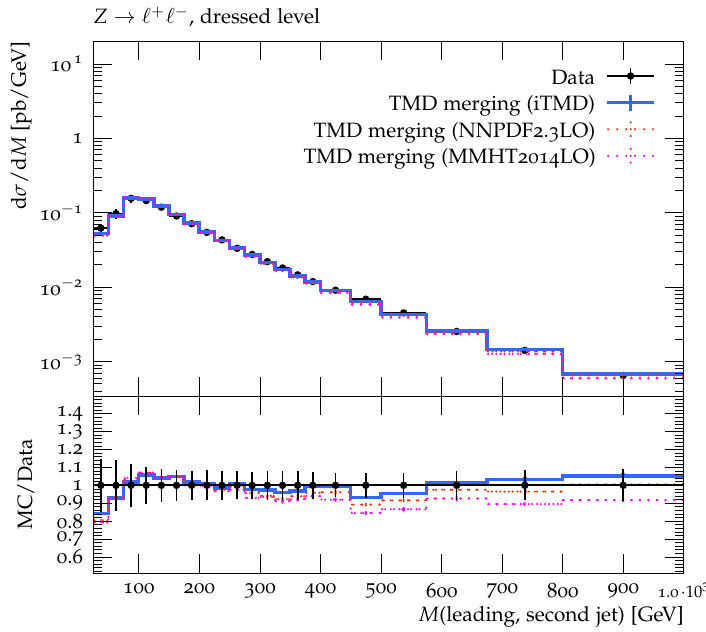} 
	\caption{Di-jet azimuthal separation (left) and di-jet mass (right) distributions for $Z+ \geq$2 jets events: PDF sensitivity.}	 
  \label{fig20}
  \end{center}
\end{figure}

\begin{figure}[hbtp]
  \begin{center}
	\includegraphics[width=.49\textwidth]{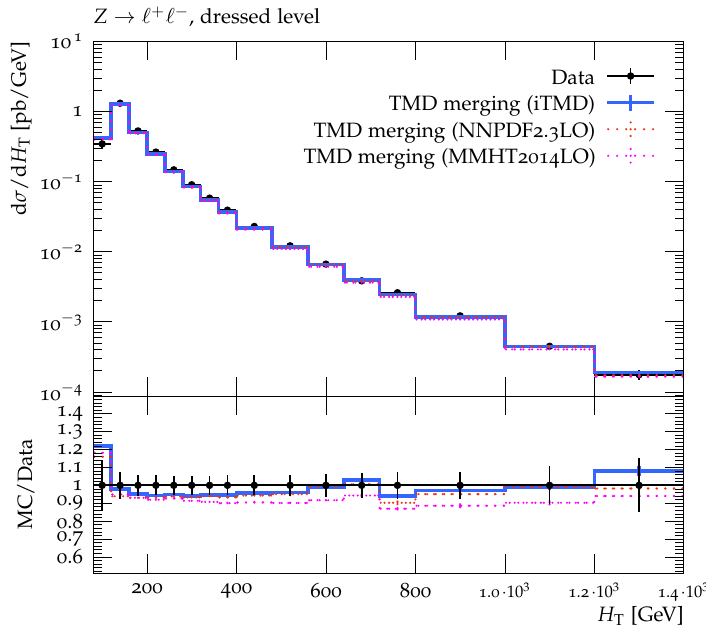}
	\includegraphics[width=.49\textwidth]{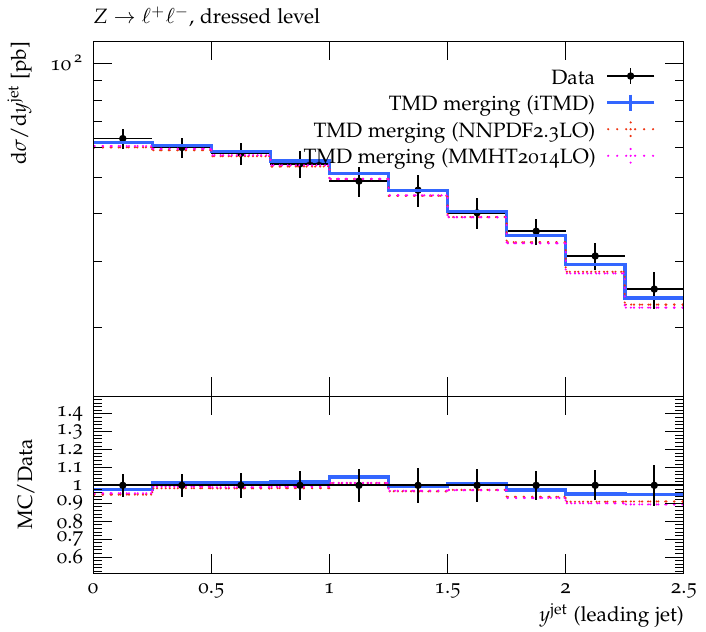} 
	\caption{Scalar sum $H_T$ of the transverse momenta of leptons and jets (left), and leading jet rapidity (right) distributions for $Z+$ jets events: PDF sensitivity.}	 
  \label{fig21}
  \end{center}
\end{figure}

We next investigate the effect of using different choices of collinear parton distributions in the calculation of the parton-level ME samples. The nominal choice of the integrated TMD parton density for the ME calculation is compared to the results obtained using the NNPDF2.3LO~\cite{Ball:2012cx}  
 and the MMHT2014LO~\cite{Harland-Lang:2014zoa} parton distributions. The comparisons are shown in Figs.~\ref{fig17}-\ref{fig21}. In general we observe 
 that a change of the collinear PDF used in the ME calculation has a small effect in the calculation compared to the experimental uncertainty.  

\clearpage

\section{Remarks on (off-shell) TMD effects in matrix elements}
\label{sec:comm-x} 

In this work we have investigated the impact of TMD evolution and 
$k_T$ broadening,  depicted in  Fig.~\ref{fig-iTMD-vs-kTmin},  on the structure 
of multi-jet production. We have not, however, studied the possible 
TMD effects which may arise at the level 
of hard-scattering matrix elements~\cite{Catani:1990eg}, as e.g.~in the 
parton-level event generator described in Ref.~\cite{vanHameren:2016kkz}. 
It is known that off-shell TMD matrix elements become important in the 
high-energy (Regge) limit, where they control  small-$x$ logarithmic 
resummations of high-energy cross sections.  
Our work has not been aimed at treating the Regge region, and thus  
the off-shell matrix elements~\cite{Catani:1990eg,vanHameren:2016kkz}   
have not been used. 
Nevertheless, in this section we provide a few remarks on this topic, and on the 
relationship between the $k_T$-dependent (off-shell) matrix elements 
and the approach used in this work. 

To this end, we recall~\cite{Catani:1990xk} that the 
small-$x$  resummation of 
high-energy cross sections is precisely achieved 
by convoluting the hard cross sections 
obtained from off-shell TMD matrix elements with the gluon Green's 
functions obtained from the solution of the 
BFKL equation~\cite{Lipatov:1976zz,Kuraev:1977fs,Balitsky:1978ic}. 
Such matrix elements, although off-shell,  
can be understood  gauge-invariantly  as 
high-energy limits of  amplitudes for $n$+2 particle production, 
representing the reggeized  initial states. 
  As 
 functions of the hard-scattering scale $\mu$ and the 
gluon's transverse momentum $k_T$ 
initiating the hard scattering, they behave as follows. For the gluon's off-shellness 
going to zero, $k_T \ll \mu$, the 
collinear matrix elements  are recovered. For finite 
$k_T / \mu$, corrections to the collinear matrix elements arise, 
which may be computed as a 
power series expansion  in $(k_T / \mu)^n$. The summation 
of these corrections leads to a ``dynamical'' cut-off of the partonic 
cross section at $k_T$ of the order of the hard scale,  $k_T \ltap \mu$, 
and to a falling transverse momentum tail for  $k_T \gtap \mu$. 
Explicit examples may be found 
e.g.~in~\cite{Catani:1992zc}, figure 4, and~\cite{Marchesini:1992jw}, figure 2.

Now let us convolute the  cross section $ {\hat \sigma} ( k_T)$  
obtained from the matrix elements described above 
 with the TMD parton distribution. As shown 
 in~\cite{Catani:1990xk}, this yields (by a suitable double Mellin 
transformation, performed both in transverse momentum and in 
energy rather than just energy as in the collinear case) the  
resummation of high-energy (Regge) logarithms if 
we take BFKL solutions for the TMD.  On the other hand, if  we 
approximate the $k_T$-dependent 
cross section  by a $\Theta$ function via   
$ {\hat \sigma} ( k_T) \sim \Theta (\mu- k_T)  {\hat \sigma} ( 0) $, we  obtain the 
collinear cross section times the integrated TMD density. 

In the latter case, TMD effects do not change the total cross section with respect to the collinearly-factorized calculation but 
affect the structure of the associated jet final states through the changes in the kinematics of the initial state due to 
the $k_T$ in the TMD density. 
This is the scenario of the present work. 
In other words, the  $k_T$ corrections to matrix 
elements~\cite{Catani:1990eg,Marchesini:1992jw,Catani:1992zc,vanHameren:2016kkz}  
are essential if we are to resum Regge logarithms in the total cross section;  but to analyze the kinematical effects of the $k_T$ generated by the   
initial-state TMD evolution, which leave the cross section  unchanged but change the jet structure of the events, we may 
work with $  {\hat \sigma} ( 0) $ (i.e., without modifying the weight with respect to collinear matrix elements) and take into account 
effects of $k_T$ in the TMD density. These are 
the effects which we illustrated earlier  in Fig.~2.  
If we, as in the present work, aim at including the contributions 
of higher multiplicity matrix elements along with the TMD evolution illustrated in 
Fig.~2, we need an appropriate TMD merging method, and this is the main 
subject of this paper.

Thus, the approach followed in this paper and the TMD matrix 
element results~\cite{Catani:1990eg,Marchesini:1992jw,vanHameren:2016kkz}
are to be regarded as  complementary. The work in this paper 
does not use the complete 
form of the matrix element but, as outlined above,  it can be viewed as 
stemming from a collinear approximation  
to it, which is sufficient for the purposes of the present study, away 
from  regions strongly affected by 
 small $x$ values, and can be regarded as  
  a starting point for a more complete treatment.  Such a treatment, 
   in order to 
address the truly Regge (high-energy) region,  would 
have to incorporate explicitly the effects of $k_T$ not only in the kinematics and  
TMD distributions but also in the  matrix elements.   An appropriate 
generalization of the TMD merging approach presented in this paper will be 
 required in this case.

\section{Conclusions}
\label{sec:concl} 
We presented in this work a new development for the description of multi-jet final states in hadronic collisions, which complements the standard approach,  relying on merging samples of different parton multiplicity showered through emissions in the collinear approximation, with the use of the TMD parton branching for the initial state evolution. We carried out an extensive study of the theoretical systematic uncertainties associated with this new approach, and applied our results to a first comparison with LHC ATLAS data on $Z$ plus multijet production. Our key findings can be summarized as follows: (i) a reduced systematic uncertainty with respect to the merging parameters and (ii) an improved description of high-order emissions, giving a better agreement for final states with jet multiplicity larger than the largest multiplicity used in the generation of the matrix-element samples. Our study used MLM matching as a merging criterion; we expect that the technique can be applied also to other LO merging schemes, such as CKKW-L. Since the introduction of the LO merging, and thanks to the automation of NLO ME calculations, more powerful NLO merging algorithms have been introduced. We hope that our work will inspire the exploration of a possible extension of the \tmd{} approach to the NLO case as well.

\vskip 0.3 cm 

\noindent 
{\bf Acknowledgments.}  We thank H.~Jung for many 
discussions and advice on Monte Carlo simulations. 

\bibliographystyle{JHEP}

\begin{thebibliography}{10}


\bibitem{Azzi:2019yne}
  P.~Azzi {\it et al.},
  %``Report from Working Group 1 : Standard Model Physics at the HL-LHC and HE-LHC,''
  CERN Yellow Rep.\  Monogr.\  {\bf 7} (2019)   1 
%  
  [arXiv:1902.04070 [hep-ph]].


\bibitem{Mangano:2016jyj}
  M.~L.~Mangano {\it et al.},
  %``Physics at a 100 TeV pp Collider: Standard Model Processes,''
  CERN Yellow Rep.\  (2017)  no.3, 1 
%  
  [arXiv:1607.01831 [hep-ph]].


\bibitem{Catani:2001cc}
  S.~Catani, F.~Krauss, R.~Kuhn and B.~R.~Webber,
  %``QCD matrix elements + parton showers,''
  JHEP {\bf 0111} (2001) 063
%  
  [hep-ph/0109231].


\bibitem{Lonnblad:2001iq}
  L.~Lonnblad,
  %``Correcting the color dipole cascade model with fixed order matrix elements,''
  JHEP {\bf 0205} (2002) 046
%  
  [hep-ph/0112284].

\bibitem{Mangano:2002} 
M.~Mangano, "Exploring theoretical systematics in the ME-to-shower MC merging for multijet process", Matrix Element/Monte Carlo Tuning Working Group, Fermilab, November 16 2002. 

%\bibitem{Mangano:2001rw}
%  M.~L.~Mangano, M.~Moretti and R.~Pittau,
  %``Multijet matrix elements and shower evolution in hadronic collisions: $W b \bar{b}$ + $n$ jets as a case study,''
%Nucl.\ Phys.\ B {\bf 632} (2002) 343
%  
%  [hep-ph/0108069 [hep-ph]].

%\cite{Mrenna:2003if}
\bibitem{Mrenna:2003if}
S.~Mrenna and P.~Richardson,
%``Matching matrix elements and parton showers with HERWIG and PYTHIA,''
JHEP \textbf{05} (2004), 040
%doi:10.1088/1126-6708/2004/05/040
[arXiv:hep-ph/0312274 [hep-ph]].

\bibitem{Mangano:2006rw}
  M.~L.~Mangano, M.~Moretti, F.~Piccinini and M.~Treccani,
  %``Matching matrix elements and shower evolution for top-quark production in hadronic collisions,''
  JHEP {\bf 0701} (2007) 013
%  
  [hep-ph/0611129 [hep-ph]].



\bibitem{Alwall:2007fs}
  J.~Alwall {\it et al.},
  %``Comparative study of various algorithms for the merging of parton showers and matrix elements in hadronic collisions,''
  Eur.\ Phys.\ J.\ C {\bf 53} (2008)   473
%  
  [arXiv:0706.2569 [hep-ph]].

\bibitem{Frederix:2012ps}
  R.~Frederix and S.~Frixione,
  %``Merging meets matching in MC@NLO,''
  JHEP {\bf 1212} (2012) 061
%  doi:10.1007/JHEP12(2012)061
  [arXiv:1209.6215 [hep-ph]].

\bibitem{Hoeche:2012yf}
  S.~Hoeche, F.~Krauss, M.~Schoenherr and F.~Siegert,
  %``QCD matrix elements + parton showers: The NLO case,''
  JHEP {\bf 1304} (2013) 027
%  doi:10.1007/JHEP04(2013)027
  [arXiv:1207.5030 [hep-ph]].
  
\bibitem{Lonnblad:2012ix}
  L.~Lonnblad and S.~Prestel,
  %``Merging Multi-leg NLO Matrix Elements with Parton Showers,''
  JHEP {\bf 1303} (2013) 166
%  doi:10.1007/JHEP03(2013)166
  [arXiv:1211.7278 [hep-ph]].

\bibitem{Bellm:2017ktr}
  J.~Bellm, S.~Gieseke and S.~Plaetzer,
  %``Merging NLO Multi-jet Calculations with Improved Unitarization,''
  Eur.\ Phys.\ J.\ C {\bf 78} (2018)   244
%  doi:10.1140/epjc/s10052-018-5723-2
  [arXiv:1705.06700 [hep-ph]].
  


\bibitem{Lavesson:2008ah}
  N.~Lavesson and L.~Lonnblad,
  %``Extending CKKW-merging to One-Loop Matrix Elements,''
  JHEP {\bf 0812} (2008) 070
%  doi:10.1088/1126-6708/2008/12/070
  [arXiv:0811.2912 [hep-ph]].

\bibitem{Hoeche:2009rj}
  S.~Hoeche, F.~Krauss, S.~Schumann and F.~Siegert,
  %``QCD matrix elements and truncated showers,''
  JHEP {\bf 0905} (2009) 053
%  
  [arXiv:0903.1219 [hep-ph]].


\bibitem{Hamilton:2009ne}
  K.~Hamilton, P.~Richardson and J.~Tully,
  %``A Modified CKKW matrix element merging approach to angular-ordered parton showers,''
  JHEP {\bf 0911} (2009) 038
% 
  [arXiv:0905.3072 [hep-ph]].


\bibitem{Lonnblad:2011xx}
  L.~Lonnblad and S.~Prestel,
  %``Matching Tree-Level Matrix Elements with Interleaved Showers,''
  JHEP {\bf 1203} (2012) 019
%  
  [arXiv:1109.4829 [hep-ph]].


\bibitem{Martinez:2021chk}
A.~Bermudez Martinez, F.~Hautmann and M.~L.~Mangano,  
%``TMD evolution and multi-jet merging,''
Phys. Lett. B \textbf{822} (2021) 136700 
%doi:10.1016/j.physletb.2021.136700
[arXiv:2107.01224 [hep-ph]]. 
  
  
  
\bibitem{Angeles-Martinez:2015sea}
  R.~Angeles-Martinez {\it et al.},
  %``Transverse Momentum Dependent (TMD) parton distribution functions: status and prospects,''
  Acta Phys.\ Polon.\ B {\bf 46} (2015)   2501
%  
  [arXiv:1507.05267 [hep-ph]].


\bibitem{Luisoni:2015xha}
  G.~Luisoni and S.~Marzani,
  %``QCD resummation for hadronic final states,''
  J.\ Phys.\ G {\bf 42} (2015)   103101
%  
  [arXiv:1505.04084 [hep-ph]].


\bibitem{Collins:1984kg}
  J.~C.~Collins, D.~E.~Soper and G.~F.~Sterman,
  %``Transverse Momentum Distribution in Drell-Yan Pair and W and Z Boson Production,''
  Nucl.\ Phys.\ B {\bf 250} (1985) 199.

\bibitem{Catani:1990eg}
  S.~Catani, M.~Ciafaloni and F.~Hautmann,
  %``High-energy factorization and small x heavy flavor production,''
  Nucl.\ Phys.\ B {\bf 366} (1991) 135.

\bibitem{Dooling:2012uw}
  S.~Dooling, P.~Gunnellini, F.~Hautmann and H.~Jung,
  %``Longitudinal momentum shifts, showering, and nonperturbative corrections in matched next-to-leading-order shower event generators,''
  Phys.\ Rev.\ D {\bf 87} (2013)   094009
%  doi:10.1103/PhysRevD.87.094009
  [arXiv:1212.6164 [hep-ph]].

\bibitem{Hautmann:2013fla}
F.~Hautmann,
%``QCD and Jets,''
Acta Phys. Polon. B \textbf{44} (2013) 761; arXiv:1304.8133 [hep-ph]. 


\bibitem{Ebert:2020yqt}
M.~A.~Ebert, B.~Mistlberger and G.~Vita,
%``Transverse momentum dependent PDFs at N$^3$LO,''
JHEP \textbf{09} (2020) 146
%doi:10.1007/JHEP09(2020)146
[arXiv:2006.05329 [hep-ph]].



\bibitem{Luo:2019szz}
M.~x.~Luo, T.~Z.~Yang, H.~X.~Zhu and Y.~J.~Zhu,
%``Quark Transverse Parton Distribution at the Next-to-Next-to-Next-to-Leading Order,''
Phys. Rev. Lett. \textbf{124} (2020) 092001
%doi:10.1103/PhysRevLett.124.092001
[arXiv:1912.05778 [hep-ph]].

\bibitem{Chen:2022cgv}
X.~Chen, T.~Gehrmann, E.~W.~N.~Glover, A.~Huss, P.~F.~Monni, E.~Re, L.~Rottoli and P.~Torrielli,
%``Third-Order Fiducial Predictions for Drell-Yan Production at the LHC,''
Phys. Rev. Lett. \textbf{128} (2022)  252001
%doi:10.1103/PhysRevLett.128.252001
[arXiv:2203.01565 [hep-ph]].


\bibitem{Neumann:2022lft}
T.~Neumann and J.~Campbell,
%``Fiducial Drell-Yan production at the LHC improved by transverse-momentum resummation at N$^4$LL+N$^3$LO,''
arXiv:2207.07056 [hep-ph].




\bibitem{Hautmann:2017fcj}
  F.~Hautmann, H.~Jung, A.~Lelek, V.~Radescu and R.~Zlebcik,
  %``Collinear and TMD Quark and Gluon Densities from Parton Branching Solution of QCD Evolution Equations,''
  JHEP {\bf 1801} (2018) 070
%  doi:10.1007/JHEP01(2018)070
  [arXiv:1708.03279 [hep-ph]].


\bibitem{Martinez:2021dwx}
A.~Bermudez Martinez, F.~Hautmann and M.~L.~Mangano,
%``Multi-jet physics at high-energy colliders and TMD parton evolution,''
arXiv:2109.08173 [hep-ph].

%\bibitem{eswbook}
%          R.~K.~Ellis, W.~J.~Stirling and B.~R.~Webber, 
%          {\it QCD and collider physics}, CUP 1996. 


\bibitem{Webber:1986mc}
    B.~R.~Webber,  Ann.\ Rev.\ Nucl.\ Part.\ Sci.\  {\bf 36} (1986) 253. 

\bibitem{Bengtsson:1987kr}
  H.~U.~Bengtsson and T.~Sjostrand,
  %``The Lund Monte Carlo for Hadronic Processes: Pythia Version 4.8,''
  Comput.\ Phys.\ Commun.\  {\bf 46} (1987) 43.

\bibitem{Marchesini:1987cf}
  G.~Marchesini and B.~R.~Webber,
  %``Monte Carlo Simulation of General Hard Processes with Coherent QCD Radiation,''
  Nucl.\ Phys.\ B {\bf 310} (1988) 461.

\bibitem{Hautmann:2017xtx}
  F.~Hautmann, H.~Jung, A.~Lelek, V.~Radescu and R.~Zlebcik,
  %``Soft-gluon resolution scale in QCD evolution equations,''
  Phys.\ Lett.\ B {\bf 772} (2017) 446
%  doi:10.1016/j.physletb.2017.07.005
  [arXiv:1704.01757 [hep-ph]].	



\bibitem{Catani:1990rr}
  S.~Catani,  G.~Marchesini   and B.~R.~Webber,
  %``QCD coherent branching and semiinclusive processes at large x,''
  Nucl.\ Phys.\ B {\bf 349} (1991) 635.



\bibitem{Hautmann:2019biw}
F.~Hautmann, L.~Keersmaekers, A.~Lelek and A.~M.~Van Kampen,
%``Dynamical resolution scale in transverse momentum distributions at the LHC,''
Nucl.\ Phys.\ B {\bf 949} (2019) 114795
%doi:10.1016/j.nuclphysb.2019.114795
[arXiv:1908.08524].




\bibitem{Gribov:1972ri}
V.~N.~Gribov and L.~N.~Lipatov, Sov.\ J.\ Nucl.\ Phys.\  {\bf 15} (1972) 438. 

\bibitem{Altarelli:1977zs}
  G.~Altarelli and G.~Parisi,      Nucl.\ Phys.\ {\bf B126} (1977) 298.

\bibitem{Dokshitzer:1977sg}
Yu.~L.~Dokshitzer,  Sov.\ J.\ Nucl.\ Phys.\  {\bf 46} (1977)  641.


\bibitem{Botje:2010ay}
  M.~Botje,
  %``QCDNUM: Fast QCD Evolution and Convolution,''
  Comput.\ Phys.\ Commun.\  {\bf 182} (2011) 490. 



\bibitem{Baranov:2021uol}
S.~Baranov \textit{et al.}, 
 Eur.\ Phys.\ J.\   {\bf C81} (2021) 425 
[arXiv:2101.10221 [hep-ph]].  


\bibitem{Bengtsson:1986gz}
M.~Bengtsson, T.~Sjostrand and M.~van Zijl,
%``Initial State Radiation Effects on $W$ and Jet Production,''
Z. Phys. C \textbf{32} (1986) 67

\bibitem{Gieseke:2003rz}
S.~Gieseke, P.~Stephens and B.~Webber,
%``New formalism for QCD parton showers,''
JHEP \textbf{12} (2003) 045  
%doi:10.1088/1126-6708/2003/12/045
[arXiv:hep-ph/0310083 [hep-ph]].

\bibitem{Platzer:2011dq}
S.~Platzer and M.~Sjodahl,
%``The Sudakov Veto Algorithm Reloaded,''
Eur. Phys. J. Plus \textbf{127} (2012) 26
%doi:10.1140/epjp/i2012-12026-x
[arXiv:1108.6180 [hep-ph]].  

\bibitem{Nagy:2020gjv}
Z.~Nagy and D.~E.~Soper,
%``Evolution of parton showers and parton distribution functions,''
Phys. Rev. D \textbf{102} (2020)  014025
%doi:10.1103/PhysRevD.102.014025
[arXiv:2002.04125 [hep-ph]].

\bibitem{Nagy:2022bph}
Z.~Nagy and D.~E.~Soper,
%``Multivariable evolution in parton showers with initial state partons,''
Phys. Rev. D \textbf{106} (2022)  014024
%doi:10.1103/PhysRevD.106.014024
[arXiv:2204.05631 [hep-ph]].

\bibitem{Gellersen:2020} 
L. Gellersen, D. Napoletano, S. Prestel, Monte Carlo studies, in {\it 11th Les Houches Workshop on Physics at TeV Colliders: Les Houches 2019: Physics at TeV Colliders: Standard Model Working Group Report}, p. 131 (2020). Also in preprint arXiv:2003.01700.  

\bibitem{Sjostrand:2014zea}
T.~Sj{\"o}strand 
%, S.~Ask, J.~R. Christiansen, R.~Corke, N.~Desai, P.~Ilten
  et~al., 
%  \emph{{An introduction to PYTHIA 8.2}},
%  \href{https://doi.org/10.1016/j.cpc.2015.01.024}{\emph
{Comput. Phys. Commun.}
  {\bfseries 191} (2015) 159   [arXiv:1410.3012 [hep-ph]].
%  }
%  [\href{https://arxiv.org/abs/1410.3012}{{\ttfamily 1410.3012}}].


\bibitem{Sjostrand:2006za}
T.~Sjostrand, S.~Mrenna and P.~Z.~Skands,
%``PYTHIA 6.4 Physics and Manual,''
JHEP \textbf{05} (2006) 026
%doi:10.1088/1126-6708/2006/05/026
[arXiv:hep-ph/0603175 [hep-ph]].


\bibitem{Bellm:2015jjp}
J.~Bellm et~al., 
%\emph{{Herwig 7.0/Herwig++ 3.0 release note}},
%  \href{https://doi.org/10.1140/epjc/s10052-016-4018-8}{\emph{Eur. Phys. J.}
%  {\bfseries C76} (2016) 196}
%  [\href{https://arxiv.org/abs/1512.01178}{{\ttfamily 1512.01178}}].
  {Eur.\ Phys.\ J.}
  {\bfseries C76} (2016) 196   [arXiv:1512.01178 [hep-ph]]. 

\bibitem{Corcella:2002jc}
G.~Corcella     \textit{et al.},  hep-ph/0210213.



\bibitem{Yang:2022qgk}
H.~Yang   \textit{et al.},  
%``Back-to-back azimuthal correlations in $Z+$jet events at high transverse momentum in the TMD parton branching method at next-to-leading order,''
arXiv:2204.01528 [hep-ph].


\bibitem{Martinez:2018jxt}
  A.~Bermudez Martinez   {\it et al.},
  %``Collinear and TMD parton densities from fits to precision DIS measurements in the parton branching method,''
  Phys.\ Rev.\ D {\bf 99} (2019)   074008
%  doi:10.1103/PhysRevD.99.074008
  [arXiv:1804.11152 [hep-ph]].

\bibitem{Abramowicz:2015mha}
%\hrefCMSnoop {}
ZEUS, H1 Collaboration, 
%``{Combination of measurements of
%  inclusive deep inelastic ${e^{\pm }p}$ scattering cross sections and QCD
%  analysis of HERA data}'',
  Eur. Phys. J.  C {\bf 75}    (2015)   580 
   [arXiv:1506.06042 [hep-ph]].
%%CITATION = ARXIV:1506.06042;%%.

\bibitem{xFitter:2022zjb}
H.~Abdolmaleki \textit{et al.} [xFitter],
%``xFitter: An Open Source QCD Analysis Framework. A resource and reference document for the Snowmass study,''
arXiv:2206.12465 [hep-ph].



\bibitem{Alekhin:2014irh}
  S.~Alekhin {\it et al.},
  %``HERAFitter,''
  Eur.\ Phys.\ J.\ C {\bf 75} 
        (2015)    304   [arXiv:1410.4412 [hep-ph]].

\bibitem{Hautmann:2014uua}
  F.~Hautmann, H.~Jung and S.~Taheri Monfared,
  %``The CCFM uPDF evolution uPDFevolv Version 1.0.00,''
  Eur.\ Phys.\ J.\ C {\bf 74} 
 (2014) 3082  [arXiv:1407.5935 [hep-ph]]. 


\bibitem{Alwall:2014hca}
  J.~Alwall {\it et al.},
  %``The automated computation of tree-level and next-to-leading order differential cross sections, and their matching to parton shower simulations,''
  JHEP {\bf 1407} (2014) 079
%  doi:10.1007/JHEP07(2014)079
  [arXiv:1405.0301 [hep-ph]].

\bibitem{Martinez:2019mwt}
  A.~Bermudez Martinez   {\it et al.},
  %``Production of Z-bosons in the parton branching method,''
  Phys.\ Rev.\ D {\bf 100} (2019)   074027
%  doi:10.1103/PhysRevD.99.074008
  [arXiv:1906.00919 [hep-ph]].

\bibitem{Martinez:2020fzs}
  A.~Bermudez Martinez {\it et al.},
  %``The transverse momentum spectrum of low mass Drell-Yan production at next-to-leading order in the parton branching method,''
  Eur.\ Phys.\ J.\ C {\bf 80} (2020)   598
%  doi:10.1140/epjc/s10052-020-8136-y
  [arXiv:2001.06488 [hep-ph]].



\bibitem{Aad:2015auj}
  G.~Aad {\it et al.} [ATLAS Collaboration],
% ``Measurement of the transverse momentum
%  and $\phi ^*_{\eta }$ distributions of Drell--Yan lepton pairs in
%  proton--proton collisions at $\sqrt{s}=8$ TeV with the ATLAS detector''
  Eur.\ Phys.\ J.\ C {\bf 76} (2016)   291
%  
  [arXiv:1512.02192 [hep-ex]].


\bibitem{Sirunyan:2019bzr}
A.M.~Sirunyan  {\it et al.} [CMS Collaboration],
%{\scshape CMS} collaboration, 
%\emph{{Measurements of differential Z boson
%  production cross sections in proton-proton collisions at $ \sqrt{s} $ = 13
%  TeV}}, \href{https://doi.org/10.1007/JHEP12(2019)061}{\emph
  {JHEP} {\bfseries  
  12} (2019) 061    [arXiv:1909.04133 [hep-ex]].
%  } [\href{https://arxiv.org/abs/1909.04133}{{\ttfamily  1909.04133}}].

\bibitem{Aidala:2018ajl}
C.~Aidala {\it et al.} [PHENIX],
%``Measurements of $\mu\mu$ pairs from open heavy flavor and Drell-Yan in $p+p$ collisions at $\sqrt{s}=200$ GeV,''
Phys.\ Rev.\ D \textbf{99} (2019)  072003
%doi:10.1103/PhysRevD.99.072003
[arXiv:1805.02448 [hep-ex]].

\bibitem{Antreasyan:1981eg}
D.~Antreasyan et~al., 
%\emph{{Dimuon Scaling Comparison at 44-{GeV} and  62-{GeV}}}, \href{https://doi.org/10.1103/PhysRevLett.48.302}{\emph
{Phys.
  Rev. Lett.} {\bfseries 48} (1982) 302. 
%  }.

\bibitem{Webb:2003ps}
{\scshape NuSea} collaboration, 
%\emph{{Absolute Drell-Yan dimuon cross sections  in 800-GeV/c p p and p d collisions}},
%  \href{https://arxiv.org/abs/hep-ex/0302019}{{\ttfamily hep-ex/0302019}}.
hep-ex/0302019. 

\bibitem{Webb:2003bj}
J.~C. Webb, 
%\emph{{Measurement of continuum dimuon production in 800-GeV/c  proton nucleon collisions}},
%  \href{https://arxiv.org/abs/hep-ex/0301031}{{\ttfamily hep-ex/0301031}}.
hep-ex/0301031. 


 
 \bibitem{Abdulhamid:2021xtt}
M.~I.~Abdulhamid  \textit{et al.}, 
%``Azimuthal correlations of high transverse momentum jets at next-to-leading order in the parton branching method,''
Eur. Phys. J. C \textbf{82} (2022)  36
%doi:10.1140/epjc/s10052-022-09997-1
[arXiv:2112.10465 [hep-ph]].


\bibitem{CMS:2017cfb}
A.~M.~Sirunyan \textit{et al.} [CMS],
%``Azimuthal correlations for inclusive 2-jet, 3-jet, and 4-jet events in pp collisions at $\sqrt{s}= $ 13 TeV,''
Eur. Phys. J. C \textbf{78} (2018)   566
%doi:10.1140/epjc/s10052-018-6033-4
[arXiv:1712.05471 [hep-ex]].

\bibitem{CMS:2019joc}
A.~M.~Sirunyan \textit{et al.} [CMS],    
%``Azimuthal separation in nearly back-to-back jet topologies in inclusive 2- and 3-jet events in pp collisions at $\sqrt{s}=$ 13 TeV,''
Eur. Phys. J. C \textbf{79} (2019)  773
%doi:10.1140/epjc/s10052-019-7276-4
[arXiv:1902.04374 [hep-ex]].

 
\bibitem{Martinez:2022gsz}
A.~Bermudez Martinez and A.~Vladimirov,
%``Determination of Collins-Soper kernel from cross-sections ratios,''
arXiv:2206.01105 [hep-ph].   
   

\bibitem{Hautmann:2021ovt}
F.~Hautmann, I.~Scimemi and A.~Vladimirov,
%``Determination of the rapidity evolution kernel from Drell-Yan data at low transverse momenta,''
arXiv:2109.12051 [hep-ph].


\bibitem{Collins:1981va}
J.~C.~Collins and D.~E.~Soper,
%``Back-To-Back Jets: Fourier Transform from B to K-Transverse,''
Nucl. Phys. B \textbf{197} (1982) 446. 

           

\bibitem{Hautmann:2020cyp}
F.~Hautmann, I.~Scimemi and A.~Vladimirov,
%``Non-perturbative contributions to vector-boson transverse momentum spectra in hadronic collisions,''
Phys. Lett. B \textbf{806} (2020)  135478
%doi:10.1016/j.physletb.2020.135478
[arXiv:2002.12810 [hep-ph]].


\bibitem{Ebert:2019tvc}
M.~A.~Ebert, I.~W.~Stewart and Y.~Zhao,
%``Renormalization and Matching for the Collins-Soper Kernel from Lattice QCD,''
JHEP \textbf{03} (2020) 099
%doi:10.1007/JHEP03(2020)099
[arXiv:1910.08569 [hep-ph]].



\bibitem{Hautmann:2022xuc}
F.~Hautmann, M.~Hentschinski, L.~Keersmaekers, A.~Kusina, K.~Kutak and A.~Lelek,
%``A parton branching with transverse momentum dependent splitting functions,''
Phys. Lett. B \textbf{833} (2022)  137276   
[arXiv:2205.15873 [hep-ph]].


\bibitem{Catani:1994sq}
S.~Catani and F.~Hautmann,
%``High-energy factorization and small x deep inelastic scattering beyond leading order,''
Nucl. Phys. B \textbf{427} (1994) 475    
%doi:10.1016/0550-3213(94)90636-X
[arXiv:hep-ph/9405388 [hep-ph]].


\bibitem{Jung:2010si}
H.~Jung   {\it et al.},  %in preparation; 
%\emph{{The CCFM  Monte Carlo generator CASCADE version 2.2.03}},
%  \href{https://doi.org/10.1140/epjc/s10052-010-1507-z}{\emph
  {Eur.\ Phys.\ J.}
  {\bfseries C70} (2010) 1237   [arXiv:1008.0152 [hep-ph]]. 
%  }
%  [\href{https://arxiv.org/abs/1008.0152}{{\ttfamily 1008.0152}}].

\bibitem{Jung:2000hk}
H.~Jung and G.~P.~Salam,
%``Hadronic final state predictions from CCFM: The Hadron level Monte Carlo generator CASCADE,''
Eur. Phys. J. C \textbf{19} (2001)  351  
%doi:10.1007/s100520100604
[arXiv:hep-ph/0012143 [hep-ph]].

\bibitem{Andersen:2011zd}
J.~R.~Andersen, L.~Lonnblad and J.~M.~Smillie,
%``A Parton Shower for High Energy Jets,''
JHEP \textbf{07} (2011) 110
%doi:10.1007/JHEP07(2011)110
[arXiv:1104.1316 [hep-ph]].

\bibitem{Hoeche:2007hlb}
S.~Hoeche, F.~Krauss and T.~Teubner,
%``Multijet events in the $k_{T}$ -factorisation scheme,''
Eur. Phys. J. C \textbf{58} (2008) 17 
%doi:10.1140/epjc/s10052-008-0735-y
[arXiv:0705.4577 [hep-ph]].

\bibitem{Golec-Biernat:2007tjf}
K.~J.~Golec-Biernat, S.~Jadach, W.~Placzek, P.~Stephens and M.~Skrzypek,
%``Markovian Monte Carlo solutions of the one-loop CCFM equations,''
Acta Phys. Polon. B \textbf{38} (2007) 3149 
[arXiv:hep-ph/0703317 [hep-ph]].

\bibitem{Chachamis:2015zzp}
G.~Chachamis and A.~Sabio Vera,
%``Monte Carlo study of double logarithms in the small x region,''
Phys. Rev. D \textbf{93} (2016)  074004
%doi:10.1103/PhysRevD.93.074004
[arXiv:1511.03548 [hep-ph]].

\bibitem{Andersson:1995jt}
B.~Andersson, G.~Gustafson, H.~Kharraziha and J.~Samuelsson,
%``Structure Functions and General Final State Properties in the Linked Dipole Chain Model,''
Z. Phys. C \textbf{71} (1996) 613. 

\bibitem{Orr:1997im}
L.~H.~Orr and W.~J.~Stirling,
%``Dijet production at hadron hadron colliders in the BFKL approach,''
Phys. Rev. D \textbf{56} (1997) 5875  
%doi:10.1103/PhysRevD.56.5875
[arXiv:hep-ph/9706529 [hep-ph]].

\bibitem{Schmidt:1996fg}
C.~R.~Schmidt,
%``A Monte Carlo solution to the BFKL equation,''
Phys. Rev. Lett. \textbf{78} (1997) 4531 
%doi:10.1103/PhysRevLett.78.4531
[arXiv:hep-ph/9612454 [hep-ph]].



\bibitem{Abdulov:2021ivr}
N.~A.~Abdulov   \textit{et al.},   
%``TMDlib2 and TMDplotter: a platform for 3D hadron structure studies,''
Eur. Phys. J. C \textbf{81} (2021)   752
%doi:10.1140/epjc/s10052-021-09508-8
[arXiv:2103.09741 [hep-ph]].

\bibitem{Hautmann:2014kza}
F.~Hautmann, H.~Jung, M.~Kr\"amer, P.~J.~Mulders, E.~R.~Nocera, T.~C.~Rogers and A.~Signori,
%``TMDlib and TMDplotter: library and plotting tools for transverse-momentum-dependent parton distributions,''
Eur. Phys. J. C \textbf{74} (2014)     3220  
%doi:10.1140/epjc/s10052-014-3220-9
[arXiv:1408.3015 [hep-ph]].    

\bibitem{Alwall:2006yp}
J.~Alwall 
%A.~Ballestrero, P.~Bartalini, S.~Belov, E.~Boos, A.~Buckley, J.~M.~Butterworth, L.~Dudko, S.~Frixione and L.~Garren, 
\textit{et al.},  
%``A Standard format for Les Houches event files,''
Comput. Phys. Commun. \textbf{176} (2007) 300
%doi:10.1016/j.cpc.2006.11.010
[arXiv:hep-ph/0609017 [hep-ph]].



%\cite{Cacciari:2008gp}
\bibitem{Cacciari:2008gp}
M.~Cacciari, G.~P.~Salam and G.~Soyez,
%``The anti-$k_t$ jet clustering algorithm,''
JHEP \textbf{04} (2008), 063
%doi:10.1088/1126-6708/2008/04/063
[arXiv:0802.1189 [hep-ph]].


\bibitem{Catani:1993hr}
  S.~Catani, Y.~L.~Dokshitzer, M.~H.~Seymour and B.~R.~Webber,
  %``Longitudinally invariant $K_t$ clustering algorithms for hadron hadron collisions,''
  Nucl.\ Phys.\ B {\bf 406} (1993) 187.
   

\bibitem{Ellis:1993tq}
S.~D.~Ellis and D.~E.~Soper,
Phys.\ Rev.\ D {\bf 48} (1993) 3160.

 
   
\bibitem{Aaboud:2017hbk}
  M.~Aaboud {\it et al.} [ATLAS Collaboration],
  %``Measurements of the production cross section of a $Z$ boson in association with jets in pp collisions at $\sqrt{s} = 13$  TeV with the ATLAS detector,''
  Eur.\ Phys.\ J.\ C {\bf 77} (2017)   361
%  
  [arXiv:1702.05725 [hep-ex]].
%\cite{Khachatryan:2016crw}
\bibitem{Khachatryan:2016crw}
V.~Khachatryan \textit{et al.} [CMS],
%``Measurements of differential production cross sections for a Z boson in association with jets in pp collisions at $ \sqrt{s}=8 $ TeV,''
JHEP \textbf{04} (2017)    022
%doi:10.1007/JHEP04(2017)022
[arXiv:1611.03844 [hep-ex]].


%\cite{Sirunyan:2018cpw}
\bibitem{Sirunyan:2018cpw}
A.~M.~Sirunyan \textit{et al.} [CMS],
%``Measurement of differential cross sections for Z boson production in association with jets in proton-proton collisions at $\sqrt{s} =$ 13 TeV,''
Eur. Phys. J. C \textbf{78} (2018)  965
%doi:10.1140/epjc/s10052-018-6373-0
[arXiv:1804.05252 [hep-ex]].  



%\cite{AbellanBeteta:2016ugk}
\bibitem{AbellanBeteta:2016ugk}
R.~Aaij \textit{et al.} [LHCb],
%``Measurement of forward $W$ and $Z$ boson production in association with jets in proton-proton collisions at $\sqrt{s}=8$ TeV,''
JHEP \textbf{05} (2016)   131
%doi:10.1007/JHEP05(2016)131
[arXiv:1605.00951 [hep-ex]].


   %\cite{Vesterinen:2008hx,Banfi:2010cf}
\bibitem{Vesterinen:2008hx}
M.~Vesterinen and T.~R.~Wyatt,
%``A Novel Technique for Studying the Z Boson Transverse Momentum Distribution at Hadron Colliders,''
Nucl. Instrum. Meth. A \textbf{602} (2009) 432  
%doi:10.1016/j.nima.2009.01.203
[arXiv:0807.4956 [hep-ex]].


%\cite{Banfi:2010cf}
\bibitem{Banfi:2010cf}
A.~Banfi, S.~Redford, M.~Vesterinen, P.~Waller and T.~R.~Wyatt,
%``Optimisation of variables for studying dilepton transverse momentum distributions at hadron colliders,''
Eur. Phys. J. C \textbf{71} (2011) 1600 
%doi:10.1140/epjc/s10052-011-1600-y
[arXiv:1009.1580 [hep-ex]].


%\bibitem{Aad:2015auj}
%{\scshape ATLAS} collaboration, \emph{{Measurement of the transverse momentum
%  and $\phi ^*_{\eta }$ distributions of Drell--Yan lepton pairs in
%  proton--proton collisions at $\sqrt{s}=8$ TeV with the ATLAS detector}},
%  \href{https://doi.org/10.1140/epjc/s10052-016-4070-4}{\emph{Eur. Phys. J.}
%  {\bfseries C76} (2016) 291}
%  [\href{https://arxiv.org/abs/1512.02192}{{\ttfamily 1512.02192}}].



%\bibitem{Alwall:2014hca}
%J.~Alwall, R.~Frederix, S.~Frixione, V.~Hirschi, F.~Maltoni et~al., \emph{{The
%  automated computation of tree-level and next-to-leading order differential
%  cross sections, and their matching to parton shower simulations}},
%  \href{https://doi.org/10.1007/JHEP07(2014)079}{\emph{JHEP} {\bfseries 1407}
%  (2014) 079} [\href{https://arxiv.org/abs/1405.0301}{{\ttfamily 1405.0301}}].
%
%\bibitem{Frederix:2015eii}
%R.~Frederix, S.~Frixione, A.~Papaefstathiou, S.~Prestel and P.~Torrielli,
%  \emph{{A study of multi-jet production in association with an electroweak
%  vector boson}}, \href{https://doi.org/10.1007/JHEP02(2016)131}{\emph{JHEP}
%  {\bfseries 02} (2016) 131}
%  [\href{https://arxiv.org/abs/1511.00847}{{\ttfamily 1511.00847}}].
%
%\bibitem{Martinez:2019mwt}
%A.~Bermudez~Martinez et~al., \emph{{Production of Z-bosons in the parton
%  branching method}},
%  \href{https://doi.org/10.1103/PhysRevD.100.074027}{\emph{Phys. Rev.}
%  {\bfseries D100} (2019) 074027}
%  [\href{https://arxiv.org/abs/1906.00919}{{\ttfamily 1906.00919}}].
%
%\bibitem{Hautmann:2017xtx}
%F.~Hautmann, H.~Jung, A.~Lelek, V.~Radescu and R.~Zlebcik, \emph{{Soft-gluon
%  resolution scale in QCD evolution equations}},
%  \href{https://doi.org/10.1016/j.physletb.2017.07.005}{\emph{Phys. Lett.}
%  {\bfseries B772} (2017) 446}
%  [\href{https://arxiv.org/abs/1704.01757}{{\ttfamily 1704.01757}}].
%
%\bibitem{Hautmann:2017fcj}
%F.~Hautmann, H.~Jung, A.~Lelek, V.~Radescu and R.~Zlebcik, \emph{{Collinear and
%  TMD quark and gluon densities from Parton Branching solution of QCD evolution
%  equations}}, \href{https://doi.org/10.1007/JHEP01(2018)070}{\emph{JHEP}
%  {\bfseries 01} (2018) 070}
%  [\href{https://arxiv.org/abs/1708.03279}{{\ttfamily 1708.03279}}].
%
%\bibitem{Angeles-Martinez:2015sea}
%R.~Angeles-Martinez et~al., \emph{{Transverse Momentum Dependent (TMD) parton
%  distribution functions: status and prospects}},
%  \href{https://doi.org/10.5506/APhysPolB.46.2501}{\emph{Acta Phys. Polon.}
%  {\bfseries B46} (2015) 2501}
%  [\href{https://arxiv.org/abs/1507.05267}{{\ttfamily 1507.05267}}].
%
%\bibitem{Martinez:2018jxt}
%A.~Bermudez~Martinez, P.~Connor, F.~Hautmann, H.~Jung, A.~Lelek, V.~Radescu
%  et~al., \emph{{Collinear and TMD parton densities from fits to precision DIS
%  measurements in the parton branching method}},
%  \href{https://doi.org/10.1103/PhysRevD.99.074008}{\emph{Phys. Rev.}
%  {\bfseries D99} (2019) 074008}
%  [\href{https://arxiv.org/abs/1804.11152}{{\ttfamily 1804.11152}}].
%
%\bibitem{Marchesini:1987cf}
%G.~Marchesini and B.~R. Webber, \emph{{Monte Carlo Simulation of General Hard
%  Processes with Coherent QCD Radiation}},
%  \href{https://doi.org/10.1016/0550-3213(88)90089-2}{\emph{Nucl. Phys.}
%  {\bfseries B310} (1988) 461}.
%
%\bibitem{Catani:1990rr}
%S.~Catani, B.~R. Webber and G.~Marchesini, \emph{{QCD coherent branching and
%  semiinclusive processes at large x}},
%  \href{https://doi.org/10.1016/0550-3213(91)90390-J}{\emph{Nucl. Phys.}
%  {\bfseries B349} (1991) 635}.
%
%\bibitem{Alekhin:2014irh}
%S.~Alekhin et~al., \emph{{HERAFitter}},
%  \href{https://doi.org/10.1140/epjc/s10052-015-3480-z}{\emph{Eur. Phys. J.}
%  {\bfseries C75} (2015) 304}
%  [\href{https://arxiv.org/abs/1410.4412}{{\ttfamily 1410.4412}}].
%
%\bibitem{Hautmann:2014uua}
%F.~Hautmann, H.~Jung and S.~T. Monfared, \emph{{The CCFM uPDF evolution
%  uPDFevolv}}, {\emph{Eur. Phys. J.} {\bfseries C74} (2014) 3082}
%  [\href{https://arxiv.org/abs/1407.5935}{{\ttfamily 1407.5935}}].
%
%\bibitem{Abramowicz:2015mha}
%{\scshape ZEUS, H1} collaboration, \emph{{Combination of measurements of
%  inclusive deep inelastic ${e^{\pm }p}$ scattering cross sections and QCD
%  analysis of HERA data}},
%  \href{https://doi.org/10.1140/epjc/s10052-015-3710-4}{\emph{Eur. Phys. J.}
%  {\bfseries C75} (2015) 580}
%  [\href{https://arxiv.org/abs/1506.06042}{{\ttfamily 1506.06042}}].


\bibitem{Buckley:2010ar}
A.~Buckley   
%, J.~Butterworth, L.~Lonnblad, D.~Grellscheid, H.~Hoeth, J.~Monk
  et~al., 
%  \emph{{Rivet user manual}},
%  \href{https://doi.org/10.1016/j.cpc.2013.05.021}{\emph
  {Comput. Phys. Commun.}
  {\bfseries 184} (2013) 2803     [arXiv:1003.0694 [hep-ph]]. 
%  }
%  [\href{https://arxiv.org/abs/1003.0694}{{\ttfamily 1003.0694}}].


%\bibitem{Hautmann:2019biw}
%F.~Hautmann, L.~Keersmaekers, A.~Lelek and A.~M. Van~Kampen, \emph{{Dynamical
%  resolution scale in transverse momentum distributions at the LHC}},   {\emph{Nucl. Phys.}
%  {\bfseries B949} (2019) 114795}
%  [\href{https://arxiv.org/abs/1908.08524}{{\ttfamily 1908.08524}}].


%\bibitem{Khachatryan:2016iob}
%  V.~Khachatryan {\it et al.} [CMS Collaboration],
%Eur.\ Phys.\ J.\ C {\bf 77} (2017)  751
%doi:10.1140/epjc/s10052-017-4900-z
%[arXiv:1611.06507 [hep-ex]].


\bibitem{CMS:2022ubq}
 CMS Collaboration,    
%``Measurement of the mass dependence of the transverse momentum of lepton pairs in Drell-Yan production in proton-proton collisions at $\sqrt{s}$ = 13 TeV,''
arXiv:2205.04897 [hep-ex].  


\bibitem{CMS:2015wcf}
V.~Khachatryan \textit{et al.} [CMS],
%``Event generator tunes obtained from underlying event and multiparton scattering measurements,''
Eur. Phys. J. C \textbf{76} (2016)  155
%doi:10.1140/epjc/s10052-016-3988-x
[arXiv:1512.00815 [hep-ex]].

\bibitem{CMS:2019csb}
A.~M.~Sirunyan \textit{et al.} [CMS],
%``Extraction and validation of a new set of CMS PYTHIA8 tunes from underlying-event measurements,''
Eur. Phys. J. C \textbf{80} (2020)  4  
%doi:10.1140/epjc/s10052-019-7499-4
[arXiv:1903.12179 [hep-ex]].   




\bibitem{Bury:2022czx}
M.~Bury   {\it et al.}, 
% , F.~Hautmann, S.~Leal-Gomez, I.~Scimemi, A.~Vladimirov and P.~Zurita,
%``PDF bias and flavor dependence in TMD distributions,''
arXiv:2201.07114 [hep-ph].  


\bibitem{Proceedings:2020eah}
Y.~Hatta  \textit{et al.}, 
%``Proceedings, Probing Nucleons and Nuclei in High Energy Collisions: Dedicated to the Physics of the Electron Ion Collider: Seattle (WA), United States, October 1 - November 16, 2018,''
%doi:10.1142/11684
arXiv:2002.12333 [hep-ph].  

\bibitem{LHeC:2020van}
P.~Agostini \textit{et al.} [LHeC and FCC-he Study Group],
%``The Large Hadron-Electron Collider at the HL-LHC,''
J. Phys. G \textbf{48} (2021)  110501
%doi:10.1088/1361-6471/abf3ba
[arXiv:2007.14491 [hep-ex]].


\bibitem{Ball:2012cx}
R.~D.~Ball  {\it et al.}, 
%``Parton distributions with LHC data,''
Nucl.\ Phys.\ B \textbf{867} (2013) 244  
%doi:10.1016/j.nuclphysb.2012.10.003
[arXiv:1207.1303 [hep-ph]].

\bibitem{Harland-Lang:2014zoa}
L.~A.~Harland-Lang, A.~D.~Martin, P.~Motylinski and R.~S.~Thorne,
%``Parton distributions in the LHC era: MMHT 2014 PDFs,''
Eur.\ Phys.\ J.\ C \textbf{75} (2015)  204
%doi:10.1140/epjc/s10052-015-3397-6
[arXiv:1412.3989 [hep-ph]].   


\bibitem{vanHameren:2016kkz}
A.~van Hameren,
%``KaTie : For parton-level event generation with $k_T$-dependent initial states,''
Comput. Phys. Commun. \textbf{224} (2018) 371  
%doi:10.1016/j.cpc.2017.11.005
[arXiv:1611.00680 [hep-ph]].


\bibitem{Catani:1990xk}
S.~Catani, M.~Ciafaloni and F.~Hautmann,
%``Gluon contributions to small x heavy flavor production,''
Phys. Lett. B \textbf{242} (1990) 97. 
 
 
\bibitem{Lipatov:1976zz}
L.~N.~Lipatov,
%``Reggeization of the Vector Meson and the Vacuum Singularity in Nonabelian Gauge Theories,''
Sov. J. Nucl. Phys. \textbf{23} (1976) 338. 
 
\bibitem{Kuraev:1977fs}
E.~A.~Kuraev, L.~N.~Lipatov and V.~S.~Fadin,
%``The Pomeranchuk Singularity in Nonabelian Gauge Theories,''
Sov. Phys. JETP \textbf{45} (1977) 199.
         
\bibitem{Balitsky:1978ic}
I.~I.~Balitsky and L.~N.~Lipatov,
%``The Pomeranchuk Singularity in Quantum Chromodynamics,''
Sov. J. Nucl. Phys. \textbf{28} (1978) 822. 

\bibitem{Catani:1992zc}
S.~Catani, M.~Ciafaloni and F.~Hautmann,
%``Production of heavy flavors at high-energies,''
CERN-TH-6398-92, in Proc. ``Physics at Hera'' (Hamburg, 1991), p.690. 

\bibitem{Marchesini:1992jw}
G.~Marchesini and B.~R.~Webber,
%``Final states in heavy quark leptoproduction at small x,''
Nucl. Phys. B \textbf{386} (1992) 215. 

\end{thebibliography}

\providecommand{\href}[2]{#2}\begingroup\raggedright\endgroup

\end{document}